\documentclass[11pt]{article}

\usepackage{color}
\usepackage{booktabs}       
\usepackage{amsfonts}       
\usepackage{graphics}
\usepackage{pgfplots}
\usepackage{pgfplotstable}
\usepgfplotslibrary{fillbetween}
\usepackage{tikz}
\usepackage{bm}
\usepackage[separate-uncertainty=true,detect-mode=true]{siunitx}    
\usepackage{color}
\usepackage{subcaption}
\usepackage[backend=bibtex,style=ieee,sortcites,doi=false,isbn=false,url=false,maxcitenames=1,mincitenames=1,citestyle=numeric-comp]{biblatex}
\usepackage{algorithm}
\usepackage{algorithmic}
\usepackage{mathtools}
\usepackage{amssymb}
\usepackage{amsmath}
\usepackage{standalone}
\usepackage[keeplastbox]{flushend}
\usepackage[capitalize]{cleveref}
\usepackage{graphicx}
\usepackage{rotating}
\usepackage{pdflscape}
\usepackage{multirow}
\usepackage{microtype}
\usepackage[toc,page]{appendix}
\usepackage{longtable,tabularx}
\usepackage[margin=1.25in,headsep=.4in]{geometry}

\usepackage{pifont}

\usetikzlibrary{arrows}
\usetikzlibrary{arrows.meta}
\usetikzlibrary{bayesnet}
\usetikzlibrary{fit}
\usetikzlibrary{matrix}
\usetikzlibrary{positioning}
\usetikzlibrary{intersections}
\usetikzlibrary{graphs}
\usetikzlibrary{calc}
\usetikzlibrary{patterns}
\usetikzlibrary{decorations.pathreplacing}
\usepgfplotslibrary{groupplots}

\pgfplotsset{compat=newest}

\pgfplotsset{every axis legend/.append style={legend cell align=left}}

\pgfmathdeclarefunction{gauss}{2}{%
  \pgfmathparse{1/(#2*sqrt(2*pi))*exp(-((x-#1)^2)/(2*#2^2))}%
}

\tikzset{
      value/.style={align=center, minimum height=8mm, minimum width=8mm, font={\small}, circle, draw=black},
      funct/.style={align=center, minimum height=8mm, minimum width=8mm, font={\small}, rectangle, draw=black, fill=white},
      arrow/.style={solid, thick, ->,>=stealth',shorten >=1pt},
      bracket/.style={decoration={brace},decorate, thick},
      lossLine/.style={dashed, red, thick, ->,>=stealth',shorten >=1pt},
      snake it/.style={decorate, decoration={snake, amplitude = .4mm, segment length = 2mm, post length=2mm}, thick, ->,>=stealth',shorten >=1pt}
    }

\definecolor{colorBluish}{RGB}{101,161,216}
\definecolor{colorBluePastel}{RGB}{240,240,255}
\definecolor{colorGreenish}{RGB}{0,130,0}
\definecolor{colorGreenPastel}{RGB}{240,250,240}
\definecolor{colorRedish}{RGB}{140,21,21}
\definecolor{colorRedPastel}{RGB}{255,240,240}
\definecolor{colorOrangish}{RGB}{237,125,49}
\definecolor{colorOrangePastel}{RGB}{213,169,143}
\definecolor{colorPurplePastel}{RGB}{216,191,216}

\newcommand{\T}{{^\intercal}}
\providecommand{\e}[1]{\ensuremath{\times 10^{#1}}}

\newcommand{\R}{\mathbb{R}}

\DeclareMathOperator*{\E}{\mathbb{E}}

\newcommand{\x}{\mathbf{x}}
\newcommand{\y}{\mathbf{y}}
\newcommand{\w}{\mathbf{w}}
\newcommand{\cs}{\mathbf{c}}

\newcommand{\z}{\mathbf{z}}
\newcommand*{\equal}{=}

\title{Parameter-Conditioned Sequential Generative Modeling of Fluid Flows}

 \author{
  Jeremy Morton\thanks{Ph.D. Candidate, Department of Aeronautics and Astronautics, jmorton2@stanford.edu} \\
  {\normalsize\itshape Stanford University, Stanford, CA 94305}
  \and 
  Freddie D. Witherden \thanks{Assistant Professor, Department of Ocean Engineering, fdw@tamu.edu} \\
  {\normalsize\itshape Texas A \& M University, College Station, TX 77843}
  \and
  Mykel J. Kochenderfer\thanks{Assistant Professor, Department of Aeronautics and Astronautics, mykel@stanford.edu} \\
  {\normalsize\itshape Stanford University, Stanford, CA 94305}
 }

\begin{document}

\maketitle

\begin{abstract}
The computational cost associated with simulating fluid flows can make it infeasible to run many simulations across multiple flow conditions.
Building upon concepts from generative modeling, we introduce a new method for learning neural network models capable of performing efficient parameterized simulations of fluid flows.
Evaluated on their ability to simulate both two-dimensional and three-dimensional fluid flows, trained models are shown to capture local and global properties of the flow fields at a wide array of flow conditions.
Furthermore, flow simulations generated by the trained models are shown to be orders of magnitude faster than the corresponding computational fluid dynamics simulations.
\end{abstract}

\section{Introduction}
Computational fluid dynamics (CFD) simulations (and other simulations that involve physics-based models) often require fine spatial and temporal discretizations to achieve high levels of fidelity.
These fine discretizations can make the corresponding simulations expensive to run due to the large number of degrees of freedom that must be modeled.
In applications such as design, which can require running many simulations across multiple design configurations, this high computational cost can be prohibitive.
Instead of running expensive simulations across the entire design space, a technique known as \emph{surrogate modeling} is often used~\cite{sacks1989designs,queipo2005surrogate,forrester2009recent}, wherein models are trained to approximate the mapping between design parameters and a scalar-valued objective that evaluates design quality.
Surrogate models can be constructed from a limited amount of high-fidelity simulation data using techniques such as polynomial or Gaussian process regression, and allow for efficient evaluation of design objectives at new design points.
However, while surrogate modeling approaches can enable faster analysis and design, they do not reduce the computational cost of running high-fidelity simulations.

In place of surrogate models, which only attempt to model the mapping from design parameters to high-level objectives, a great deal of recent attention has focused on using learning-based techniques to facilitate more efficient high-fidelity simulation.
Many proposed techniques seek a low-dimensional, reduced representation of the modeled system that can be simulated more efficiently than the full system dynamics.
This generally requires first learning a mapping from the high-dimensional dynamical system states to reduced states, and subsequently devising a method for integrating the reduced states forward in time.
Projection-based approaches~\cite{willcox2002balanced,carlberg2011efficient} learn a mapping from high-dimensional states to a low-dimensional subspace, and then determine the reduced state dynamics through a projection onto known governing equations.
The learned mappings are generally obtained through proper orthogonal decomposition (POD), which yields linear embeddings, although some recent work has explored the use of nonlinear embeddings~\cite{lee2018model,lee2019deep}.
One drawback of projection-based approaches is that they must be integrated into existing simulation software, which can be challenging and time intensive~\cite{white2019neural}.

An alternative approach for determining the reduced-state dynamics is to \emph{learn} the dynamics directly from data.
Provided the learned models are accurate, such a data-driven approach requires no integration with existing simulation software and no knowledge of the system governing equations.
A number of recent works have applied two-stage (dimensionality reduction and dynamics learning) deep learning procedures to the task of modeling fluid flows~\cite{otto2017linearly,puligilla2018deep,morton2018deep,lusch2018deep,carlberg2019recovering,wiewel2019latent,kim2019deep}.
These approaches generally require complex multi-stage training procedures, and often can only be applied to simulation at a single flow condition.
In contrast, this paper proposes a new method, grounded in variational inference, for parameter-conditioned modeling of fluid flows, in which the state mapping and reduced state dynamics are learned \emph{jointly}.
The derived learning procedure allows the trained model to perform parameterized simulations, thus enabling the efficient simulation of modeled systems at a wide range of conditions.
To motivate the proposed method, the discussion now proceeds by reviewing relevant background material related to generative modeling.

\section{Generative Modeling}
Generative modeling refers to modeling the process by which data is generated in the real world~\cite{kingma2019introduction}.
As such, physical models like the Navier--Stokes equations can be viewed as a form of generative model.
In the absence of known governing equations, \emph{learned} generative models can attempt to model real-world systems by approximating the distribution over data, $p_\text{data}(\x)$.
Several powerful generative modeling paradigms have emerged in recent years, including variational autoencoders~\cite{kingma2013autoencoding}, generative adversarial networks~\cite{goodfellow2014generative}, and flow-based models~\cite{rezende2015variational,dinh2016density}.
The following section provides an introduction to variational autoencoders.

\subsection{Variational Autoencoders}\label{sec:vae}
The goal in generative modeling is to learn $p_{\bm{\theta}}(\x)$, a distribution over (possibly high-dimensional) data $\x \in \mathcal D$, where $\bm{\theta}$ represents the parameters that govern the learned distribution.
An accurate learned distribution will be close to the true data distribution, $p_\text{data}(\x)$, where the difference between the distributions can be measured through similarity measures such as the KL-divergence:
\begin{equation}
\mathcal D_\text{KL}(p_\text{data}(\x) \mid \mid p_{\bm{\theta}}(\x)) = \E_{\x \sim p_\text{data}(\x)} \left[\log \frac{p_\text{data}(\x)}{p_{\bm{\theta}}(\x)}   \right].
\end{equation}
It can be shown that minimizing the KL-divergence between these distributions is equivalent to maximizing the log-likelihood the model assigns to the data~\cite{kingma2019introduction}:
\begin{equation}
\log p_{\bm{\theta}}(\mathcal D) = \sum_{\x \in \mathcal D} \log p_{\bm{\theta}}(\x).
\end{equation}

Success in generative modeling relies upon a model that (1) can be optimized efficiently (i.e. the gradients $\nabla_{\bm{\theta}} \log p_{\bm{\theta}}(\x)$ are known and easy to calculate) and (2) is sufficiently expressive in order to accurately approximate the true data distribution.
For example, a joint Gaussian distribution over pixel color intensities could be used to represent a distribution over image data, but, if the true data distribution is non-Gaussian, then such a model would be incapable of representing $p_\text{data}(\x)$ even under the optimal parameter settings.

The expressiveness of modeled distributions can be enhanced by introducing \emph{latent variables}, denoted by $\z$.
Latent variables are unobserved, but are assumed to summarize high-level information about the data, and are often assumed to be (much) lower-dimensional than the data.
As an example, in image data the latent variable $\z$ may be taken to represent the class of an image (e.g. dog or human or car), and the data $\x$ would then represent a particular instantiation of that class.
Under the presence of latent variables, the modeled distribution becomes:
\begin{equation}
\begin{split}
p_{\bm{\theta}}(\x) & = \int p_{\bm{\theta}}(\x, \z) d\z\\
& = \int p_{\bm{\theta}}(\x \mid \z) p_{\bm{\theta}}(\z) d\z.
\label{eq:marg_int}
\end{split}
\end{equation}
When the distribution is modeled in this manner then, even if $p_{\bm{\theta}}(\x \mid \z)$ and $p_{\bm{\theta}}(\z)$ are Gaussian distributions, the resulting marginal distribution $p_{\bm{\theta}}(\x)$ can be quite complex.

Unfortunately, calculating (and in turn optimizing) the likelihood as defined in \cref{eq:marg_int} is generally intractable due to the difficulty of evaluating the integral over $\z$.
Additionally, finding the posterior over $\z$ given $\x$, defined as:
\begin{equation}
p_{\bm{\theta}}(\z \mid \x) = \frac{p_{\bm{\theta}}(\x \mid \z) p_{\bm{\theta}}(\z)}{p_{\bm{\theta}}(\x)}
\end{equation}
is also intractable due to the presence of $p_{\bm{\theta}}(\x)$ in the denominator of the expression.
In place of this intractable posterior distribution, variational inference~\cite{blei2017variational} introduces an approximate posterior distribution $q_{\bm{\phi}}(\z \mid \x)$, which is defined by a set of parameters $\bm{\phi}$.
Multiplying and dividing the likelihood expression by this approximate posterior distribution yields:
\begin{equation}
\begin{split}
\log p_{\bm{\theta}}(\x) & = \log \int{p_{\bm{\theta}}(\x \mid \z) p_{\bm{\theta}}(\z) \frac{q_{\bm{\phi}}(\z \mid \x)}{q_{\bm{\phi}}(\z \mid \x)}} d \z\\\\
& = \log \E_{\z \sim q_{\bm{\phi}}(\z \mid \x)} \left[\frac{p_{\bm{\theta}}(\x \mid \z) p_{\bm{\theta}}(\z)}{q_{\bm{\phi}}(\z \mid \x)}  \right].
\end{split}
\end{equation}
Through Jensen's inequality the following expression can be derived:
\begin{equation}
\begin{split}
\log p_{\bm{\theta}}(\x) & \geq \E_{\z \sim q_{\bm{\phi}}(\z \mid \x)} \left[\log p_{\bm{\theta}}(\x \mid \z) - \log \frac{q_{\bm{\phi}}(\z \mid \x)}{p_{\bm{\theta}}(\z)}    \right]\\
& \geq \E_{\z \sim q_{\bm{\phi}}(\z \mid \x)} \log p_{\bm{\theta}}(\x \mid \z) - \mathcal D_\text{KL}(q_{\bm{\phi}}(\z \mid \x) \mid \mid p_{\bm{\theta}}(\z)).
\label{eq:elbo}
\end{split}
\end{equation}
The right-hand side of the above expression is often referred to as the \emph{evidence lower bound} (ELBO), and serves as a lower bound on the log-likelihood objective.
It can be shown that the ELBO will be equal to $\log p_{\bm{\theta}}(\x)$ when:
\begin{equation}
D_\text{KL}(q_{\bm{\phi}}(\z \mid \x) \mid \mid p_{\bm{\theta}}(\z \mid \x)) = 0,
\end{equation}
i.e. the approximate posterior distribution matches the true posterior distribution over $\z$.
Thus, optimizing the evidence lower bound with respect to parameters $\bm{\theta}$ and $\bm{\phi}$ serves the dual purpose of (1) approximately maximizing the true objective $\log p_{\bm{\theta}}(\x)$ and (2) driving the approximate posterior closer to the true posterior distribution~\cite{kingma2019introduction}.

The variational autoencoder (VAE)~\cite{kingma2013autoencoding} presents a method for performing variational inference with neural network models.
Much like standard autoencoders, VAEs consist of encoder and decoder components.
The encoder (or \emph{recognition model}) is denoted by $q_{\bm{\phi}}(\z \mid \x)$, and maps an input $\x$ to a distribution over the latent variable $\z$.
The decoder (or \emph{generative model}) is denoted by $p_{\bm{\theta}}(\x \mid \z)$, and maps samples from $q_{\bm{\phi}}(\z \mid \x)$ to a distribution over $\x$.
The outputs of the neural networks are the distribution parameters, which are often assumed to be Gaussian with diagonal or constant covariance matrices.

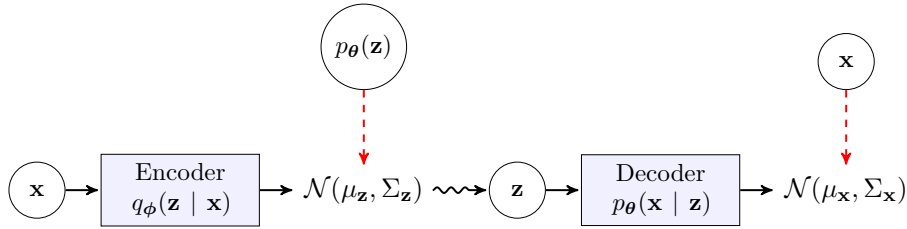
\begin{figure}[t]
\centering
\begin{tikzpicture}
        \node[value] (x) {$\x$};
        \node[funct] (encoder)[right=0.5 of x, fill=colorBluePastel, text width=2cm]   {Encoder $q_{\bm{\phi}}(\z \mid \x)$};
        \draw[arrow] (x) to (encoder);
        \node[draw=none, anchor=center] (dist_z) [right=0.5 of encoder] {$\mathcal{N}(\mu_{\z}, \Sigma_{\z})$};
        \draw[arrow] (encoder) to (dist_z);
        
        \node[value] (z)[right=0.8 of dist_z] {$\z$};
        \draw[snake it] (dist_z)  to (z);
        \node[funct] (decoder)[right=0.5 of z, fill=colorBluePastel, text width=2cm]   {Decoder $p_{\bm{\theta}}(\x \mid \z)$};
        \draw[arrow] (z) to (decoder);
        \node[draw=none, anchor=center] (dist_x) [right=0.5 of decoder] {$\mathcal{N}(\mu_{\x}, \Sigma_{\x})$};
        \draw[arrow] (decoder) to (dist_x);
        
        \node[value] (prior) [above=1.1 of dist_z] {$p_{\bm{\theta}}(\z)$};
        \node[value] (xt) [above=1.1 of dist_x] {$\x$};
        \draw[lossLine] (prior) to (dist_z);
        \draw[lossLine] (xt) to (dist_x);
\end{tikzpicture}
\caption[Illustration of VAE training procedure]{Illustration of the training procedure for the variational autoencoder. Red lines indicate relationships relevant to the optimization objective.}
\label{fig:train_vae}
\end{figure}

Training a variational autoencoder consists of optimizing the set of parameters $\bm{\theta}$ and $\bm{\phi}$ in order to maximize the evidence lower bound presented in \cref{eq:elbo}.
\Cref{fig:train_vae} illustrates the VAE training procedure.
Training inputs are fed into the encoder, which outputs parameters to a distribution over the latent variable $\z$.
Subsequently, samples from that distribution are mapped through the decoder to obtain a distribution over $\x$.
The optimization objective balances maximizing the likelihood assigned to $\x$ with minimizing the KL-divergence between the approximate posterior and the prior distribution.
For simplicity, the prior is often assumed to be a standard Gaussian distribution.
Evaluating the gradients with respect to the encoder parameters requires the use of a technique known as  the reparameterization trick~\cite{kingma2015variational}, which treats the encoder as a deterministic mapping subject to exogenous noise, and allows for the propagation of gradients through the sampling operation.
A VAE that has been trained to maximize the evidence lower bound can subsequently be employed in a generative fashion, where samples are drawn directly from the prior $p_{\bm{\theta}}(\z)$ and passed through the decoder.
This process can result in the generation of new data that appears as if it was drawn from $p_\text{data}(\x)$.

\subsection{Conditional Variational Autoencoders}
\begin{figure}[t]
\centering
\begin{subfigure}[b]{0.2\linewidth}
\centering
\begin{tikzpicture}
        \node[value] (z) {$\z$};
        \node[value] (x)[below=0.8 of z] {$\x$};
        \draw[arrow] (z) to (x);        
\end{tikzpicture}
\caption{VAE}
\label{fig:vae}
\end{subfigure}\hfill
\begin{subfigure}[b]{0.2\linewidth}
\centering
\begin{tikzpicture}
        \node[value] (z) {$\z$};
        \node[value] (x)[below=0.8 of z] {$\x$};
        \node[value] (c)[left=0.8 of z] {$\cs$};
        \draw[arrow] (z) to (x); 
        \draw[arrow] (c) to (x); 
        \draw[arrow] (c) to (z);        
\end{tikzpicture}
\caption{CVAE}
\label{fig:cvae}
\end{subfigure}
\begin{subfigure}[b]{0.55\linewidth}
\centering
\begin{tikzpicture}
        \node[value] (z1) {$\z_1$};
        \node[value] (x1)[below=0.8 of z1] {$\x_1$};
        \draw[arrow] (z1) to (x1); 

        \node[value] (z2) [right=0.8 of z1] {$\z_2$};
        \node[value] (x2)[below=0.8 of z2] {$\x_2$};
        \draw[arrow] (z2) to (x2); 

        \node[value] (dots) [right=0.8 of z2, draw=none] {$\cdots$};

        \node[value] (zT) [right=0.8 of dots] {$\z_T$};
        \node[value] (xT)[below=0.8 of zT] {$\x_T$};
        \draw[arrow] (zT) to (xT); 

        \draw[arrow] (z1) to (z2); 
        \draw[arrow] (z2) to (dots); 
        \draw[arrow] (dots) to (zT);    
\end{tikzpicture}
\caption{Sequential generative model}
\label{fig:seq}
\end{subfigure}
\caption[Graphical models]{Graphical models for VAE, CVAE, and sequential generative model.}
\end{figure}
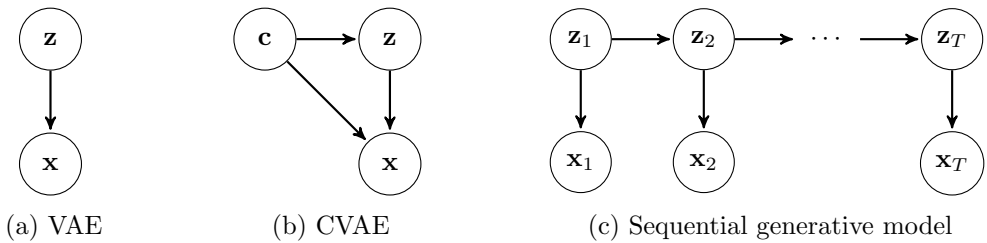

The conditional variational autoencoder (CVAE)~\cite{sohn2015learning} extends the VAE by attempting to learn $p_{\bm{\theta}}(\x \mid \cs)$, a \emph{conditional} model for the data distribution.
Rather than modeling the distribution over the entire dataset, a CVAE models data distributions conditioned on context parameters $\cs$.
The graphical model for the CVAE can be found in \cref{fig:cvae}.

As with the variational autoencoder, a lower bound to the log-likelihood objective can be derived for the CVAE:
\begin{equation}
\log p_{\bm{\theta}}(\x \mid \cs) \geq \E_{\z \sim q_{\bm{\phi}}(\z \mid \x, \cs)} \log p_{\bm{\theta}}(\x \mid \z, \cs) - \mathcal D_\text{KL}(q_{\bm{\phi}}(\z \mid \x, \cs) \mid \mid p_{\bm{\theta}}(\z \mid \cs)).
\end{equation}
Optimizing for this objective requires relatively few modifications to the VAE training procedure outlined in \cref{sec:vae}.
Most notably, the encoder and decoder networks must be modified to accept the context parameters as inputs.
Furthermore, the prior over $\z$ must be modeled as a \emph{conditional} prior distribution, which is typically accomplished by introducing an auxiliary neural network that is trained to take $\cs$ as an input and output the prior distribution parameters.

\subsection{Generative Modeling of Sequential Data}\label{sec:seq_gen}
Generative models for sequential data attempt to learn the joint distribution $p_{\bm{\theta}}(\x_{1:T})$, where $\x_1, \ldots, \x_T$ represent data with temporal relationships; here it is assumed that each $\x_t$ represents the state of a dynamical system at a given time step.
In constructing sequential generative models, it is common to assume the existence of (lower-dimensional) latent variables $\z_{1:T}$, also referred to as latent states, that govern the time evolution of the modeled system.
Furthermore, the dynamics of the latent states are often assumed to be \emph{Markovian}, meaning that a latent state $\z_t$ is conditionally independent of $\z_{1:t-2}$ given $\z_{t-1}$ (i.e. $\z_{t-1}$ summarizes all relevant past information).
The graphical model that defines the problem structure can be found in \cref{fig:seq}; a model of this form is commonly referred to as a \emph{state space model} (SSM)~\cite{durbin2012time}.

Analogous to the material presented in \cref{sec:vae}, the likelihood expression can be written in a form that explicitly accounts for the presence of the latent variables:
\begin{equation}
\begin{split}
p_{\bm{\theta}}(\x_{1:T}) &= \int p_{\bm{\theta}}(\x_{1:T}, \z_{1:T}) d \z_{1:T}\\
&= \int p_{\bm{\theta}}(\x_{1:T} \mid \z_{1:T}) p_{\bm{\theta}}(\z_{1:T}) d \z_{1:T}.
\end{split}
\end{equation}
Evaluating this integral is generally intractable due to the need to marginalize over the latent states.
Likewise, evaluating the posterior distribution $p_{\bm{\theta}}(\z_{1:T} \mid \x_{1:T})$ is generally intractable, once again motivating the introduction of an approximate posterior distribution $q_{\bm{\phi}}(\z_{1:T} \mid \x_{1:T})$.

Given the conditional independence assumptions encoded by the graphical model, the conditional distribution $p_{\bm{\theta}}(\x_{1:T} \mid \z_{1:T})$ can be rewritten as:
\begin{equation}
\begin{split}
p_{\bm{\theta}}(\x_{1:T} \mid \z_{1:T}) & = p_{\bm{\theta}}(\x_1 \mid \z_{1:T}) \prod_{t=2}^T p_{\bm{\theta}}(\x_t \mid \x_{1:t-1}, \z_{1:T})\\
& = \prod_{t=1}^T p_{\bm{\theta}}(\x_t \mid \z_t).
\end{split}
\end{equation}
Likewise, the distribution $p_{\bm{\theta}}(\z_{1:T})$ can be simplified to:
\begin{equation}
\begin{split}
p_{\bm{\theta}}(\z_{1:T}) & = p_{\bm{\theta}}(\z_1) \prod_{t=2}^T p_{\bm{\theta}}(\z_t \mid \z_{1:t-1})\\
& = p_{\bm{\theta}}(\z_1) \prod_{t=2}^T p_{\bm{\theta}}(\z_t \mid \z_{t-1}).
\end{split}
\end{equation}
Finally, the approximate posterior distribution can be expressed as:
\begin{equation}
\begin{split}
q_{\bm{\phi}}(\z_{1:T} \mid \x_{1:T}) & = q_{\bm{\phi}}(\z_1 \mid \x_{1:T}) \prod_{t=2}^T q_{\bm{\phi}}(\z_t \mid \z_{1:t-1}, \x_{1:T})\\
& = q_{\bm{\phi}}(\z_1 \mid \x_{1:T}) \prod_{t=2}^T q_{\bm{\phi}}(\z_t \mid \z_{t-1}, \x_{t:T}).
\end{split}
\end{equation}

Given these simplified expressions, a lower bound to the log-likelihood, $\log p_{\bm{\theta}}(\x_{1:T})$, can be derived by following the procedures outlined in \cref{sec:vae}.
This lower bound is given by:
\begin{equation}
\begin{split}
\log p_{\bm{\theta}}(\x_{1:T}) \geq \sum_{t=1}^T & \left[\E_{\z_t \sim q_{\bm{\phi}}} \log p_{\bm{\theta}}(\x_t \mid \z_t)\right] + \mathcal D_\text{KL}\left(q_{\bm{\phi}}(\z_1 \mid \x_{1:T}) \mid \mid p_{\bm{\theta}}(\z_1) \right)\\
& + \sum_{t=2}^T D_\text{KL}\left(q_{\bm{\phi}}(\z_t \mid \z_{t-1}, \x_{t:T}) \mid \mid p_{\bm{\theta}}(\z_t \mid \z_{t-1}) \right).
\label{eq:elbo_seq}
\end{split}
\end{equation}
The ELBO thus consists of $T$ likelihood components, representing the likelihood of states $\x_t$ given latent states $\z_t$, and $T$ KL-divergence components, representing the difference between the approximate posterior and (conditional) prior distributions.
If the latent states are assumed to follow non-Markovian dynamics, then the ELBO becomes:
\begin{equation}
\begin{split}
\log p_{\bm{\theta}}(\x_{1:T}) \geq \sum_{t=1}^T & \left[\E_{\z_t \sim q_{\bm{\phi}}} \log p_{\bm{\theta}}(\x_t \mid \z_t)\right] + \mathcal D_\text{KL}\left(q_{\bm{\phi}}(\z_1 \mid \x_{1:T}) \mid \mid p_{\bm{\theta}}(\z_1) \right)\\
& + \sum_{t=2}^T D_\text{KL}\left(q_{\bm{\phi}}(\z_t \mid \textcolor{blue}{\z_{1:t-1}}, \x_{t:T}) \mid \mid p_{\bm{\theta}}(\z_t \mid \textcolor{blue}{\z_{1:t-1}}) \right),
\label{eq:elbo_nomark}
\end{split}
\end{equation}
where the elements that differentiate \cref{eq:elbo_nomark} from \cref{eq:elbo_seq} are highlighted in blue.

As with the variational autoencoder, the prior distribution $p_{\bm{\theta}}(\z_1)$ can be defined to be a standard Gaussian.
Additionally, the parameters of the conditional prior distributions can be output by auxiliary neural networks.
In contrast with the ELBOs for the VAE and CVAE, the ELBOs derived in this section contain distributions that require conditioning on \emph{sequences} of variables.
These distributions can be represented by standard feedforward neural networks, but such architectures are generally incapable of accommodating exceedingly long or variable-length sequences as inputs.
This consideration motivates the use of \emph{recurrent} neural networks, which allow for learning from arbitrarily long input sequences.
Common recurrent neural network architectures include long short term memory (LSTM)~\cite{hochreiter1997long} and gated recurrent unit (GRU)~\cite{chung2014empirical} networks.

\subsection{Parameter-Conditioned Generative Modeling}
This work proposes an extension to generative models for sequential data that is akin to the CVAE, wherein the latent state dynamics are assumed to be governed by parameters $\cs$.
In the context of dynamical systems, $\cs$ may represent physical parameters that can affect system behavior.
Examples of such parameters include the mass and length of a pendulum, or the Reynolds number of a fluid flow.
Thus, a \emph{parameter-conditioned} generative model attempts to model the density $p_{\bm{\theta}}(\x_{1:T} \mid \cs)$.

\begin{figure}[t]
\centering
\begin{tikzpicture}
        \node[value, minimum size=1.25cm] (z1) {$\z_1$};
        \node[value, minimum size=1.25cm] (x1)[below=0.8 of z1] {$\x_1$};
        \draw[arrow] (z1) to (x1); 

        \node[value, minimum size=1.25cm] (z2) [right=2 of z1] {$\z_2$};
        \node[value, minimum size=1.25cm] (x2)[below=0.8 of z2] {$\x_2$};
        \draw[arrow] (z2) to (x2); 

        \node[value] (dots) [right=2 of z2, draw=none] {$\cdots$};

        \node[value, minimum size=1.25cm] (zTm1) [right=2 of dots] {$\z_{T-1}$};
        \node[value, minimum size=1.25cm] (xTm1)[below=0.8 of zTm1] {$\x_{T-1}$};
        \draw[arrow] (zTm1) to (xTm1);

        \node[value, minimum size=1.25cm] (zT) [right=2 of zTm1] {$\z_T$};
        \node[value, minimum size=1.25cm] (xT)[below=0.8 of zT] {$\x_T$};
        \draw[arrow] (zT) to (xT); 

        \draw[arrow] (z1) to (z2); 
        \draw[arrow] (z2) to (dots); 
        \draw[arrow] (dots) to (zTm1);
        \draw[arrow] (zTm1) to (zT); 

        \node[value, minimum size=1.25cm] (c) [above=0.8 of dots, fill=colorRedPastel] {$\cs$};
        \draw[arrow] (c) to (z1); 
        \draw[arrow] (c) to (z2); 
        \draw[arrow] (c) to (zTm1); 
        \draw[arrow] (c) to (zT); 
\end{tikzpicture}
\caption[Graphical model for parameter-conditioned generative model]{Graphical model for parameter-conditioned generative model. Parameters $\cs$ are assumed to influence the time evolution of the latent states.}
\label{fig:pc_seq}
\end{figure}
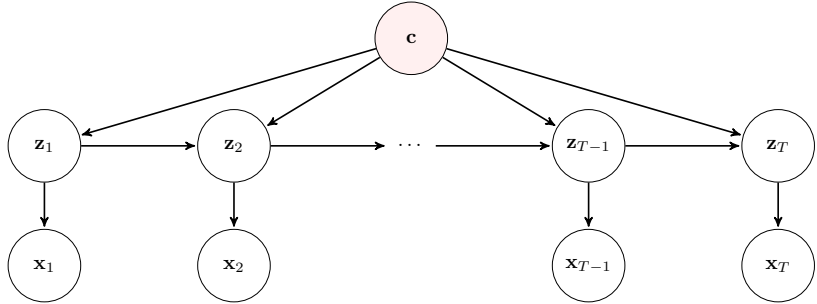

The graphical model for this problem can be found in \cref{fig:pc_seq}.
Following the same procedure as in \cref{sec:seq_gen}, the evidence lower bound for the parameter-conditioned sequential generative model can be derived.
For the case of non-Markovian latent state dynamics, this lower bound is found to be:
\begin{equation}
\begin{split}
\log p_{\bm{\theta}}(\x_{1:T} \mid \cs) \geq \sum_{t=1}^T & \left[\E_{\z_t \sim q_{\bm{\phi}}} \log p_{\bm{\theta}}(\x_t \mid \z_t)\right]\\
& + \mathcal D_\text{KL}\left(q_{\bm{\phi}}(\z_1 \mid \x_{1:T}, \cs) \mid \mid p_{\bm{\theta}}(\z_1 \mid \cs) \right)\\
& + \sum_{t=2}^T D_\text{KL}\left(q_{\bm{\phi}}(\z_t \mid \z_{1:t-1}, \x_{t:T}, \cs) \mid \mid p_{\bm{\theta}}(\z_t \mid \z_{1:t-1}, \cs) \right).
\label{eq:elbo_pc}
\end{split}
\end{equation}
Hence, all prior and approximate posterior distributions must now condition on the parameters $\cs$.

A trained parameter-conditioned generative model provides the opportunity to generate sequential data that evolves according to the prescribed parameters $\cs$.
In the context of dynamical systems, this could enable the efficient study of how such parameters affect a system's behavior.
Given a parameter space of interest, training data can be collected by simulating (or experimentally evaluating) a system at a handful of parameter values.
Subsequently, a generative model trained on the collected data could potentially provide insight into the nature of the studied system throughout the remainder of the parameter space.
A generative model's ability to accurately simulate a system throughout the parameter space is partly dependent on how well it learns to account for the manner in which the parameters affect the system's dynamics.
A description of a new objective that encourages the model to learn these relationships can be found in the next subsection.

\subsection{Mutual Information Objective}\label{sec:mi}
\emph{Entropy}, $H(\y)$, serves as a measure of uncertainty about a random variable $\y$, and is defined as:
\begin{equation}
H(\y) = -\E_{\y \sim p(\y)} \log p(\y).
\end{equation}
Intuitively, a random variable will have high entropy if its probability mass (or density) is spread evenly across its support.
The mutual information $I(\w, \y)$ between random variables $\w$ and $\y$ is defined as:
\begin{equation}
I(\w, \y) = H(\w) - H(\w \mid \y).
\end{equation}
Thus, $I(\w, \y)$ represents the reduction in uncertainty about $\w$ given knowledge of $\y$; if $\w$ and $\y$ are independent, then $I(\w, \y) = 0$.

To ensure that parameter-conditioned generative models can be used to perform simulations throughout the parameter space, high mutual information is desired between the prescribed parameters $\cs$ and the latent states $\z_{1:T}$.
This can be explicitly enforced by training the generative model to maximize a mutual information objective, $\sum_{t=1}^T I(\cs, \z_t)$, in addition to the evidence lower bound from \cref{eq:elbo_pc}.
Unfortunately, maximizing the mutual information objective directly is challenging because it requires access to the generally intractable posterior distributions $p_{\bm{\theta}}(\cs \mid \z_t)$.

In a procedure similar to variational inference, variational mutual information maximization~\cite{barber2003algorithm,chen2016infogan,li2017infogail} introduces a lower bound to the mutual information objective that can be tractably optimized.
Let $r_{\bm{\psi}}(\cs \mid \z_t)$ represent an approximate posterior distribution over $\cs$, which is defined by parameters $\bm{\psi}$.
Then the following lower bound to the mutual information objective can be derived:
\begin{equation}
I(\cs, \z_t) \geq H(\cs) + \E_{\cs \sim p(\cs),\, \z_t \sim p_{\bm{\theta}}} \log r_{\bm{\psi}}(\cs \mid \z_t).
\end{equation}
It is assumed that $p(\cs)$ represents the distribution over parameter values within the dataset, and thus $H(\cs)$ is treated as a constant.
Hence, raising the lower bound on the mutual information objective requires maximizing the log-likelihood assigned to $\cs$ by the approximate posterior distribution.
This can be accomplished in practice by defining a \emph{reconstruction model} with parameters $\bm{\psi}$, which takes in values of $\z_t$ sampled from the prior distributions and outputs parameters to a distribution over $\cs$.
All model parameters, $\bm{\theta}$, $\bm{\phi}$, and $\bm{\psi}$, can then be jointly optimized to maximize the objective:
\begin{equation}
\mathcal L = \mathcal L_\text{ELBO} + \lambda \mathcal L_\text{Info},
\label{eq:gen_obj}
\end{equation}
where $\mathcal L_\text{ELBO}$ represents the lower bound to the likelihood objective, $\mathcal L_\text{Info}$ represents the lower bound to the mutual information objective, and $\lambda$ is a scalar value that trades off between the objectives.

The next section presents experimental results that evaluate the performance of parameter-conditioned generative models on simulating fluid flows.

\section{Experiments}
This section applies parameter-conditioned generative modeling to the task of simulating fluid flows.
Evaluations are performed in two problem domains: (1) two-dimensional airflow over a pair of counter-rotating cylinders and (2) three-dimensional airflow over a half-cylinder.
The next section studies the ability of learned models to simulate two-dimensional airflow, evaluating the accuracy of generated solutions at a range of flow conditions.
A discussion of generative modeling performance for three-dimensional flow follows thereafter.

\subsection{Counter-Rotating Cylinders}\label{sec:double_cylinder}
The flow over non-streamlined, bluff bodies such as cylinders has been the subject of extensive experimental and numerical study.
The scientific interest in such flows is motivated by a phenomenon known as vortex-induced vibration, where wake instabilities lead to unsteady vortex shedding over the surface of bluff bodies.
Vortex-induced vibrations can arise in many engineered structures, such as buildings, bridges, and offshore piers.
Vortex shedding gives rise to strong transverse forces, and is associated with higher drag and unsteady lift forces~\cite{roussopoulos1993feedback,illingworth2016model}.
The collapse of the Tacoma Narrows Bridge in 1940 serves as a particularly stark illustration of the destructive potential of the induced structural vibrations~\cite{chan2012control}.

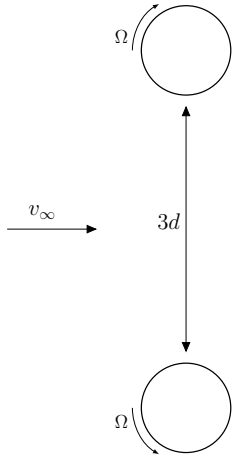
\begin{figure}[t]
\centering
\begin{tikzpicture}
	\coordinate (A) at (0,0);
        
    \draw[thick] (A) circle (1);
\draw[-latex] (-1.2,0) arc (180:120:1.2cm) node[near start,left] {\large $\Omega$};
\draw[thick] (0,-8) circle (1);
\draw[-latex] (-1.2,-8) arc (180:240:1.2cm) node[near start,left] {\large $\Omega$};

\draw[<->] (0,-6.75) to (0, -1.25);
\node[draw=none](label)[below left=3.5 and 0 of A]{\Large$3d$};

\draw[->] (-4,-4) to (-2, -4);
\node[draw=none](u)[above left=-0.4 and 2 of label]{\Large$v_\infty$};

\end{tikzpicture}
\caption{Counter-rotating cylinder configuration.}
\label{fig:config}
\end{figure}

This section considers the test case of airflow over counter-rotating cylinders, as depicted in \cref{fig:config}.
The cylinder surfaces are assumed to be separated by a distance of $3d$, where $d$ is the cylinder diameter, and rotate at a fixed non-dimensional rotation speed given by:
\begin{equation}
\Omega = \frac{\omega d}{2 v_{\infty}},
\end{equation}
where $\omega$ is the angular velocity of the cylinder and $v_{\infty}$ is the velocity of the incident flow.
As with flow over a single bluff body, flow over a pair of cylinders has been shown to exhibit unsteady vortex shedding~\cite{williamson1985evolution,kang2003characteristics,peschard1996coupled}.
Experimental and numerical results have demonstrated that there exists a critical rotation speed, $\Omega_\text{crit}$, above which the wake instabilities disappear and the flow becomes steady, leaving a constant transverse load acting upon the cylinders~\cite{yoon2007laminar,chan2010suppression,chan2011vortex}.
The thesis of Chan~\cite{chan2012control} performs a wide array of numerical experiments that characterize the behavior of the double cylinder system as the cylinder separation, Reynolds number, rotation speed, and rotation direction are varied.
The experiments in this work only allow for variation in rotation speed and Reynolds number (i.e. $\cs = [\Omega, \,\text{Re}]\T$); the cylinder separation and rotation direction are assumed to be fixed.

\begin{figure}[t]
\centering
\begin{subfigure}[b]{0.47\linewidth}
\centering
\includegraphics[width=0.9\linewidth]{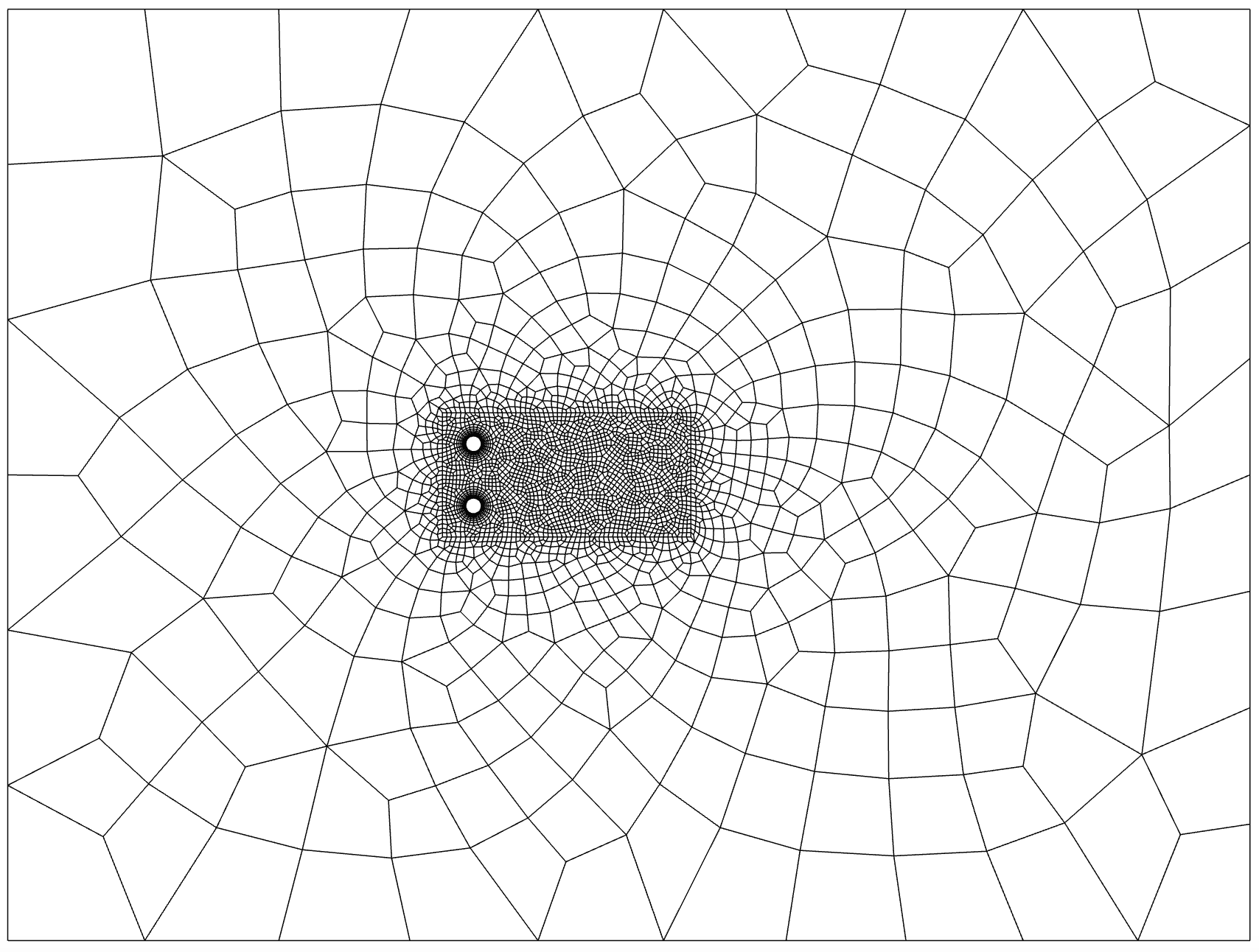}
\end{subfigure}\hfill
\begin{subfigure}[b]{0.47\linewidth}
\centering
\includegraphics[width=0.9\linewidth]{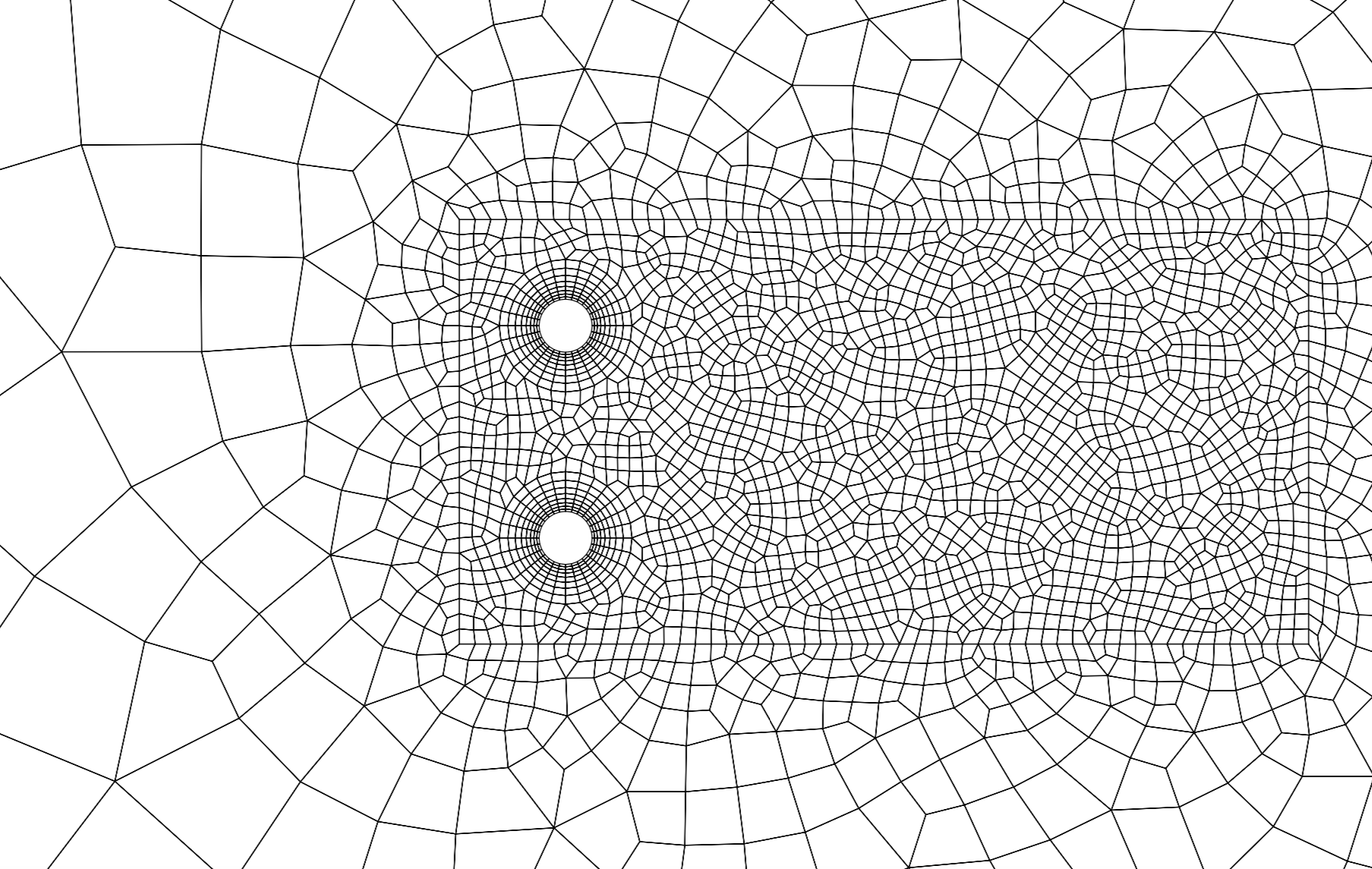}
\end{subfigure}
\caption[Mesh used for simulation of counter-rotating cylinders]{Left: View of mesh used for simulation of counter-rotating cylinders over entire computational domain. Right: Zoomed-in view of the mesh in the vicinity of the cylinders. Generative models are trained to only simulate flow in the region with fine mesh resolution.}
\label{fig:double_cyl_mesh}
\end{figure}

All simulations are performed with the PyFR solver~\cite{witherden2014pyfr}, which uses the flux reconstruction (FR) approach~\cite{huynh2007flux}, a high-order numerical method for unstructured grids.
The computational domain is defined as $[-30, 50]$ in the stream-wise direction and $[-30, 30]$ in the cross-wise direction.
The diameter of each cylinder is set to a value of one; the upper cylinder is centered at $(0, 2)$, while the lower cylinder is centered at $(0, -2)$. 
A spatially varying velocity is applied to the surface of each cylinder to model cylinder rotation.
The full mesh, depicted in the left figure of \cref{fig:double_cyl_mesh}, consists of \num{3148} unstructured, quadratically curved quadrilateral elements.
To align with previous work, the Mach number is set to $M = 0.1$, which implies a free-stream velocity of $v_{\infty} \simeq 0.118$. 

\begin{figure}[t]
\centering
    \begin{subfigure}[b]{0.24 \linewidth}
  \centering
    \includegraphics[width=0.96\linewidth, height=0.48\linewidth]{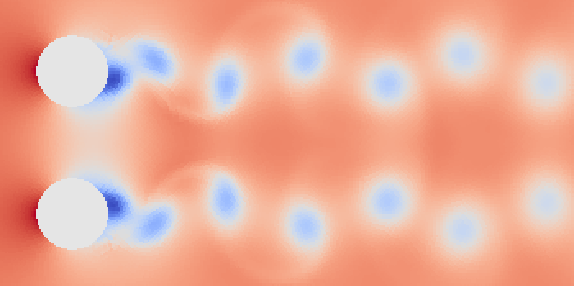}
    \caption{Density} 
  \end{subfigure}  
  \begin{subfigure}[b]{0.24 \linewidth}
  \centering
    \includegraphics[width=0.96\linewidth, height=0.48\linewidth]{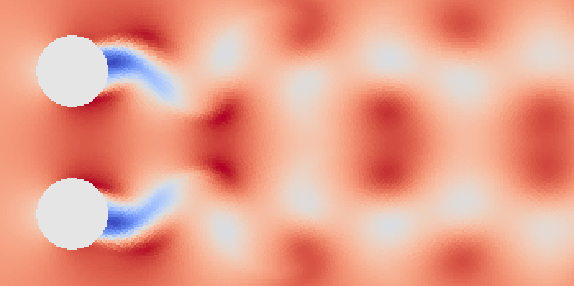}
    \caption{$x$-velocity}
  \end{subfigure}
  \begin{subfigure}[b]{0.24 \linewidth}
  \centering
    \includegraphics[width=0.96\linewidth, height=0.48\linewidth]{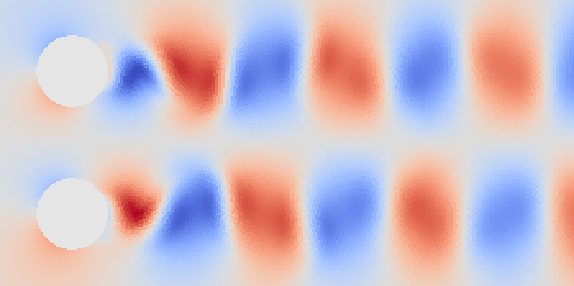}
    \caption{$y$-velocity}
  \end{subfigure}  
  \begin{subfigure}[b]{0.24 \linewidth}
  \centering
    \includegraphics[width=0.96\linewidth, height=0.48\linewidth]{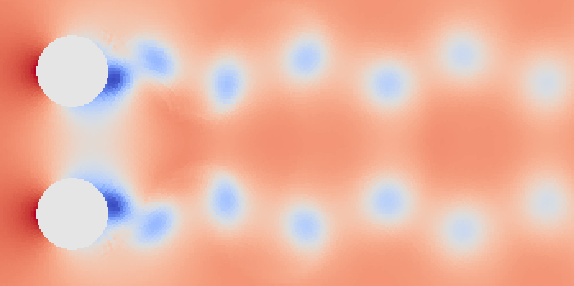}
    \caption{Pressure}
  \end{subfigure}
  \caption[Format of inputs to the generative model]{Format of inputs to the generative model for $\text{Re}=150$ and $\Omega = 0$. Different physical quantities are treated as different channels in the input.}
  \label{fig:inputs_gen}
\end{figure} 

Training a model to generate flow quantities across the entire region shown in the left figure of \cref{fig:double_cyl_mesh} would be challenging due to the limitations that memory constraints impose upon the size of neural network inputs.
Furthermore, the non-uniform spacing between the solution points in the CFD simulations also presents a challenge, because many of the most powerful neural network architectures are designed to handle image data, and thus assume uniform spacing between points.
To account for these challenges, the generative models are only trained to simulate fluid flow on a grid of equispaced points in the vicinity of the cylinders.
Formatting the data in this manner allows for training on a set of image-like inputs of dimension $175 \times 350 \times 4$, where the first two input dimensions correspond to spatial dimensions and the final input dimension corresponds to different flow quantities.
The modeled region spans from \num{-2} to \num{14} in the stream-wise direction and \num{-4} to \num{4} in the cross-wise direction, which is selected to coincide with the area of fine mesh resolution depicted in the right image of \cref{fig:double_cyl_mesh}.
\Cref{fig:inputs_gen} shows an example training input at Reynolds number $\text{Re} = 150$ and rotation speed $\Omega = 0$; the modeled flow quantities are density, $x$-velocity, $y$-velocity, and pressure.
In all neural network inputs, flow quantities inside the cylinders are set to be free-stream values, and the network predictions for those points are overwritten with free-stream values during training and evaluation.
The training data is generated by running CFD simulations at a range of flow conditions and saving image-like inputs every \num{500} solver steps.

\begin{figure}[t]
\centering
\begin{tikzpicture}
\begin{axis}[xlabel=\large Reynolds Number,
    ylabel=\large $\Omega$,
    xmin=70, xmax=205,
    ymin=-0.2,
    ymax=2.7,
    ytick={0,1,2},
    yticklabel style={
        /pgf/number format/fixed,
        /pgf/number format/precision=5
    }]

    \addplot +[only marks, mark options={black}] 
        table[x=re,y=om, col sep=comma] {re_om.csv};

    \addplot +[draw=none, no markers, name path=A] 
        coordinates { (75,0) (75,2.5) };
    \addplot +[draw=none, no markers, name path=B] 
        coordinates { (75,0) (200,0) };
    \addplot +[draw=none, no markers, name path=C] 
        coordinates { (200,0) (200,2.5) };
    \addplot +[draw=none, no markers, name path=D] 
        coordinates { (75,2.5) (200,2.5) };
    \addplot[colorPurplePastel] fill between[of=A and B];
    \addplot[colorPurplePastel] fill between[of=B and D];

\end{axis}
\end{tikzpicture}
\caption[Illustration of parameter space for counter-rotating cylinder simulations]{Illustration of parameter space for counter-rotating cylinder simulations. The points at which CFD simulations are run are represented by the black dots.}
\label{fig:re_om_scatter}
\end{figure}
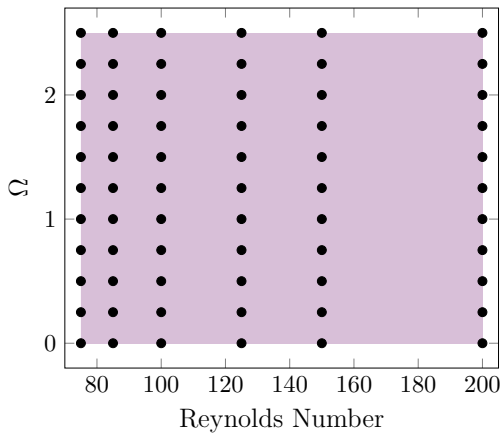

Training data is generated at a range of flow conditions, corresponding to a variety of Reynolds numbers and rotation speeds.
The set of Reynolds numbers considered is $\text{Re} \in \{75, 85, 100, 125, 150, 200\}$.
This set is designed to span the range of Reynolds numbers over which vortex shedding occurs and the flow can still be considered two-dimensional.
Because viscous forces are inversely proportional to Reynolds number for a fixed flow velocity, a change in Reynolds number has a more pronounced effect on flow behavior at $\text{Re}=75$ than it does at e.g. $\text{Re} = 150$, and hence a finer spacing between simulation points is used at lower Reynolds numbers.
At each Reynolds number, simulations are run at rotation speeds $\Omega \in \{0, 0.25, 0.5, \ldots, 2.5\}$.
This range of rotation speeds is selected such that the training data contains information about the system behavior with no cylinder rotation, as well as the behavior of the flow as $\Omega$ is increased, including rotation speeds in excess of $\Omega_\text{crit}$.
\Cref{fig:re_om_scatter} shows the points within the parameter space at which simulations are performed.
The hope is that a generative model could be trained on CFD data from each point in the plot, and subsequently be capable of simulating fluid flow anywhere in the shaded parameter domain in \cref{fig:re_om_scatter}.

\begin{figure}[t!]
\centering
\begin{tikzpicture}
\begin{groupplot}[group style={group size= 3 by 3},height=7cm,width=8cm]
\nextgroupplot[xlabel=\large Time Step,
    ylabel=\large $C_L$,
    xmin=0, xmax=128,
    xtick={0,20,40,...,120},
    ymin=-6,
    ymax=6,
    ylabel shift=-5pt,
    title=\Large$\Omega\equal0$]

    \addplot +[blue,thick, no markers] 
        table[x=t,y expr=\thisrow{om1}*2/1.4/0.1^2, col sep=comma] {lift_top_cfd.csv};

    \addplot +[red,thick, no markers]
        table[x=t,y expr=\thisrow{om1}*2/1.4/0.1^2, col sep=comma] {lift_bottom_cfd.csv};

    \legend{Top, Bottom}

\nextgroupplot[xlabel=\large Time Step,
    ylabel=\large $C_L$,
    xmin=0, xmax=128,
    xtick={0,20,40,...,120},
    ymin=-6,
    ymax=6,
    ylabel shift=-5pt,
    title=\Large$\Omega\equal0.25$,
    xshift=2em]

    \addplot +[blue,thick, no markers] 
        table[x=t,y expr=\thisrow{om2}*2/1.4/0.1^2, col sep=comma] {lift_top_cfd.csv};

    \addplot +[red,thick, no markers]
        table[x=t,y expr=\thisrow{om2}*2/1.4/0.1^2, col sep=comma] {lift_bottom_cfd.csv};

\nextgroupplot[xlabel=\large Time Step,
    ylabel=\large $C_L$,
    xmin=0, xmax=128,
    xtick={0,20,40,...,120},
    ymin=-6,
    ymax=6,
    ylabel shift=-5pt,
    title=\Large$\Omega\equal0.5$,
    xshift=2em]

    \addplot +[blue,thick, no markers] 
        table[x=t,y expr=\thisrow{om3}*2/1.4/0.1^2, col sep=comma] {lift_top_cfd.csv};

    \addplot +[red,thick, no markers]
        table[x=t,y expr=\thisrow{om3}*2/1.4/0.1^2, col sep=comma] {lift_bottom_cfd.csv};

\nextgroupplot[xlabel=\large Time Step,
    ylabel=\large $C_L$,
    xmin=0, xmax=128,
    xtick={0,20,40,...,120},
    ymin=-6,
    ymax=6,
    ylabel shift=-5pt,
    title=\Large$\Omega\equal0.75$,
    yshift=-5em]

    \addplot +[blue,thick, no markers] 
        table[x=t,y expr=\thisrow{om4}*2/1.4/0.1^2, col sep=comma] {lift_top_cfd.csv};

    \addplot +[red,thick, no markers]
        table[x=t,y expr=\thisrow{om4}*2/1.4/0.1^2, col sep=comma] {lift_bottom_cfd.csv};

\nextgroupplot[xlabel=\large Time Step,
    ylabel=\large $C_L$,
    xmin=0, xmax=128,
    xtick={0,20,40,...,120},
    ymin=-6,
    ymax=6,
    ylabel shift=-5pt,
    title=\Large$\Omega\equal1$,
    yshift=-5em]

    \addplot +[blue,thick, no markers] 
        table[x=t,y expr=\thisrow{om5}*2/1.4/0.1^2, col sep=comma] {lift_top_cfd.csv};

    \addplot +[red,thick, no markers]
        table[x=t,y expr=\thisrow{om5}*2/1.4/0.1^2, col sep=comma] {lift_bottom_cfd.csv};

\nextgroupplot[xlabel=\large Time Step,
    ylabel=\large $C_L$,
    xmin=0, xmax=128,
    xtick={0,20,40,...,120},
    ymin=-6,
    ymax=6,
    ylabel shift=-5pt,
    title=\Large$\Omega\equal1.25$,
    yshift=-5em]

    \addplot +[blue,thick, no markers] 
        table[x=t,y expr=\thisrow{om6}*2/1.4/0.1^2, col sep=comma] {lift_top_cfd.csv};

    \addplot +[red,thick, no markers]
        table[x=t,y expr=\thisrow{om6}*2/1.4/0.1^2, col sep=comma] {lift_bottom_cfd.csv};

\nextgroupplot[xlabel=\large Time Step,
    ylabel=\large $C_L$,
    xmin=0, xmax=128,
    xtick={0,20,40,...,120},
    ymin=-6,
    ymax=6,
    ylabel shift=-5pt,
    title=\Large$\Omega\equal1.5$,
    yshift=-5em]

    \addplot +[blue,thick, no markers] 
        table[x=t,y expr=\thisrow{om7}*2/1.4/0.1^2, col sep=comma] {lift_top_cfd.csv};

    \addplot +[red,thick, no markers]
        table[x=t,y expr=\thisrow{om7}*2/1.4/0.1^2, col sep=comma] {lift_bottom_cfd.csv};

\nextgroupplot[xlabel=\large Time Step,
    ylabel=\large $C_L$,
    xmin=0, xmax=128,
    xtick={0,20,40,...,120},
    ymin=-6,
    ymax=6,
    ylabel shift=-5pt,
    title=\Large$\Omega\equal1.75$,
    yshift=-5em]

    \addplot +[blue,thick, no markers] 
        table[x=t,y expr=\thisrow{om8}*2/1.4/0.1^2, col sep=comma] {lift_top_cfd.csv};

    \addplot +[red,thick, no markers]
        table[x=t,y expr=\thisrow{om8}*2/1.4/0.1^2, col sep=comma] {lift_bottom_cfd.csv};

\nextgroupplot[xlabel=\large Time Step,
    ylabel=\large $C_L$,
    xmin=0, xmax=128,
    xtick={0,20,40,...,120},
    ymin=-6,
    ymax=6,
    ylabel shift=-5pt,
    title=\Large$\Omega\equal2$,
    yshift=-5em]

    \addplot +[blue,thick, no markers] 
        table[x=t,y expr=\thisrow{om9}*2/1.4/0.1^2, col sep=comma] {lift_top_cfd.csv};

    \addplot +[red,thick, no markers]
        table[x=t,y expr=\thisrow{om9}*2/1.4/0.1^2, col sep=comma] {lift_bottom_cfd.csv};




\end{groupplot}
\end{tikzpicture}
\caption[Variation of lift with rotation speed]{Lift coefficients on top and bottom cylinder over time as a function of rotation speed.}
\label{fig:lift_plot}
\end{figure}
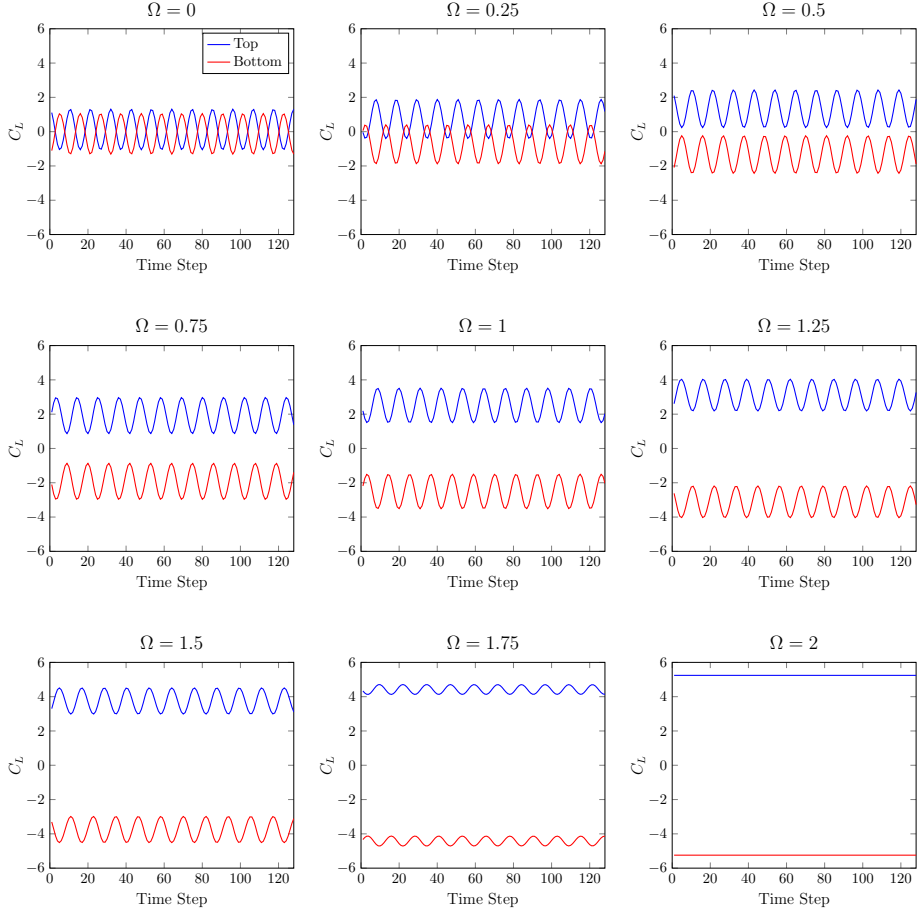

\begin{figure}[h]
\centering
    \begin{subfigure}[b]{0.24 \linewidth}
  \centering
    \includegraphics[width=0.96\linewidth, height=0.48\linewidth]{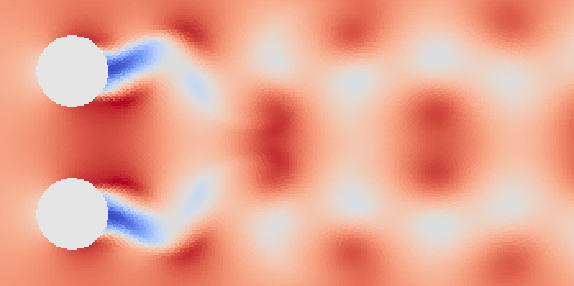}
    \caption{$\Omega = 0$} 
  \end{subfigure}
  \begin{subfigure}[b]{0.24 \linewidth}
  \centering
    \includegraphics[width=0.96\linewidth, height=0.48\linewidth]{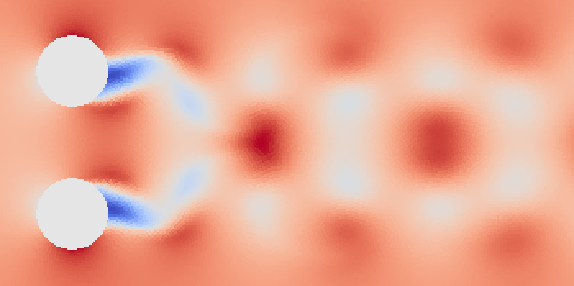}
    \caption{$\Omega = 1$}
  \end{subfigure}
  \begin{subfigure}[b]{0.24 \linewidth}
  \centering
    \includegraphics[width=0.96\linewidth, height=0.48\linewidth]{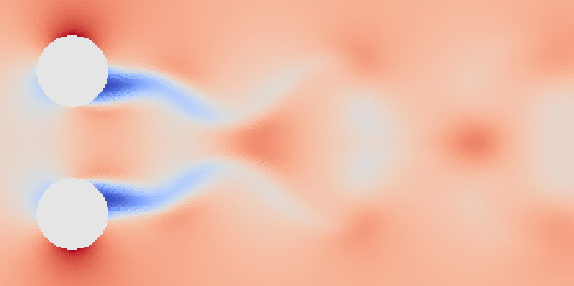}
    \caption{$\Omega = 1.75$}
  \end{subfigure}
  \begin{subfigure}[b]{0.24 \linewidth}
  \centering
    \includegraphics[width=0.96\linewidth, height=0.48\linewidth]{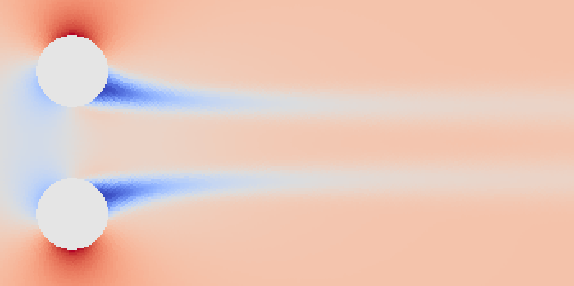}
    \caption{$\Omega = 2$}
  \end{subfigure}
  \caption[Variation of $x$-velocity with rotation speed]{Variation of $x$-velocity components of training inputs as a function of rotation speed at Reynolds number of $\text{Re} = 150$.}
  \label{fig:xvel_om}
\end{figure}

\begin{figure}[h]
\centering
    \begin{subfigure}[b]{0.24 \linewidth}
  \centering
    \includegraphics[width=0.96\linewidth, height=0.48\linewidth]{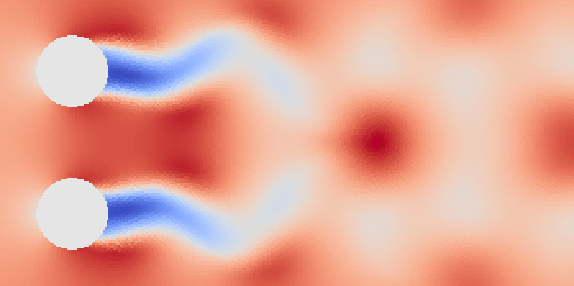}
    \caption{$\text{Re} = 75$} 
  \end{subfigure}
  \begin{subfigure}[b]{0.24 \linewidth}
  \centering
    \includegraphics[width=0.96\linewidth, height=0.48\linewidth]{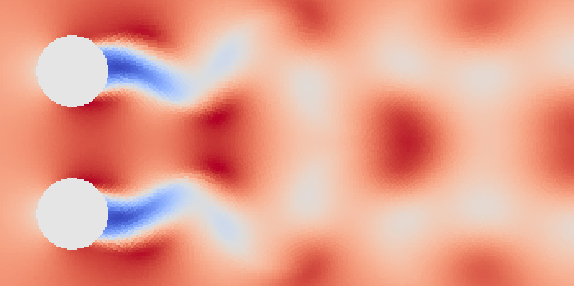}
    \caption{$\text{Re} = 100$}
  \end{subfigure}
  \begin{subfigure}[b]{0.24 \linewidth}
  \centering
    \includegraphics[width=0.96\linewidth, height=0.48\linewidth]{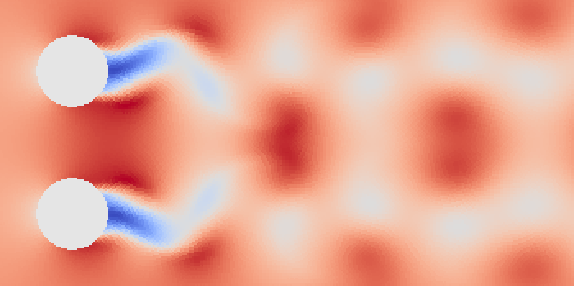}
    \caption{$\text{Re} = 125$}
  \end{subfigure}
  \begin{subfigure}[b]{0.24 \linewidth}
  \centering
    \includegraphics[width=0.96\linewidth, height=0.48\linewidth]{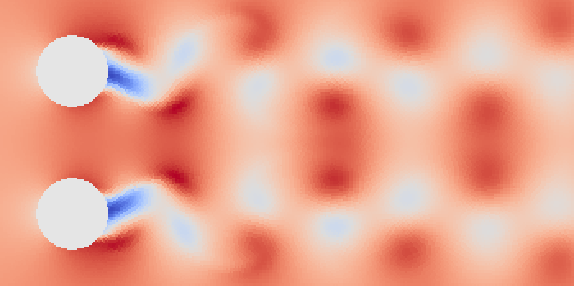}
    \caption{$\text{Re} = 200$}
  \end{subfigure}
  \caption[Variation of $x$-velocity with Reynolds number]{Variation of $x$-velocity components of training inputs as a function of Reynolds number at a rotation speed of $\Omega = 0$.}
  \label{fig:xvel_re}
\end{figure} 

By examining the training data, the effects of varying the rotation speed and Reynolds number can be observed.
\Cref{fig:lift_plot} shows the time evolution of the lift coefficients for the top and bottom cylinders as a function of rotation speed at Reynolds number $\text{Re} = 150$.
The lift coefficient is defined as:
\begin{equation}
C_L = \frac{L}{\frac{1}{2} \rho_{\infty} v_{\infty}^2 d},
\end{equation}
where $L$ is the lift force, $\rho_{\infty}$ is the free-stream density, $v_{\infty}$ is the free-stream velocity, and $d$ is the cylinder diameter.
From the individual plots, two trends can be observed.
First, the lift curves separate as the rotation speed is increased, with the mean lift coefficient becoming more positive on the top cylinder and more negative on the bottom cylinder.
Second, the amplitude of oscillation in the lift curves is damped out at higher rotation speeds, with a constant transverse force acting on the cylinders at a rotation speed of $\Omega = 2$.
From examining these plots, it is apparent that the critical rotation speed, at which the flow transitions from unsteady to steady flow, is somewhere in the interval $\Omega_\text{crit} \in [1.75, \, 2]$.

\Cref{fig:xvel_om,fig:xvel_re} show the change in behavior of the cylinder system as the rotation speed and Reynolds number are varied.
\Cref{fig:xvel_om} visualizes the $x$-velocity component of the training inputs for four different rotation speeds at a Reynolds number of $\text{Re} = 150$.
It is clear from these images that the curved wake regions of low $x$-velocity begin to flatten out and emanate from different locations on the cylinders as the rotation speed increases.
\Cref{fig:xvel_re} likewise illustrates the variation in flow behavior for four Reynolds numbers with a fixed rotation speed of $\Omega = 0$.
It can be seen that the extent of the region of low $x$-velocity decreases with increasing Reynolds number.
Although it cannot be observed from these static images, an increase in Reynolds number also corresponds to a decreased period of vortex shedding.

\subsection{Two-Dimensional Flow Training Details}\label{sec:training_2d}
This section provides details about training parameter-conditioned generative models on data from the counter-rotating cylinder system.
All components of the generative model are jointly defined and trained in Tensorflow~\cite{abadi2015tensorflow}.
The parameters to the prior and approximate posterior distributions over $\z_1$, denoted by $p_{\bm{\theta}}(\z_1 \mid \cs)$ and $q_{\bm{\phi}}(\z_1 \mid \x_{1:T}, \cs)$, are output by feedforward neural networks with a single hidden layer containing \num{128} units.
The prior and approximate posterior distributions over all remaining latent states, denoted by $p_{\bm{\theta}}(\z_t \mid \z_{1:t-1}, \cs)$ and $q_{\bm{\phi}}(\z_t \mid \z_{1:t-1}, \x_{t:T}, \cs)$, are represented by two-layer feedforward neural networks with \num{256} and \num{128} units in each hidden layer.
Long short term memory recurrent neural networks with \num{256} hidden units are used to summarize information about the (latent) state sequences upon which these distributions are conditioned.
As explained in \cref{sec:seq_gen}, defining the distributions in this manner implicitly assumes that the dynamics of the latent states are non-Markovian; treating the latent states this way was empirically found to yield better performance than assuming Markovian dynamics in the latent space.

As defined above, the networks used to represent the approximate posterior distributions may have an insufficient number of parameters to learn effectively from the very high-dimensional flow states.
Therefore, a feature extractor network is introduced, which is trained to provide a mapping from the network inputs, $\x_t \in \R^{(175 \times 350 \times 4)}$, to a low-dimensional set of features.
The approximate posterior distributions over the latent states are then conditioned upon these lower-dimensional features rather than the full flow state.

The feature extractor is comprised of a series of two-dimensional convolutional layers with ResNet~\cite{he2016resnet} skip connections and \num{256}, \num{128}, \num{64}, \num{32}, \num{16}, \num{8}, and \num{4} filters in each layer.
Convolutions are performed with a stride of two in the first five layers, and a striding of one is employed in the remaining layers.
An affine transformation maps the output of the convolutional layers to a set of features with the same dimensionality as the latent states; based on performance in numerical experiments, the latent states are defined to be four-dimensional in all studies involving the counter-rotating cylinder system.
The decoder network, denoted by $p_{\bm{\theta}}(\x_t \mid \z_t)$, is constructed to invert all operations performed by the feature extractor, which allows for the mapping of latent state samples back to the state space.
Finally, the reconstruction network introduced in \cref{sec:mi}, denoted by $r_{\bm{\psi}}(\cs \mid \z_t)$, is defined to have a single \num{128}-unit hidden layer.
Rectified linear units~\cite{nair2010rectified} are used as the activation functions across all feedforward layers in the model.

During training, all network parameters are concurrently optimized to maximize the objective defined by \cref{eq:gen_obj}.
The likelihood components of the evidence lower bound are optimized by minimizing the $L_1$ error between the true flow states, $\x_t$, and the output of the decoder network.
This is equivalent to assuming that $p_{\bm{\theta}}(\x_t \mid \z_t)$ is a Laplacian distribution, where the mean of the distribution is output by the decoder network and the variance is constant.
$L_1$ error is used rather than $L_2$ error because the $L_1$ norm is more sensitive to small errors, which encourages more accurate simulation of the subtle temporal variations in the modeled flow quantities.
To allow for the ELBO and mutual information objectives to be of the same scale, a weight of $\lambda = 10^4$ is applied to the mutual information objective.

The generative model is trained on \num{20} distinct \num{32}-step sequences from each flow condition present within the training data.
Training is initialized with a learning rate of $6 \e{-4}$, which is decayed by a factor of \num{0.75} whenever the loss fails to decrease on a validation set.
Training terminates once the learning rate falls below a value of $1 \e{-5}$.
The Adam optimizer~\cite{kingma2014adam} is used to perform updates to the model parameters.
In total, training takes approximately \num{30} hours with the neural network parameters divided across two NVIDIA K40 GPUs.
While this represents a significant upfront cost, flow simulations generated by the trained model are found to be approximately \num{120} times faster than the CFD simulations used to generate the training data.
This speedup suggests that an immense amount of time could be saved by using generative models in place of CFD solvers, assuming that the generative models prove capable of performing accurate simulations at a wide range of flow conditions.
The next section presents numerical experiments that evaluate the learned model's ability to generate flow solutions for the counter-rotating cylinder system.

\subsection{Two-Dimensional Flow Results}

\begin{figure}
\centering
\begin{subfigure}[h]{0.9\linewidth}
    \begin{subfigure}[b]{0.24 \linewidth}
  \centering
    \includegraphics[width=0.96\linewidth, height=0.48\linewidth]{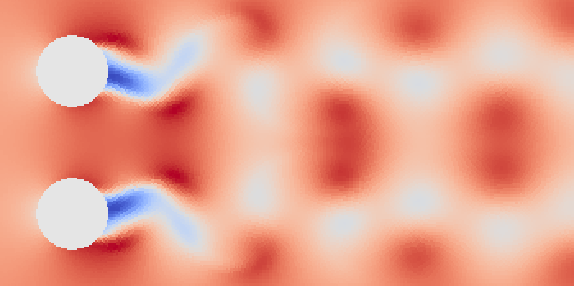}
  \end{subfigure}
  \begin{subfigure}[b]{0.24 \linewidth}
  \centering
    \includegraphics[width=0.96\linewidth, height=0.48\linewidth]{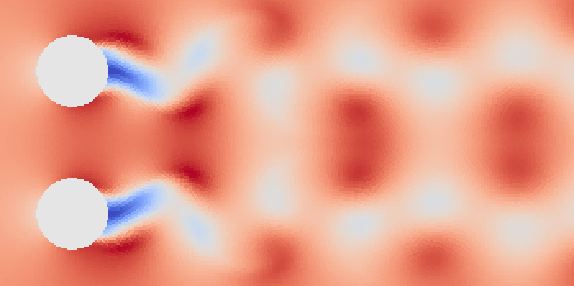}
  \end{subfigure}
 \begin{subfigure}[b]{0.24 \linewidth}
 \centering
   \includegraphics[width=0.96\linewidth, height=0.48\linewidth]{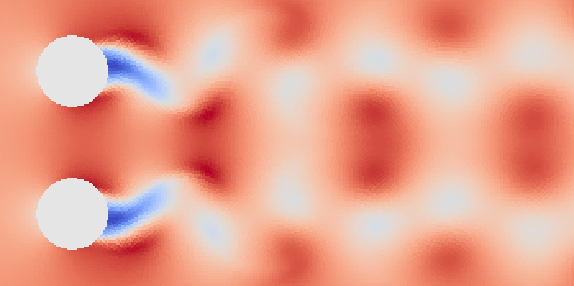}
 \end{subfigure}
 \begin{subfigure}[b]{0.24 \linewidth}
 \centering
   \includegraphics[width=0.96\linewidth, height=0.48\linewidth]{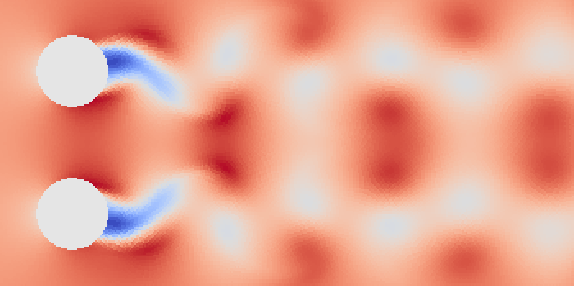}
 \end{subfigure}\vspace{1em}

  \begin{subfigure}[b]{0.24 \linewidth}
  \centering
    \includegraphics[width=0.96\linewidth, height=0.48\linewidth]{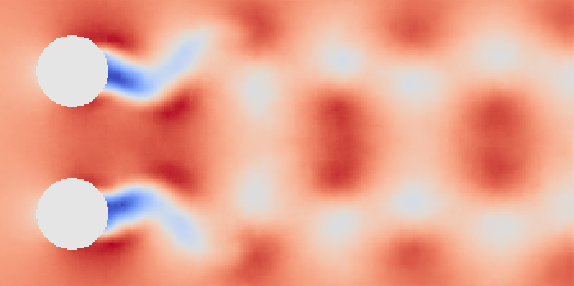}
  \end{subfigure}
 \begin{subfigure}[b]{0.24 \linewidth}
 \centering
   \includegraphics[width=0.96\linewidth, height=0.48\linewidth]{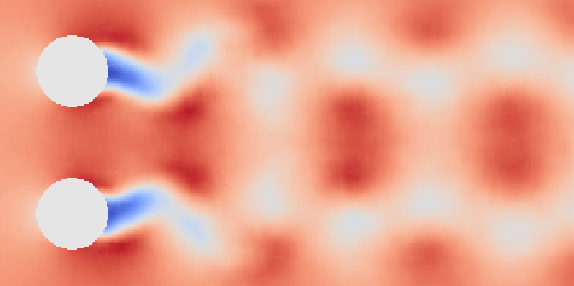}
 \end{subfigure}
 \begin{subfigure}[b]{0.24 \linewidth}
 \centering
   \includegraphics[width=0.96\linewidth, height=0.48\linewidth]{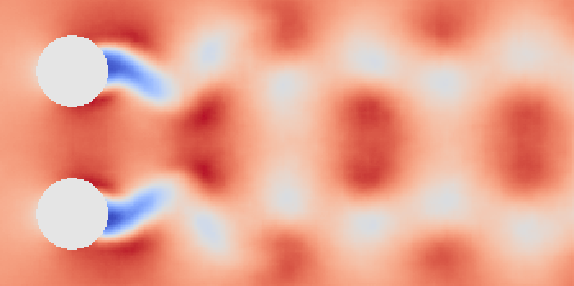}
 \end{subfigure}
  \begin{subfigure}[b]{0.24 \linewidth}
  \centering
    \includegraphics[width=0.96\linewidth, height=0.48\linewidth]{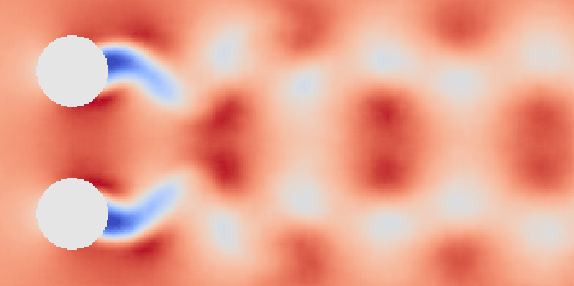}
  \end{subfigure}
  \caption{$\text{Re} = 150$, $\Omega = 0$}
  \label{fig:comp_standard}
\end{subfigure}\vspace{1em}

\begin{subfigure}[h]{0.9\linewidth}
    \begin{subfigure}[b]{0.24 \linewidth}
  \centering
    \includegraphics[width=0.96\linewidth, height=0.48\linewidth]{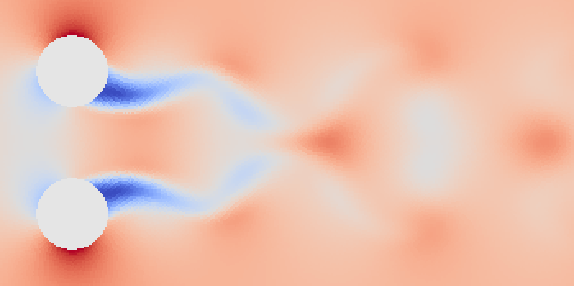}
  \end{subfigure}
  \begin{subfigure}[b]{0.24 \linewidth}
  \centering
    \includegraphics[width=0.96\linewidth, height=0.48\linewidth]{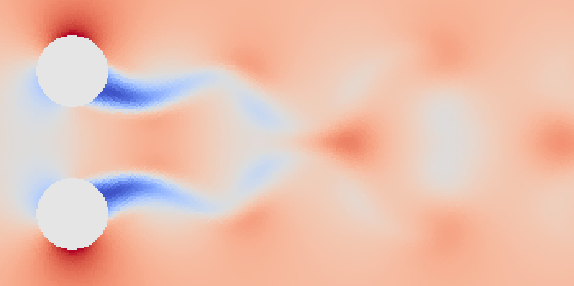}
  \end{subfigure}
 \begin{subfigure}[b]{0.24 \linewidth}
 \centering
   \includegraphics[width=0.96\linewidth, height=0.48\linewidth]{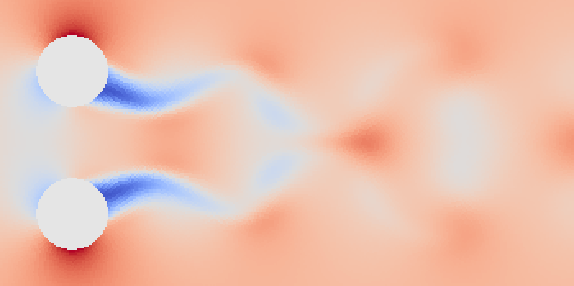}
 \end{subfigure}
 \begin{subfigure}[b]{0.24 \linewidth}
 \centering
   \includegraphics[width=0.96\linewidth, height=0.48\linewidth]{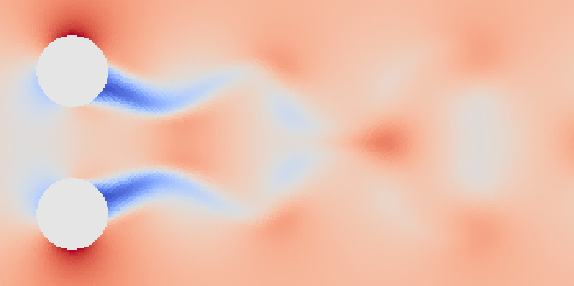}
 \end{subfigure}\vspace{1em}

  \begin{subfigure}[b]{0.24 \linewidth}
  \centering
    \includegraphics[width=0.96\linewidth, height=0.48\linewidth]{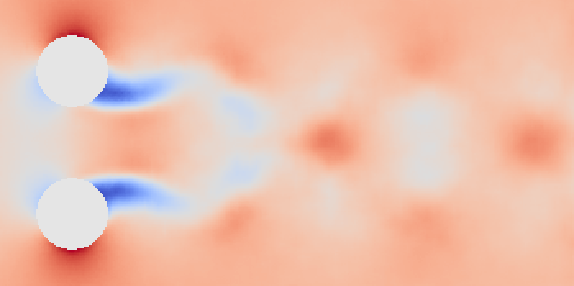}
  \end{subfigure}
 \begin{subfigure}[b]{0.24 \linewidth}
 \centering
   \includegraphics[width=0.96\linewidth, height=0.48\linewidth]{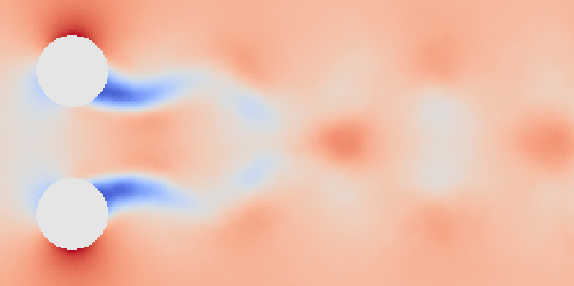}
 \end{subfigure}
 \begin{subfigure}[b]{0.24 \linewidth}
 \centering
   \includegraphics[width=0.96\linewidth, height=0.48\linewidth]{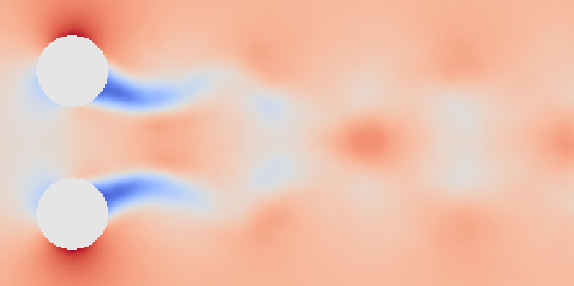}
 \end{subfigure}
  \begin{subfigure}[b]{0.24 \linewidth}
  \centering
    \includegraphics[width=0.96\linewidth, height=0.48\linewidth]{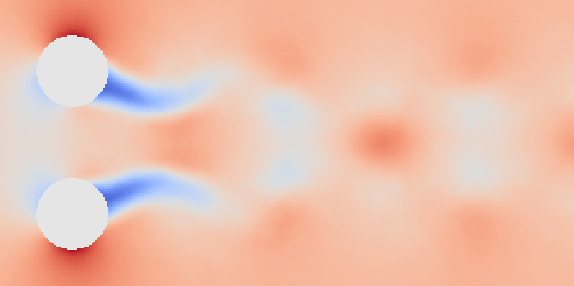}
  \end{subfigure}
  \caption{$\text{Re} = 150$, $\Omega = 1.75$}
  \label{fig:comp_om175}
\end{subfigure}\vspace{1em}

\begin{subfigure}[h]{0.9\linewidth}
    \begin{subfigure}[b]{0.24 \linewidth}
  \centering
    \includegraphics[width=0.96\linewidth, height=0.48\linewidth]{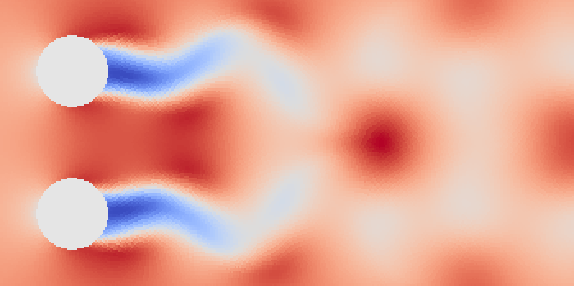}
  \end{subfigure}
  \begin{subfigure}[b]{0.24 \linewidth}
  \centering
    \includegraphics[width=0.96\linewidth, height=0.48\linewidth]{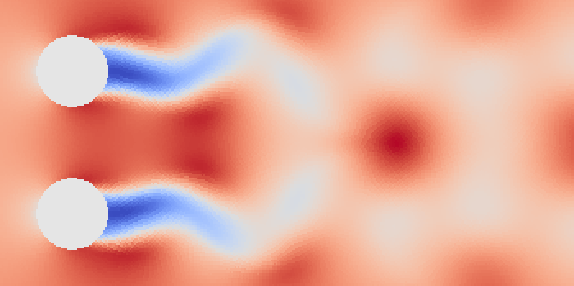}
  \end{subfigure}
 \begin{subfigure}[b]{0.24 \linewidth}
 \centering
   \includegraphics[width=0.96\linewidth, height=0.48\linewidth]{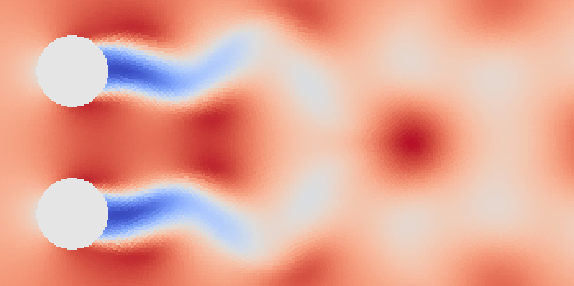}
 \end{subfigure}
 \begin{subfigure}[b]{0.24 \linewidth}
 \centering
   \includegraphics[width=0.96\linewidth, height=0.48\linewidth]{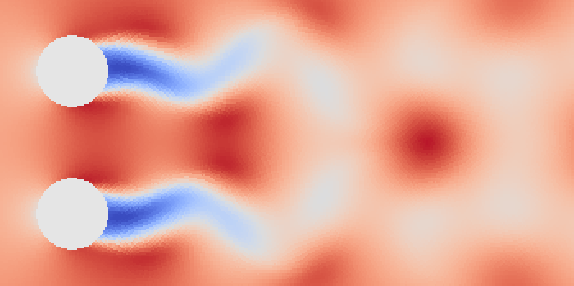}
 \end{subfigure}\vspace{1em}

  \begin{subfigure}[b]{0.24 \linewidth}
  \centering
    \includegraphics[width=0.96\linewidth, height=0.48\linewidth]{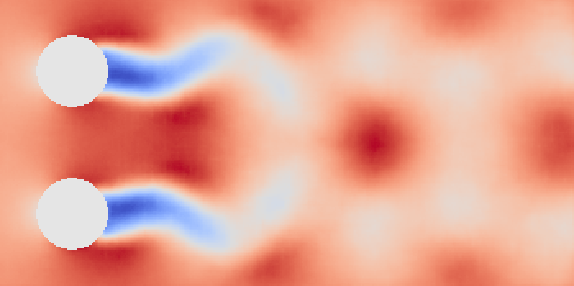}
  \end{subfigure}
 \begin{subfigure}[b]{0.24 \linewidth}
 \centering
   \includegraphics[width=0.96\linewidth, height=0.48\linewidth]{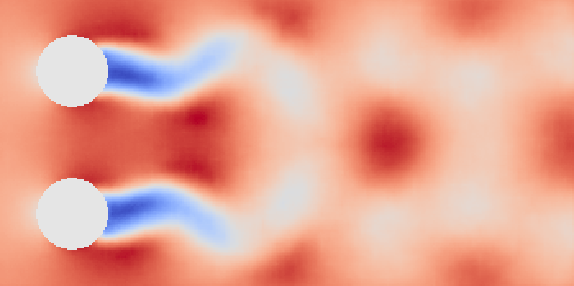}
 \end{subfigure}
 \begin{subfigure}[b]{0.24 \linewidth}
 \centering
   \includegraphics[width=0.96\linewidth, height=0.48\linewidth]{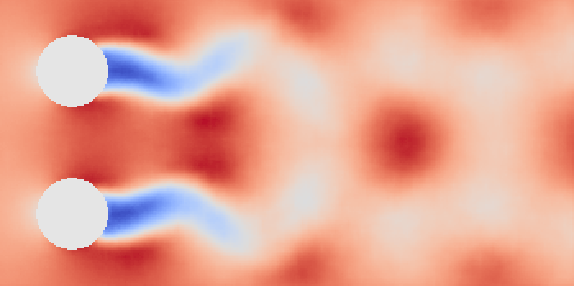}
 \end{subfigure}
  \begin{subfigure}[b]{0.24 \linewidth}
  \centering
    \includegraphics[width=0.96\linewidth, height=0.48\linewidth]{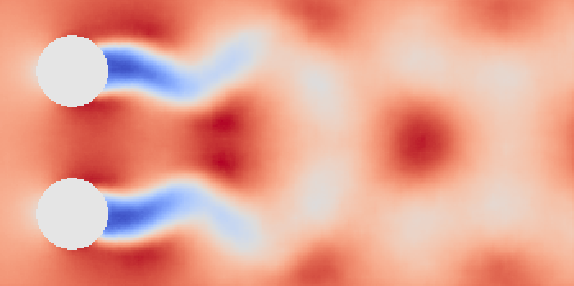}
  \end{subfigure}
  \caption{$\text{Re} = 75$, $\Omega = 0$}
  \label{fig:comp_re75}
\end{subfigure}\vspace{1em}

\begin{subfigure}[h]{0.9\linewidth}
    \begin{subfigure}[b]{0.24 \linewidth}
  \centering
    \includegraphics[width=0.96\linewidth, height=0.48\linewidth]{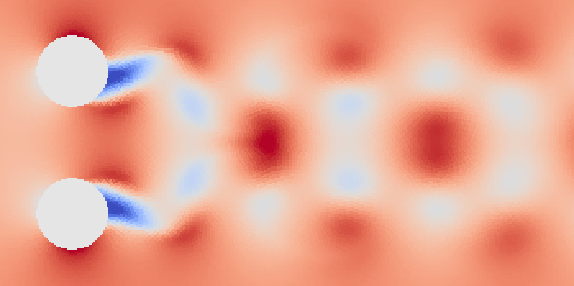}
  \end{subfigure}
  \begin{subfigure}[b]{0.24 \linewidth}
  \centering
    \includegraphics[width=0.96\linewidth, height=0.48\linewidth]{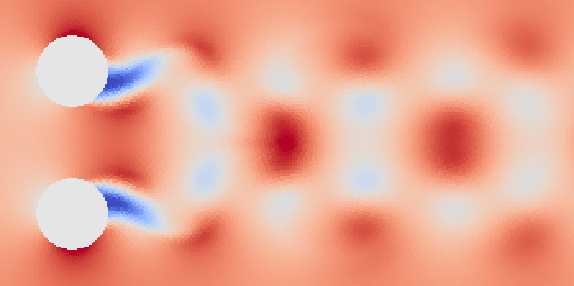}
  \end{subfigure}
 \begin{subfigure}[b]{0.24 \linewidth}
 \centering
   \includegraphics[width=0.96\linewidth, height=0.48\linewidth]{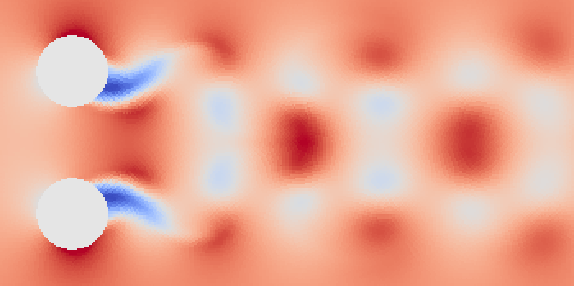}
 \end{subfigure}
 \begin{subfigure}[b]{0.24 \linewidth}
 \centering
   \includegraphics[width=0.96\linewidth, height=0.48\linewidth]{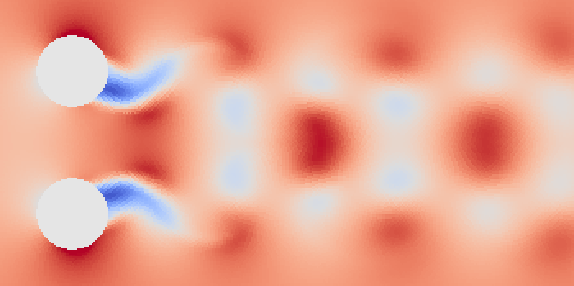}
 \end{subfigure}\vspace{1em}

  \begin{subfigure}[b]{0.24 \linewidth}
  \centering
    \includegraphics[width=0.96\linewidth, height=0.48\linewidth]{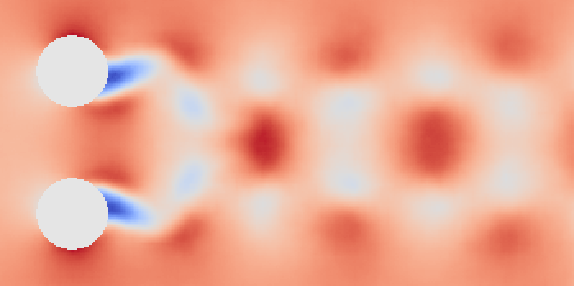}
  \end{subfigure}
 \begin{subfigure}[b]{0.24 \linewidth}
 \centering
   \includegraphics[width=0.96\linewidth, height=0.48\linewidth]{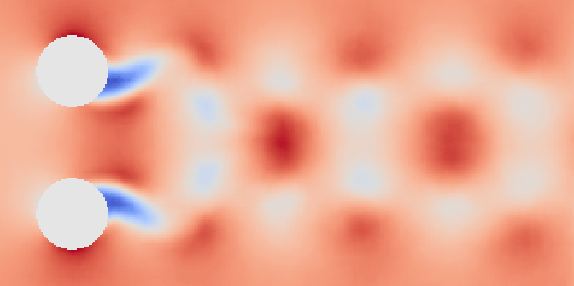}
 \end{subfigure}
 \begin{subfigure}[b]{0.24 \linewidth}
 \centering
   \includegraphics[width=0.96\linewidth, height=0.48\linewidth]{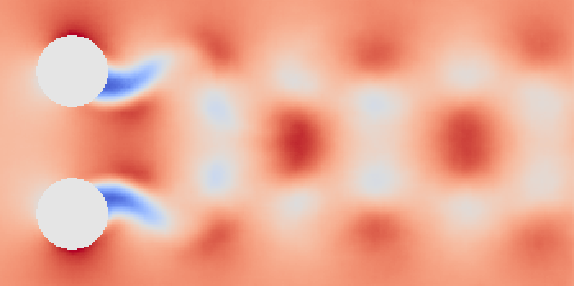}
 \end{subfigure}
  \begin{subfigure}[b]{0.24 \linewidth}
  \centering
    \includegraphics[width=0.96\linewidth, height=0.48\linewidth]{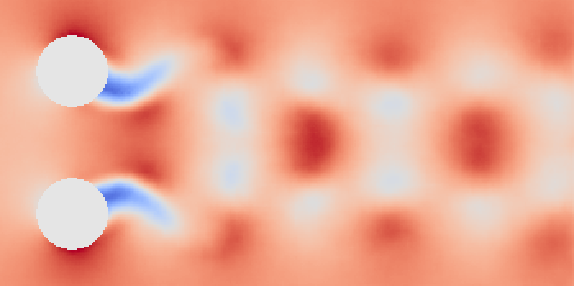}
  \end{subfigure}
  \caption{$\text{Re} = 175$, $\Omega = 1$}
  \label{fig:comp_re175}
\end{subfigure}
\caption[Visual comparison of $x$-velocity over time at various flow conditions]{Visual comparison of $x$-velocity over time at four flow conditions. Solutions generated by the CFD solver are shown in the top row, while flows sampled from the generative model are shown in the bottom row.}
\label{fig: comparison}
\end{figure}

The quality of generated solutions can be evaluated by first qualitatively studying whether the trained model is capable of generating realistic solutions for multiple settings of the prescribed parameter $\cs$.
\Cref{fig: comparison} presents a visual comparison of $x$-velocity values over time at four distinct flow conditions.
For each flow condition, the top row of images corresponds to the ground-truth CFD solutions, while the bottom row contains solutions that were sampled from the generative model.
To facilitate easier comparison, care was taken to approximately align the CFD and generated solutions in time such that both simulations show the system at similar moments within the vortex-shedding cycle.
However, it should be noted that the initial conditions for the respective solutions are not the same, so discrepancies that arise should not necessarily be attributed to modeling errors by the generative model.

The results in \cref{fig: comparison} provide confidence that the generative model effectively captures the changes in flow behavior that occur as the Reynolds number and rotation speed are varied.
Additionally, the generative model is shown to not just generate individual time snapshots that are visually similar to the corresponding fluid flow, it also generates solutions that appear to \emph{evolve} in the same manner as the fluid flow.
This temporal consistency is critical if the generative model is to be used to simulate the system for long time periods.

The most noteworthy takeaway from \cref{fig: comparison} is the visual similarity between the CFD and generated solutions at Reynolds number $\text{Re} = 175$ and rotation speed $\Omega = 1$.
These flow conditions are not contained within the training data, and yet the generative model is still able to generate solutions that are qualitatively similar to the ground-truth fluid flow.
Thus, these results imply that it may be possible to use the trained generative model for simulation at many flow conditions, including flow conditions not contained in the training data.

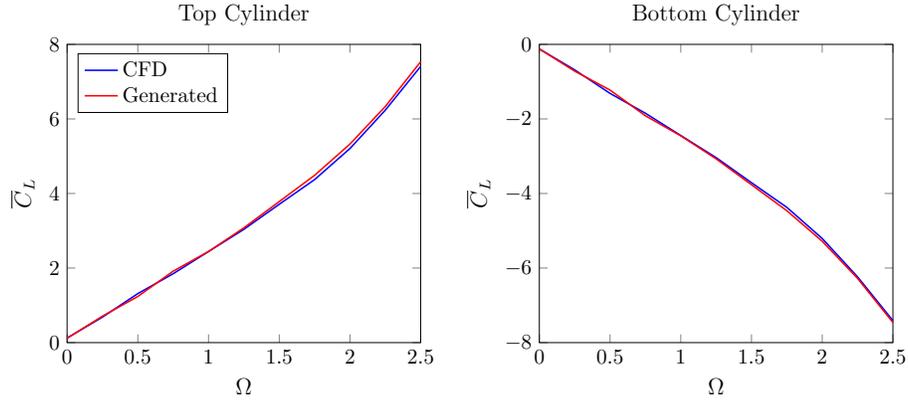
\begin{figure}[t]
\centering
\begin{tikzpicture}
\begin{groupplot}[group style={group size= 2 by 1},height=7cm,width=8cm]
\nextgroupplot[xlabel=\large $\Omega$,
    ylabel=\large $\overline C_L$,
    xmin=0, xmax=2.5,
    ymin=0,
    ymax=8,
    yticklabel style={
        /pgf/number format/fixed,
        /pgf/number format/precision=5
    },
    title=\large Top Cylinder,
    legend pos = north west]

    \addplot +[blue,thick, no markers] 
        table[x=om,y expr=\thisrow{top_cfd}*2/1.4/0.1^2, col sep=comma] {mean_lift.csv};

    \addplot +[red, thick, no markers]
        table[x=om,y expr=\thisrow{top_gen}*2/1.4/0.1^2, col sep=comma] {mean_lift.csv};

    \legend{CFD, Generated}

\nextgroupplot[xlabel=\large $\Omega$,
    ylabel=\large $\overline C_L$,
    xmin=0, xmax=2.5,
    ymin=-8,
    ymax=0,
    yticklabel style={
        /pgf/number format/fixed,
        /pgf/number format/precision=5
    },
    title=\large Bottom Cylinder,
    xshift=3em]

    \addplot +[blue,thick, no markers] 
        table[x=om,y expr=\thisrow{bottom_cfd}*2/1.4/0.1^2, col sep=comma] {mean_lift.csv};

    \addplot +[red, thick, no markers]
        table[x=om,y expr=\thisrow{bottom_gen}*2/1.4/0.1^2, col sep=comma] {mean_lift.csv};

\end{groupplot}
\end{tikzpicture}
\caption[Mean lift coefficient as a function of rotation speed]{Mean lift coefficient as a function of rotation speed for top and bottom cylinders with Reynolds number $\text{Re} = 150$.}
\label{fig:mean_lift}
\end{figure}

\begin{figure}[th!]
\centering
\begin{tikzpicture}
\begin{groupplot}[group style={group size= 4 by 3},height=7cm,width=8cm]
\nextgroupplot[xlabel=\large Reynolds Number,
    ylabel=\large $f_{\max}$,
    xmin=70, xmax=205,
    ymin=0,
    ymax=0.15,
    yticklabel style={
        /pgf/number format/fixed,
        /pgf/number format/precision=5
    },
    title=\Large$\Omega\equal0$]

    \addplot +[blue,very thick, no markers] 
        table[x=t,y=om1, col sep=comma] {lift_freq_cfd.csv};

    \addplot +[red,very thick, no markers]
        table[x=t,y=om1, col sep=comma] {lift_freq_gen.csv};

    \legend{CFD, Generated}

\nextgroupplot[xlabel=\large Reynolds Number,
    ylabel=\large $f_{\max}$,
    xmin=70, xmax=205,
    ymin=0,
    ymax=0.15,
    yticklabel style={
        /pgf/number format/fixed,
        /pgf/number format/precision=5
    },
    title=\Large$\Omega\equal0.25$,
    xshift=3em]

    \addplot +[blue,very thick, no markers] 
        table[x=t,y=om2, col sep=comma] {lift_freq_cfd.csv};

    \addplot +[red,very thick, no markers]
        table[x=t,y=om2, col sep=comma] {lift_freq_gen.csv};

\nextgroupplot[xlabel=\large Reynolds Number,
    ylabel=\large $f_{\max}$,
    xmin=70, xmax=205,
    ymin=0,
    ymax=0.15,
    yticklabel style={
        /pgf/number format/fixed,
        /pgf/number format/precision=5
    },
    title=\Large$\Omega\equal0.5$,
    xshift=3em]

    \addplot +[blue,very thick, no markers] 
        table[x=t,y=om3, col sep=comma] {lift_freq_cfd.csv};

    \addplot +[red,very thick, no markers]
        table[x=t,y=om3, col sep=comma] {lift_freq_gen.csv};

\nextgroupplot[xlabel=\large Reynolds Number,
    ylabel=\large $f_{\max}$,
    xmin=70, xmax=205,
    ymin=0,
    ymax=0.15,
    yticklabel style={
        /pgf/number format/fixed,
        /pgf/number format/precision=5
    },
    title=\Large$\Omega\equal0.75$,
    xshift=3em]

    \addplot +[blue,very thick, no markers] 
        table[x=t,y=om4, col sep=comma] {lift_freq_cfd.csv};

    \addplot +[red,very thick, no markers]
        table[x=t,y=om4, col sep=comma] {lift_freq_gen.csv};

\nextgroupplot[xlabel=\large Reynolds Number,
    ylabel=\large $f_{\max}$,
    xmin=70, xmax=205,
    ymin=0,
    ymax=0.15,
    yticklabel style={
        /pgf/number format/fixed,
        /pgf/number format/precision=5
    },
    title=\Large$\Omega\equal1$,
    yshift=-4em]

    \addplot +[blue,very thick, no markers] 
        table[x=t,y=om5, col sep=comma] {lift_freq_cfd.csv};

    \addplot +[red,very thick, no markers]
        table[x=t,y=om5, col sep=comma] {lift_freq_gen.csv};

\nextgroupplot[xlabel=\large Reynolds Number,
    ylabel=\large $f_{\max}$,
    xmin=70, xmax=205,
    ymin=0,
    ymax=0.15,
    yticklabel style={
        /pgf/number format/fixed,
        /pgf/number format/precision=5
    },
    title=\Large$\Omega\equal1.25$,
    yshift=-4em]

    \addplot +[blue,very thick, no markers] 
        table[x=t,y=om6, col sep=comma] {lift_freq_cfd.csv};

    \addplot +[red,very thick, no markers]
        table[x=t,y=om6, col sep=comma] {lift_freq_gen.csv};

\nextgroupplot[xlabel=\large Reynolds Number,
    ylabel=\large $f_{\max}$,
    xmin=70, xmax=205,
    ymin=0,
    ymax=0.15,
    yticklabel style={
        /pgf/number format/fixed,
        /pgf/number format/precision=5
    },
    title=\Large$\Omega\equal1.5$,
    yshift=-4em]

    \addplot +[blue,very thick, no markers] 
        table[x=t,y=om7, col sep=comma] {lift_freq_cfd.csv};

    \addplot +[red,very thick, no markers]
        table[x=t,y=om7, col sep=comma] {lift_freq_gen.csv};

\end{groupplot}
\end{tikzpicture}
\caption[Variation of dominant frequency in lift signal]{Variation of dominant frequency in lift signal as a function of rotation speed and Reynolds number for CFD and generated solutions.}
\label{fig:freq_lift}
\end{figure}

A more quantitative manner of assessing the accuracy of the generated solutions is to look at how the lift forces acting upon the cylinder vary with prescribed Reynolds number and rotation speed.
As noted previously in the discussion about \cref{fig:lift_plot}, the lift curves for the top and bottom cylinders tend to separate as the rotation speed is increased, with the mean lift on the top cylinder becoming more positive and the mean lift on the bottom cylinder becoming more negative.
To examine if the generated solutions are capable of recreating this trend, \cref{fig:mean_lift} presents the mean lift coefficient for the top and bottom cylinders as a function of rotation speed for flow at Reynolds number $\text{Re} = 150$.
It is clear from these plots that the mean lift coefficients calculated from the generated solutions are nearly identical to those determined from the CFD solutions.

In addition to trends in the mean lift coefficient, it can also be studied whether the generative model is capable of capturing variations in the \emph{frequency content} of the lift curves.
The time variation of the lift coefficient values presented in \cref{fig:lift_plot} is roughly sinusoidal, meaning that the corresponding power spectral density (PSD) should show a single peak at a dominant frequency denoted by $f_{\max}$.
\Cref{fig:freq_lift} shows how the dominant frequency in the lift signal for the top cylinder varies with rotation speed and Reynolds number in both the CFD and generated solutions.
Results are only shown up to a rotation speed of $\Omega = 1.5$ because $\Omega =1.75$ is beyond the critical rotation speed for lower Reynolds numbers.
The curves associated with CFD solutions show an increase in the dominant frequency with Reynolds number and a slight decrease with rotation speed.
The curves associated with the generated solutions appear to capture these trends quite well, with only minor discrepancies visible between the dominant frequencies identified from the CFD and generated solutions.

\begin{figure}[t]
\centering
\includegraphics[width=0.4\linewidth]{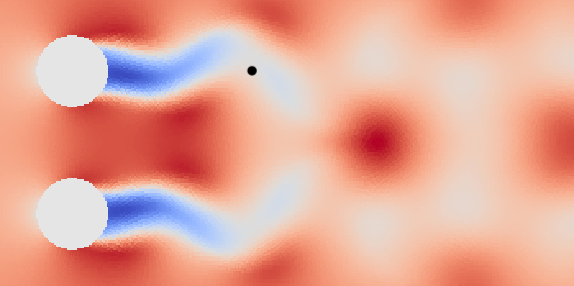}
\caption[Depiction of wake measurement location]{Depiction of wake measurement location.}
\label{fig:wake_meas}
\end{figure}

Lift is determined by integrating pressure values over the surface of an object.
As such, lift values can be largely insensitive to local errors in the pressure distribution, as long as errors at one location on the object surface are balanced out by errors elsewhere.
Therefore, the next set of numerical experiments evaluates how accurately the generative model is able to simulate the time evolution of flow quantities at a single point in space.
The selected point, represented by the black dot in \cref{fig:wake_meas}, is positioned directly in the wake of the top cylinder, located approximately two cylinder diameters downstream.
The following discussion is restricted to the modeling of pressure values at this measurement location; similar results were obtained for density, $x$-velocity, and $y$-velocity.

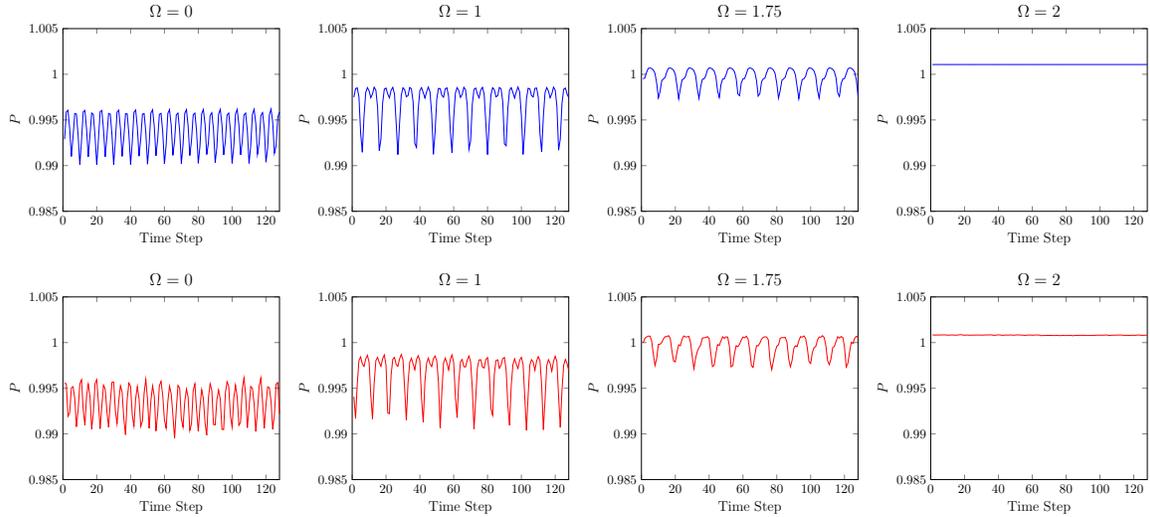
\begin{figure}[t]
\centering
\begin{tikzpicture}
\begin{groupplot}[group style={group size= 4 by 2},height=7cm,width=8cm]
\nextgroupplot[xlabel=\large Time Step,
    ylabel=\large $P$,
    xmin=0, xmax=128,
    xtick={0,20,40,...,120},
    ymin=0.985,
    ymax=1.005,
    yticklabel style={
        /pgf/number format/fixed,
        /pgf/number format/precision=5
    },
    title=\Large$\Omega\equal0$]

    \addplot +[blue,thick, no markers] 
        table[x=t,y=om1, col sep=comma] {press_cfd.csv};

\nextgroupplot[xlabel=\large Time Step,
    ylabel=\large $P$,
    xmin=0, xmax=128,
    xtick={0,20,40,...,120},
    ymin=0.985,
    ymax=1.005,
    yticklabel style={
        /pgf/number format/fixed,
        /pgf/number format/precision=5
    },
    title=\Large$\Omega\equal1$,
    xshift=3em]

    \addplot +[blue,thick, no markers] 
        table[x=t,y=om5, col sep=comma] {press_cfd.csv};

\nextgroupplot[xlabel=\large Time Step,
    ylabel=\large $P$,
    xmin=0, xmax=128,
    xtick={0,20,40,...,120},
    ymin=0.985,
    ymax=1.005,
    yticklabel style={
        /pgf/number format/fixed,
        /pgf/number format/precision=5
    },
    title=\Large$\Omega\equal1.75$,
    xshift=3em]

    \addplot +[blue,thick, no markers] 
        table[x=t,y=om8, col sep=comma] {press_cfd.csv};

\nextgroupplot[xlabel=\large Time Step,
    ylabel=\large $P$,
    xmin=0, xmax=128,
    xtick={0,20,40,...,120},
    ymin=0.985,
    ymax=1.005,
    yticklabel style={
        /pgf/number format/fixed,
        /pgf/number format/precision=5
    },
    title=\Large$\Omega\equal2$,
    xshift=3em]

    \addplot +[blue,thick, no markers] 
        table[x=t,y=om9, col sep=comma] {press_cfd.csv};

\nextgroupplot[xlabel=\large Time Step,
    ylabel=\large $P$,
    xmin=0, xmax=128,
    xtick={0,20,40,...,120},
    ymin=0.985,
    ymax=1.005,
    yticklabel style={
        /pgf/number format/fixed,
        /pgf/number format/precision=5
    },
    title=\Large$\Omega\equal0$,
    yshift=-4em]

    \addplot +[red,thick, no markers] 
        table[x=t,y=om1, col sep=comma] {press_gen.csv};

\nextgroupplot[xlabel=\large Time Step,
    ylabel=\large $P$,
    xmin=0, xmax=128,
    xtick={0,20,40,...,120},
    ymin=0.985,
    ymax=1.005,
    yticklabel style={
        /pgf/number format/fixed,
        /pgf/number format/precision=5
    },
    title=\Large$\Omega\equal1$,
    yshift=-4em]

    \addplot +[red,thick, no markers] 
        table[x=t,y=om5, col sep=comma] {press_gen.csv};

\nextgroupplot[xlabel=\large Time Step,
    ylabel=\large $P$,
    xmin=0, xmax=128,
    xtick={0,20,40,...,120},
    ymin=0.985,
    ymax=1.005,
    yticklabel style={
        /pgf/number format/fixed,
        /pgf/number format/precision=5
    },
    title=\Large$\Omega\equal1.75$,
    yshift=-4em]

    \addplot +[red,thick, no markers] 
        table[x=t,y=om8, col sep=comma] {press_gen.csv};

\nextgroupplot[xlabel=\large Time Step,
    ylabel=\large $P$,
    xmin=0, xmax=128,
    xtick={0,20,40,...,120},
    ymin=0.985,
    ymax=1.005,
    yticklabel style={
        /pgf/number format/fixed,
        /pgf/number format/precision=5
    },
    title=\Large$\Omega\equal2$,
    yshift=-4em]

    \addplot +[red,thick, no markers] 
        table[x=t,y=om9, col sep=comma] {press_gen.csv};

\end{groupplot}
\end{tikzpicture}
\caption[Pressure over time for CFD and generated solutions]{Comparison of pressure values over time for CFD solutions (top) and generated solutions (bottom) as a function of rotation speed at Reynolds number $\text{Re} = 150$.}
\label{fig:press_cfd_plot}
\end{figure}

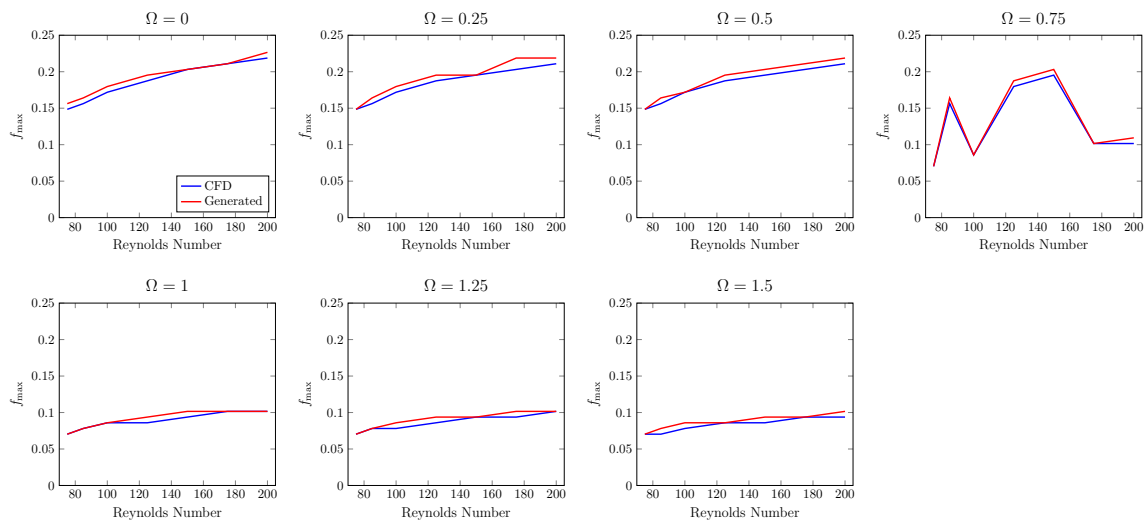
\begin{figure}[th]
\centering
\begin{tikzpicture}
\begin{groupplot}[group style={group size= 4 by 3},height=7cm,width=8cm]
\nextgroupplot[xlabel=\large Reynolds Number,
    ylabel=\large $f_{\max}$,
    xmin=70, xmax=205,
    ymin=0,
    ymax=0.25,
    yticklabel style={
        /pgf/number format/fixed,
        /pgf/number format/precision=5
    },
    legend pos = south east,
    title=\Large$\Omega\equal0$]

    \addplot +[blue,very thick, no markers] 
        table[x=t,y=om1, col sep=comma] {press_freq_cfd.csv};

    \addplot +[red,very thick, no markers]
        table[x=t,y=om1, col sep=comma] {press_freq_gen.csv};

    \legend{CFD, Generated}

\nextgroupplot[xlabel=\large Reynolds Number,
    ylabel=\large $f_{\max}$,
    xmin=70, xmax=205,
    ymin=0,
    ymax=0.25,
    yticklabel style={
        /pgf/number format/fixed,
        /pgf/number format/precision=5
    },
    title=\Large$\Omega\equal0.25$,
    xshift=3em]

    \addplot +[blue,very thick, no markers] 
        table[x=t,y=om2, col sep=comma] {press_freq_cfd.csv};

    \addplot +[red,very thick, no markers]
        table[x=t,y=om2, col sep=comma] {press_freq_gen.csv};

\nextgroupplot[xlabel=\large Reynolds Number,
    ylabel=\large $f_{\max}$,
    xmin=70, xmax=205,
    ymin=0,
    ymax=0.25,
    yticklabel style={
        /pgf/number format/fixed,
        /pgf/number format/precision=5
    },
    title=\Large$\Omega\equal0.5$,
    xshift=3em]

    \addplot +[blue,very thick, no markers] 
        table[x=t,y=om3, col sep=comma] {press_freq_cfd.csv};

    \addplot +[red,very thick, no markers]
        table[x=t,y=om3, col sep=comma] {press_freq_gen.csv};

\nextgroupplot[xlabel=\large Reynolds Number,
    ylabel=\large $f_{\max}$,
    xmin=70, xmax=205,
    ymin=0,
    ymax=0.25,
    yticklabel style={
        /pgf/number format/fixed,
        /pgf/number format/precision=5
    },
    title=\Large$\Omega\equal0.75$,
    xshift=3em]

    \addplot +[blue,very thick, no markers] 
        table[x=t,y=om4, col sep=comma] {press_freq_cfd.csv};

    \addplot +[red,very thick, no markers]
        table[x=t,y=om4, col sep=comma] {press_freq_gen.csv};

\nextgroupplot[xlabel=\large Reynolds Number,
    ylabel=\large $f_{\max}$,
    xmin=70, xmax=205,
    ymin=0,
    ymax=0.25,
    yticklabel style={
        /pgf/number format/fixed,
        /pgf/number format/precision=5
    },
    title=\Large$\Omega\equal1$,
    yshift=-4em]

    \addplot +[blue,very thick, no markers] 
        table[x=t,y=om5, col sep=comma] {press_freq_cfd.csv};

    \addplot +[red,very thick, no markers]
        table[x=t,y=om5, col sep=comma] {press_freq_gen.csv};

\nextgroupplot[xlabel=\large Reynolds Number,
    ylabel=\large $f_{\max}$,
    xmin=70, xmax=205,
    ymin=0,
    ymax=0.25,
    yticklabel style={
        /pgf/number format/fixed,
        /pgf/number format/precision=5
    },
    title=\Large$\Omega\equal1.25$,
    yshift=-4em]

    \addplot +[blue,very thick, no markers] 
        table[x=t,y=om6, col sep=comma] {press_freq_cfd.csv};

    \addplot +[red,very thick, no markers]
        table[x=t,y=om6, col sep=comma] {press_freq_gen.csv};

\nextgroupplot[xlabel=\large Reynolds Number,
    ylabel=\large $f_{\max}$,
    xmin=70, xmax=205,
    ymin=0,
    ymax=0.25,
    yticklabel style={
        /pgf/number format/fixed,
        /pgf/number format/precision=5
    },
    title=\Large$\Omega\equal1.5$,
    yshift=-4em]

    \addplot +[blue,very thick, no markers] 
        table[x=t,y=om7, col sep=comma] {press_freq_cfd.csv};

    \addplot +[red,very thick, no markers]
        table[x=t,y=om7, col sep=comma] {press_freq_gen.csv};

\end{groupplot}
\end{tikzpicture}
\caption[Variation of dominant frequency in pressure signal]{Variation of dominant frequency in pressure signal as a function of rotation speed and Reynolds number for CFD and generated solutions.}
\label{fig:freq_press}
\end{figure}

\Cref{fig:press_cfd_plot} shows the time evolution of pressure values at the measurement location as a function of rotation speed for flow at Reynolds number $\text{Re} = 150$.
The top row shows the pressure signal obtained from the CFD simulations, while the bottom row shows the corresponding results for the generated solutions.
These plots show that the nature of the pressure signal changes dramatically as the rotation speed increases, and the generative model is able to track these changes quite well.
Furthermore, note that the generative model correctly predicts a steady solution at $\Omega = 2$.

Once again, the similarity between these signals can be evaluated quantitatively by considering the peaks in their power spectral density.
In contrast with the PSD for the lift signals, the PSD for the pressure signals will generally have more than one peak.
The comparison here thus considers only the frequency corresponding to the largest peak.
\Cref{fig:freq_press} shows the variation of $f_{\max}$ with rotation speed and Reynolds number, and demonstrates strong agreement between the curves associated with the CFD and generated solutions.
The largest relative error across all flow quantities, Reynolds numbers, and rotation speeds is $11\%$, suggesting that the generative model is quite effective at modeling the local properties of the flow.
Furthermore, note that these curves include results from Reynolds number $\text{Re} = 175$, which is not present in the training data, implying that the model could be effective at modeling the flow properties even at unobserved flow conditions.

\begin{figure}[t]
\centering
\begin{subfigure}[b]{0.4\linewidth}
\centering
\begin{tikzpicture}
\begin{axis}[xlabel=\large $\Omega$,
    ylabel=\large Norm of Time Derivative,
    xmin=1.1, xmax=2.1,
    ymin=-0,
    yticklabel style={
        /pgf/number format/fixed,
        /pgf/number format/precision=5
    }]

    \addplot +[blue, very thick, no markers] 
        table[x=om,y=diff, col sep=comma] {time_diffs.csv};

    \addplot +[black, thick, dashed, no markers] 
        coordinates { (1.1,3) (2.1,3) };

      \addplot +[black, thick, dashed, no markers] 
        coordinates { (1.1,1.5) (2.1,1.5) };

\end{axis}
\end{tikzpicture}
\end{subfigure}\hspace{4em}
\begin{subfigure}[b]{0.4\linewidth}
\centering
\begin{tikzpicture}
\begin{axis}[xlabel=\large Reynolds Number,
    ylabel=\large $\Omega$,
    xmin=70, xmax=205,
    ymin=-0.2,
    ymax=2.7,
    ytick={0,1,2},
    yticklabel style={
        /pgf/number format/fixed,
        /pgf/number format/precision=5
    }]

    \addplot +[draw=none, no markers, name path=A] 
        table[x=re,y=om_lo, col sep=comma] {om_crit_bounds.csv};
    \addplot +[draw=none, no markers, name path=B] 
        table[x=re,y=om_hi, col sep=comma] {om_crit_bounds.csv};
    \addplot[colorPurplePastel] fill between[of=A and B];

    \addplot +[only marks, mark options={black, thick}, mark=x] 
        table[x=re,y=om_crit, col sep=comma] {om_crit.csv};

\end{axis}
\end{tikzpicture}
\end{subfigure}
\caption[Identification of critical rotation speed]{Left: Norm of time derivative in generated solutions as a function of $\Omega$ at $\text{Re} = 200$. Right: Plot of parameter space with points representing identified critical rotation speeds and the plausible region represented by the shaded area.}
\label{fig:diffs_crit}
\end{figure}
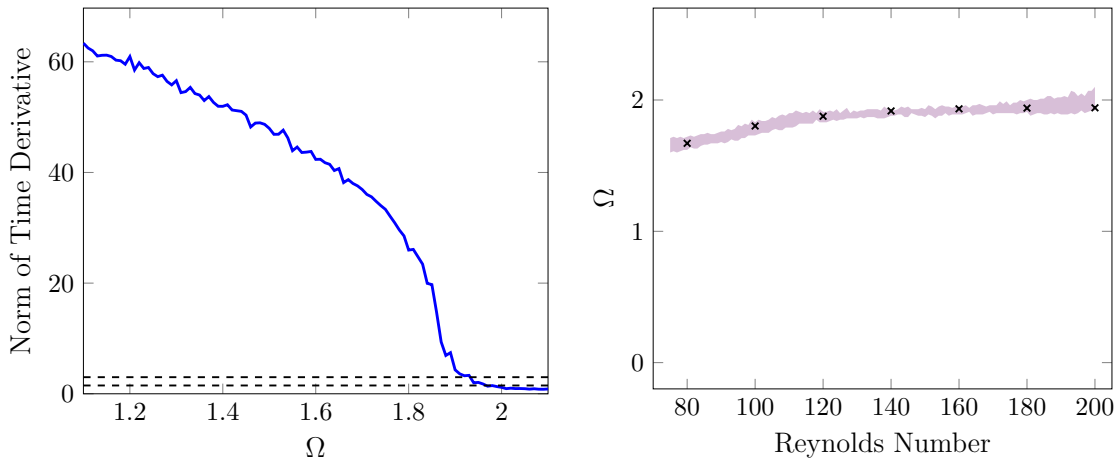

Having established confidence that the generative model can be used for accurate simulation at multiple flow conditions, the model is now used to carry out an experiment intended to mimic a design procedure.
A design problem typically consists of a set of parameters and objectives, and the goal is to identify design parameter values that perform well according to the design objectives.
Exhaustive searches of the design space can be impractical if evaluations of individual design parameters are expensive, as is the case with CFD simulations.
However, given the relative low cost of sampling solutions from the generative model, it may be possible to rapidly simulate a given system at a variety of design parameters with the generative model in order to identify promising regions of the design space.
In this experiment, the parameter-conditioned generative model is incorporated into a design-like procedure, where an exhaustive search across the parameter space is performed in order to estimate how the critical rotation speed, $\Omega_\text{crit}$, varies with Reynolds number.

The goal in this experiment is to identify, as a function of Reynolds number, a plausible range of rotation speeds within which the true critical rotation speed is likely to lie.
This is accomplished by first using the generative model to generate solutions at a large number of Reynolds numbers and rotation speeds: $\text{Re} \in \{75, 76, \ldots, 200\}$ and $\Omega \in \{1.10,1.11,\ldots, 2.10\}$.
This requires the generative model to simulate the cylinder system at nearly \num{13000} different flow conditions.
In total, running these simulations with the generative model takes a little under \num{10} hours; in contrast, running the same number of simulations with a CFD solver would take nearly two months.

The critical rotation speed corresponds to the transition from unsteady to steady flow.
Thus, the critical rotation speed can be identified by considering the time variation between successive snapshots in generated solutions; this time variation should go to zero when steady flow is achieved.
The blue line in the left plot of \cref{fig:diffs_crit} shows the norm of the numerically estimated time derivative of generated solutions as a function of rotation speed for $\text{Re} = 200$.
Note that the time derivatives do tend to decrease with rotation speed, with a sharp drop-off observed at $\Omega \approx 1.9$.
Given the smooth variation in the time derivatives, it is somewhat difficult to identify definitively which rotation speed should correspond to $\Omega_\text{crit}$.
Instead, two time-derivative thresholds, represented by the black dotted lines in \cref{fig:diffs_crit}, are considered, which are assumed to yield a lower- and upper-bound on the true critical rotation speed.
By considering these two thresholds, a plausible range for the critical rotation speed:
\begin{equation}
\Omega_\text{low} \leq \Omega_\text{crit} \leq \Omega_\text{high}
\end{equation}
can be identified.

Furthermore, by identifying the plausible range at each Reynolds number, a plausible \emph{region} of the parameter space can be determined, wherein all critical rotation speeds are assumed to lie.
The shaded area in the right plot of \cref{fig:diffs_crit} represents the plausible region identified from the simulations generated by the parameter-conditioned generative model.
The size of this plausible region represents a significant reduction relative to the full parameter space considered in \cref{fig:re_om_scatter}.
To determine if this plausible region is reasonable, a bisection search was run at Reynolds numbers $\text{Re} \in \{80, 100, 120, 140, 160, 180, 200\}$ to find the true critical rotation speeds according to the CFD solver.
These identified values are represented by the plotted points in \cref{fig:diffs_crit}, and, importantly, are all found to lie within the plausible region.

This section demonstrated that parameter-conditioned generative models can be used to efficiently and accurately simulate two-dimensional fluid flows.
The next section studies whether the same techniques can be applied to modeling three-dimensional flows.

\subsection{Three-Dimensional Flow over a Half-Cylinder}
The test case under consideration is three-dimensional flow over an extruded half-cylinder at an effectively incompressible Mach number of $M = 0.2$.
This flow exhibits several complex flow features, including separated shear layers, turbulent transition, and a fully turbulent wake.
The dynamics of the half-cylinder are noteworthy relative to the circular cylinder because the flow separation point remains fixed at the boundary between the curved and flat surfaces.
Flow over this configuration has been the subject of several numerical and experimental studies~\cite{nakamura1996vortex,kumarasamy1997computation,santa2005characterization,carlberg2019recovering}.

\begin{figure}[t]
\begin{center}
\includegraphics[width=0.5\linewidth]{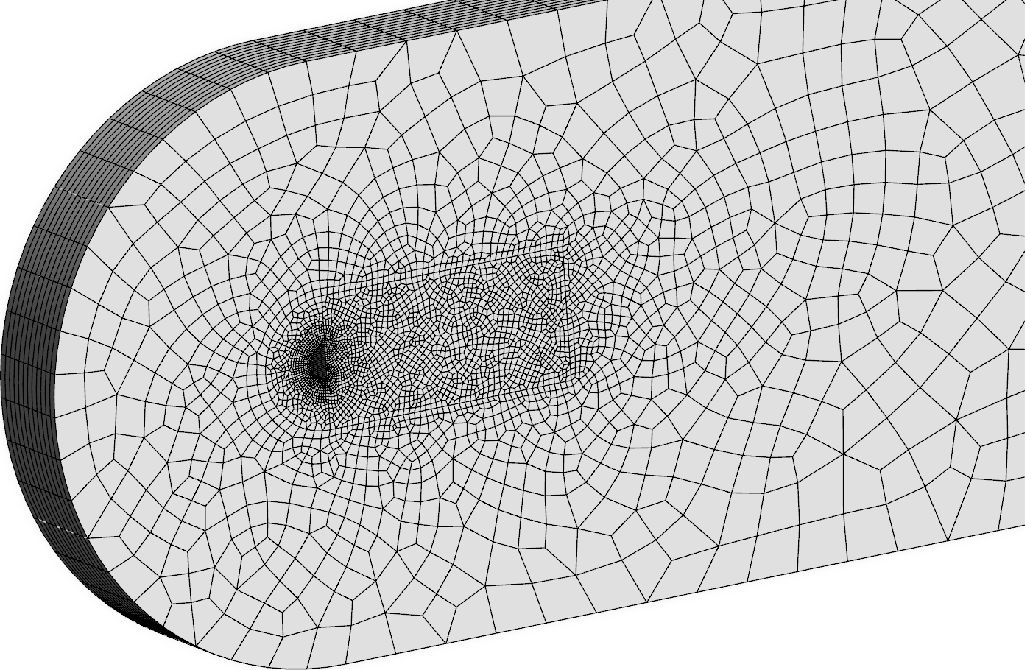}
\caption[Mesh visualization for the half-cylinder domain]{Visualization of the mesh for the half-cylinder domain.}
\label{fig:mesh}
\end{center}
\end{figure}

In the given test case, the cylinder is taken to have a diameter $d = 1$ along its major axis.
The domain is taken to be $[-9, 25]$, $[-9, 9]$, and $[0, \pi]$ in the stream-, cross-, and span-wise directions, and the cylinder is positioned such that the back surface is centered at $(0, 0, 0)$.
A visualization of the mesh can be found in \cref{fig:mesh}.
The domain is divided into $N_\text{el} = 40584$ non-overlapping, conforming hexahedral elements.
As in \cref{sec:double_cylinder}, the system is simulated with the PyFR solver.

Because it is infeasible to train a model on data from the entire computational domain, the models are once again trained only on data sampled from equispaced points in the vicinity of the cylinder.
The data at these points is obtained through multi-linear interpolation, where the ParaView software~\cite{ahrens2005paraview} is used to perform both the interpolation and flow visualizations.
The sampling must be performed on a grid with sufficient fineness such that the vortical structures in the flow can be recovered.
Thus, samples are drawn at a grid of points that is $128 \times 64 \times 32$, spanning a domain of $[-1, 11]$, $[-3.75, 3.75]$, and $[0, \pi]$ in the stream-, cross-, and span-wise directions, respectively.
This sampling procedure yields training inputs that are $128 \times 64 \times 32 \times 5$, where the first three input dimensions correspond to spatial dimensions and the final input dimension corresponds to the five modeled flow quantities: density, $x$-momentum, $y$-momentum, $z$-momentum, and pressure.

\begin{figure}[t]
\centering
\begin{subfigure}[b]{0.45\linewidth}
\centering
\includegraphics[height=0.6\linewidth]{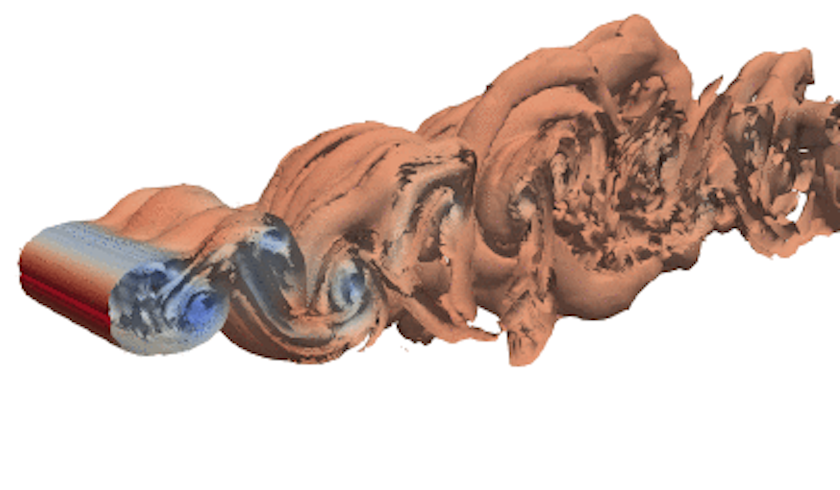}
\caption{Full solution}
\end{subfigure}\hfill
\begin{subfigure}[b]{0.45\linewidth}
\centering
\includegraphics[height=0.6\linewidth]{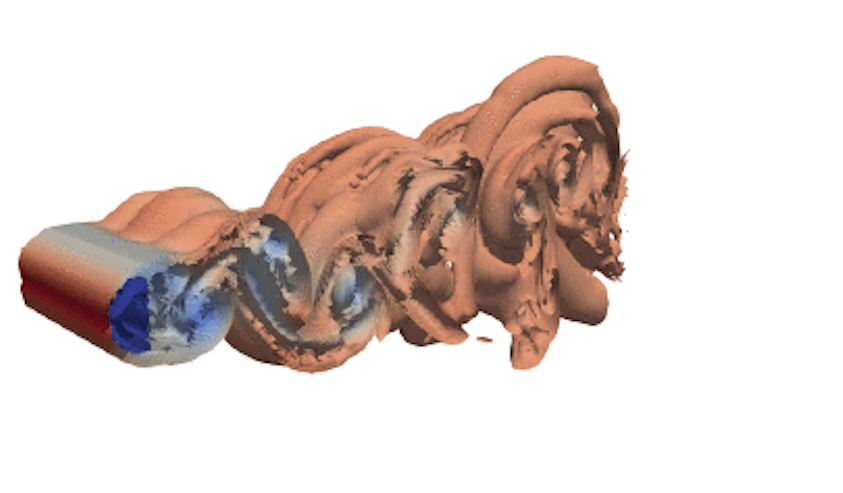}
\caption{Solution at equispaced points}
\end{subfigure}
\caption[Visualization of CFD-simulated and sampled flow]{Visual comparison between flow generated by CFD simulation of entire computational domain and solution obtained by sampling flow quantities at a grid of equispaced points. The visualizations show iso-surfaces of vorticity magnitude, colored according to pressure.}
\label{fig:sampling}
\end{figure}

\Cref{fig:sampling} shows a qualitative comparison of the flow obtained by running CFD simulations across the entire computational domain against the flow obtained by sampling the flow quantities on the specified grid of points.
The visualizations show iso-surfaces of vorticity magnitude, colored according to pressure.
It can be seen that the sampled solution retains many of the key features of the original flow, but does not contain any information about the fluid flow far downstream of the cylinder.
Hence, the sampled solution represents an approximation to the flow field obtained through CFD simulations, but such an approximation is necessary for training to be tractable.

\subsection{Three-Dimensional Flow Training Details}
In this experiment, only the Reynolds number is varied.
Training data is generated by simulating the half-cylinder system at Reynolds numbers of $\text{Re} \in \{175, 200, 250, 300\}$.
The flow at these Reynolds numbers exhibits a range of behaviors, with solutions at $\text{Re} = 175$ showing mostly laminar flow with some turbulence downstream, and solutions at $\text{Re} = 300$ exhibiting a significant amount of turbulence in the cylinder wake.
The neural network architectures for the generative model are largely identical to those described for the two-dimensional flow in \cref{sec:training_2d}, with one notable exception: the two-dimensional convolutions in the feature extractor and decoder network are replaced with three-dimensional convolutions. 
Additionally, to account for the added complexities inherent to modeling three-dimensional, turbulent flow, \num{32}-dimensional latent states are used in place of the four-dimensional latent states employed previously.

Parameter-conditioned generative models are trained on \num{165} distinct sequences from each flow condition.
Training a single model takes approximately three days with the model parameters divided across two NVIDIA K40 GPUs and one NVIDIA GeForce GTX 1070 GPU.
The longer training time can be attributed to the larger size of the training inputs and the additional computational cost associated with performing three-dimensional convolutions.
Flow simulations generated by the trained model are once again found to be approximately \num{120} times faster than the corresponding CFD simulations.
The next section studies how well the generative models are able to model the fluid flow.

\begin{figure}
\centering
\begin{subfigure}[h]{0.9\linewidth}
    \begin{subfigure}[b]{0.24 \linewidth}
  \centering
    \includegraphics[width=0.96\linewidth, height=0.48\linewidth]{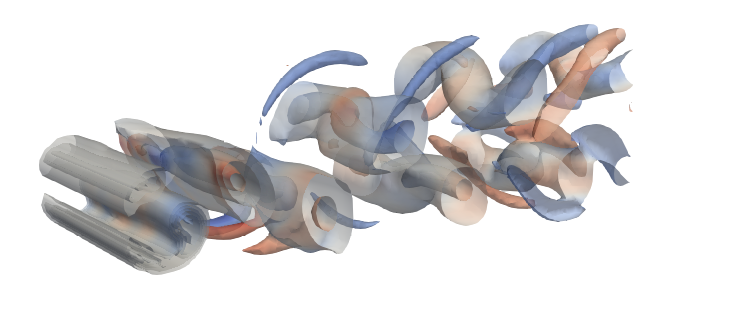}
  \end{subfigure}
  \begin{subfigure}[b]{0.24 \linewidth}
  \centering
    \includegraphics[width=0.96\linewidth, height=0.48\linewidth]{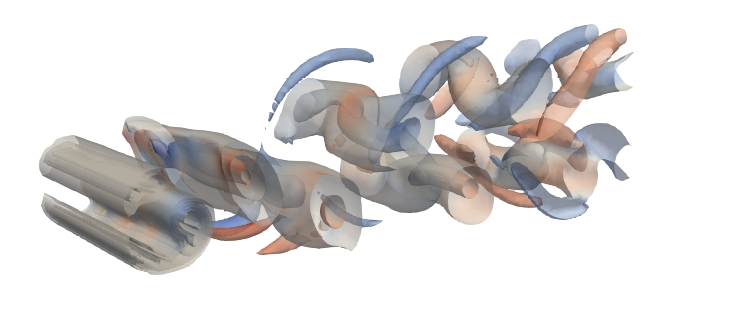}
  \end{subfigure}
 \begin{subfigure}[b]{0.24 \linewidth}
 \centering
   \includegraphics[width=0.96\linewidth, height=0.48\linewidth]{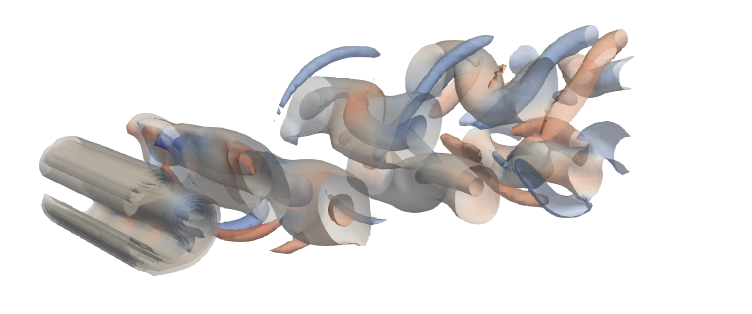}
 \end{subfigure}
 \begin{subfigure}[b]{0.24 \linewidth}
 \centering
   \includegraphics[width=0.96\linewidth, height=0.48\linewidth]{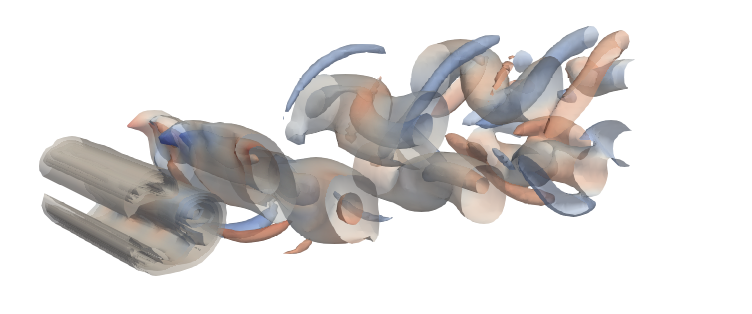}
 \end{subfigure}\vspace{1em}

  \begin{subfigure}[b]{0.24 \linewidth}
  \centering
    \includegraphics[width=0.96\linewidth, height=0.48\linewidth]{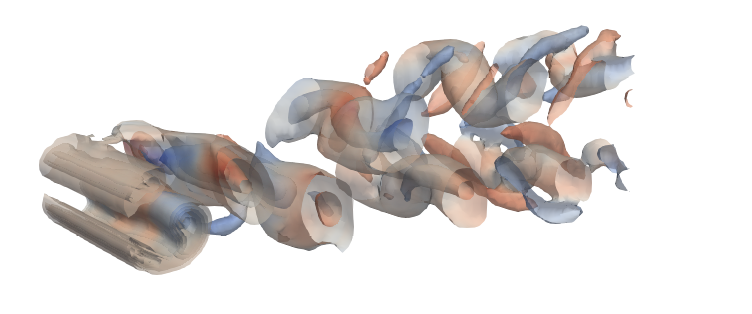}
  \end{subfigure}
 \begin{subfigure}[b]{0.24 \linewidth}
 \centering
   \includegraphics[width=0.96\linewidth, height=0.48\linewidth]{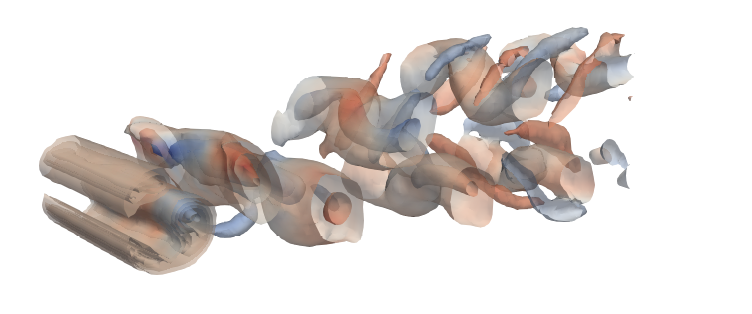}
 \end{subfigure}
 \begin{subfigure}[b]{0.24 \linewidth}
 \centering
   \includegraphics[width=0.96\linewidth, height=0.48\linewidth]{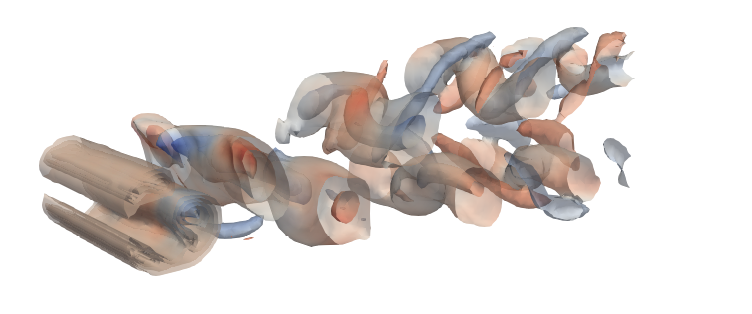}
 \end{subfigure}
  \begin{subfigure}[b]{0.24 \linewidth}
  \centering
    \includegraphics[width=0.96\linewidth, height=0.48\linewidth]{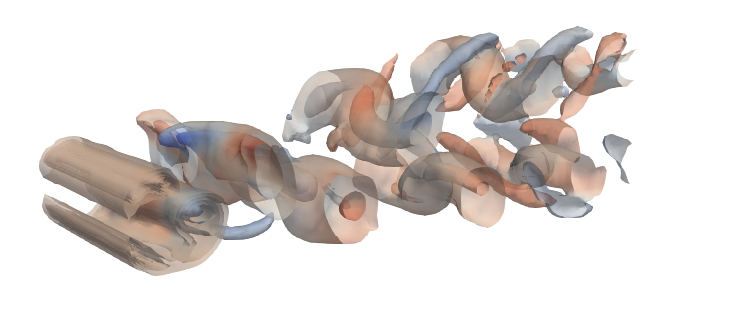}
  \end{subfigure}
  \caption{$\text{Re} = 175$}
\end{subfigure}\vspace{1em}

\begin{subfigure}[h]{0.9\linewidth}
    \begin{subfigure}[b]{0.24 \linewidth}
  \centering
    \includegraphics[width=0.96\linewidth, height=0.48\linewidth]{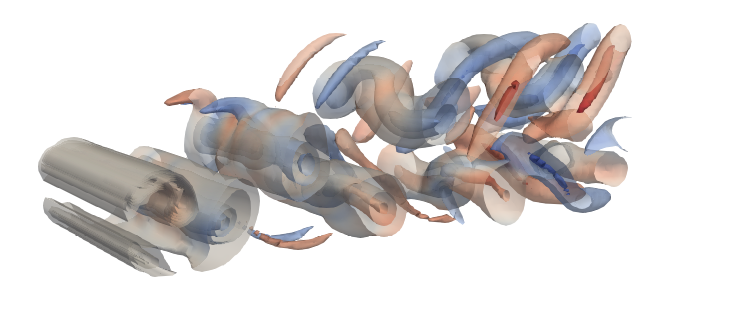}
  \end{subfigure}
  \begin{subfigure}[b]{0.24 \linewidth}
  \centering
    \includegraphics[width=0.96\linewidth, height=0.48\linewidth]{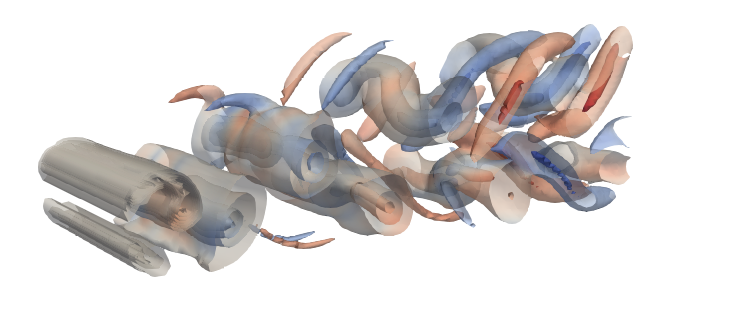}
  \end{subfigure}
 \begin{subfigure}[b]{0.24 \linewidth}
 \centering
   \includegraphics[width=0.96\linewidth, height=0.48\linewidth]{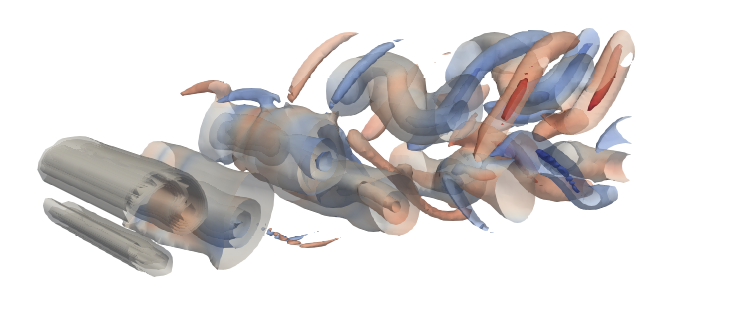}
 \end{subfigure}
 \begin{subfigure}[b]{0.24 \linewidth}
 \centering
   \includegraphics[width=0.96\linewidth, height=0.48\linewidth]{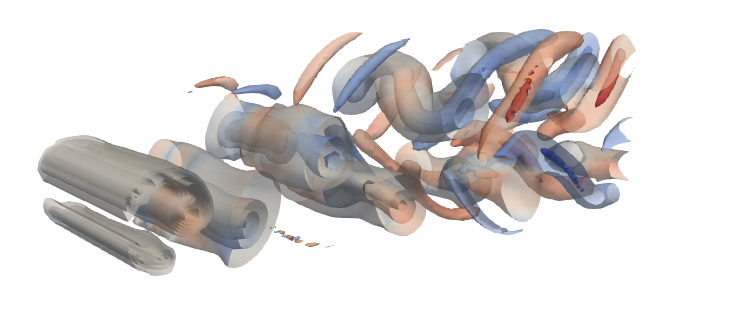}
 \end{subfigure}\vspace{1em}

  \begin{subfigure}[b]{0.24 \linewidth}
  \centering
    \includegraphics[width=0.96\linewidth, height=0.48\linewidth]{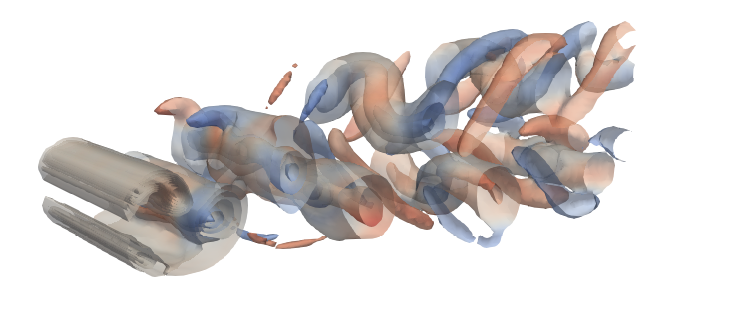}
  \end{subfigure}
 \begin{subfigure}[b]{0.24 \linewidth}
 \centering
   \includegraphics[width=0.96\linewidth, height=0.48\linewidth]{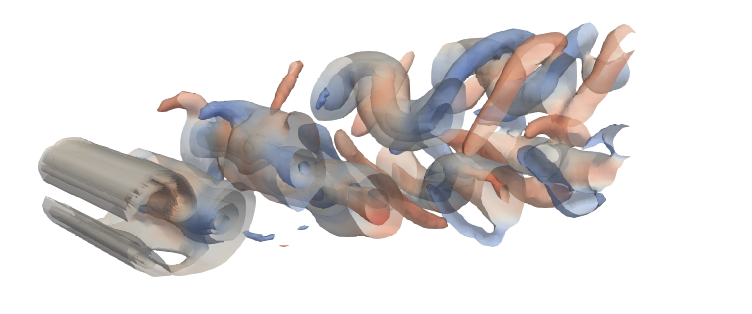}
 \end{subfigure}
 \begin{subfigure}[b]{0.24 \linewidth}
 \centering
   \includegraphics[width=0.96\linewidth, height=0.48\linewidth]{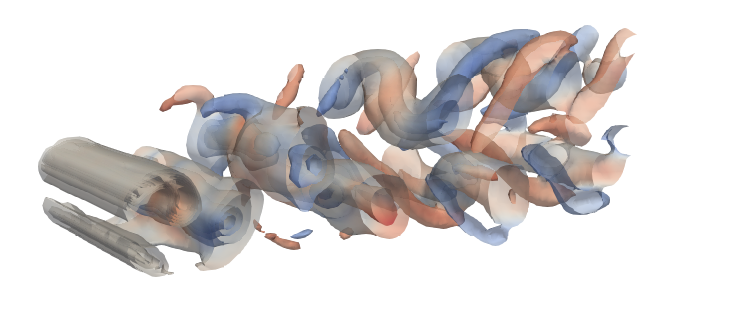}
 \end{subfigure}
  \begin{subfigure}[b]{0.24 \linewidth}
  \centering
    \includegraphics[width=0.96\linewidth, height=0.48\linewidth]{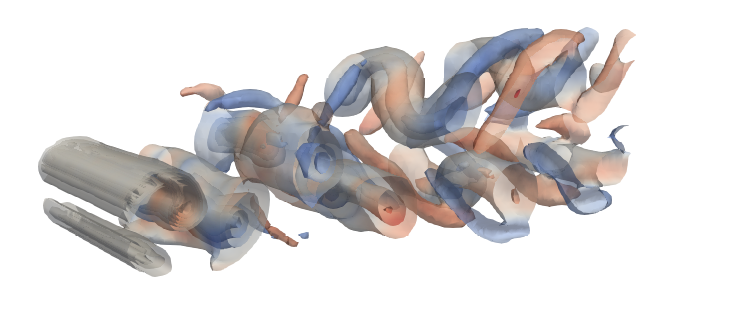}
  \end{subfigure}
  \caption{$\text{Re} = 200$}
\end{subfigure}\vspace{1em}

\begin{subfigure}[h]{0.9\linewidth}
    \begin{subfigure}[b]{0.24 \linewidth}
  \centering
    \includegraphics[width=0.96\linewidth, height=0.48\linewidth]{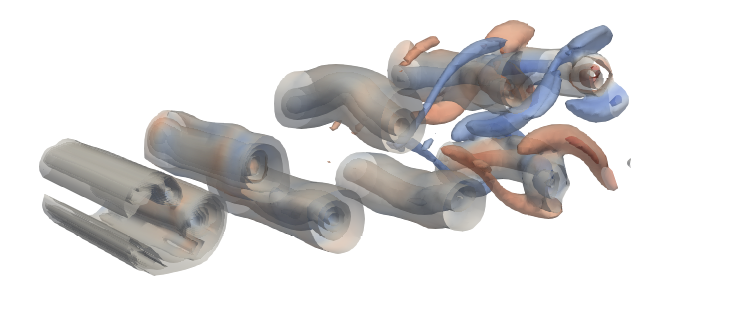}
  \end{subfigure}
  \begin{subfigure}[b]{0.24 \linewidth}
  \centering
    \includegraphics[width=0.96\linewidth, height=0.48\linewidth]{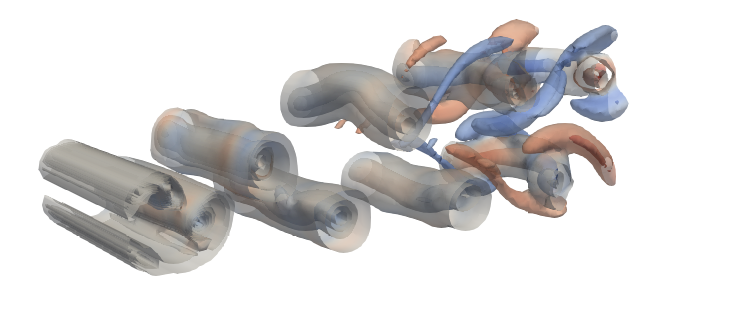}
  \end{subfigure}
 \begin{subfigure}[b]{0.24 \linewidth}
 \centering
   \includegraphics[width=0.96\linewidth, height=0.48\linewidth]{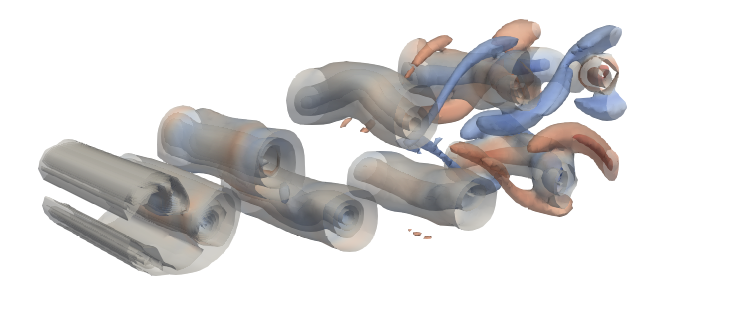}
 \end{subfigure}
 \begin{subfigure}[b]{0.24 \linewidth}
 \centering
   \includegraphics[width=0.96\linewidth, height=0.48\linewidth]{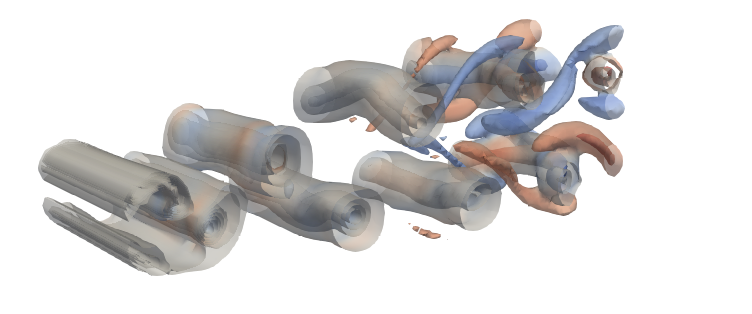}
 \end{subfigure}\vspace{1em}

  \begin{subfigure}[b]{0.24 \linewidth}
  \centering
    \includegraphics[width=0.96\linewidth, height=0.48\linewidth]{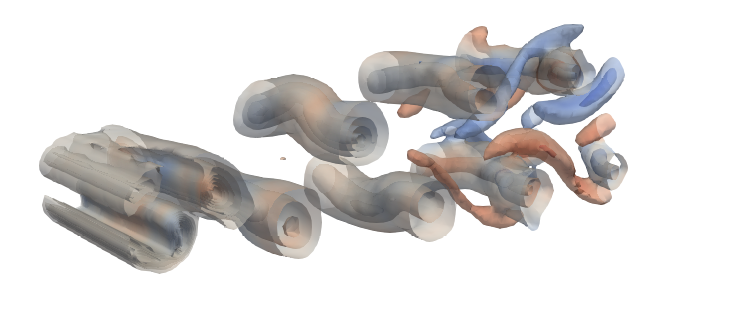}
  \end{subfigure}
 \begin{subfigure}[b]{0.24 \linewidth}
 \centering
   \includegraphics[width=0.96\linewidth, height=0.48\linewidth]{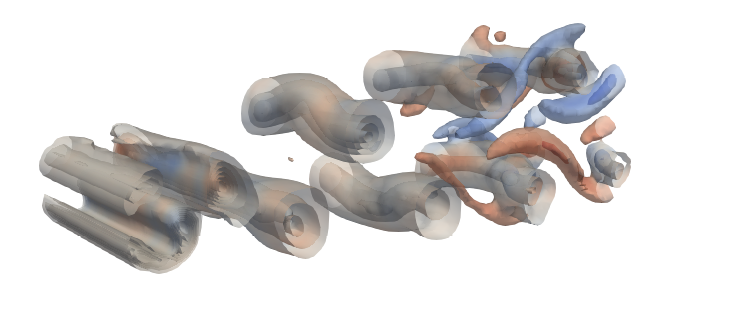}
 \end{subfigure}
 \begin{subfigure}[b]{0.24 \linewidth}
 \centering
   \includegraphics[width=0.96\linewidth, height=0.48\linewidth]{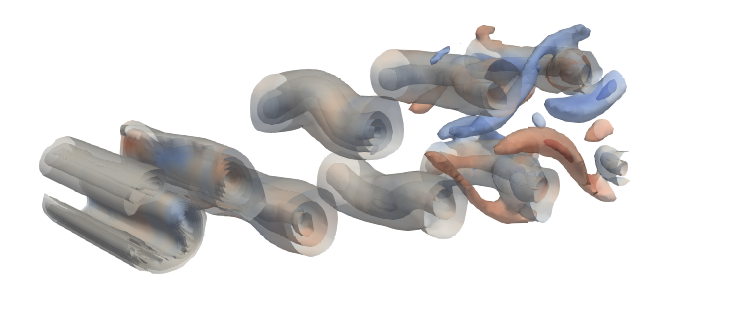}
 \end{subfigure}
  \begin{subfigure}[b]{0.24 \linewidth}
  \centering
    \includegraphics[width=0.96\linewidth, height=0.48\linewidth]{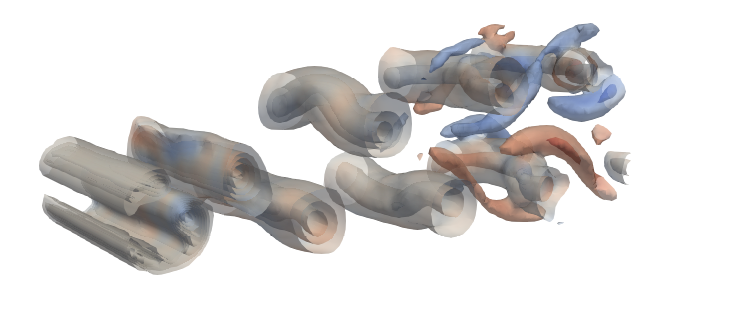}
  \end{subfigure}
  \caption{$\text{Re} = 250$}
\end{subfigure}\vspace{1em}

\begin{subfigure}[h]{0.9\linewidth}
    \begin{subfigure}[b]{0.24 \linewidth}
  \centering
    \includegraphics[width=0.96\linewidth, height=0.48\linewidth]{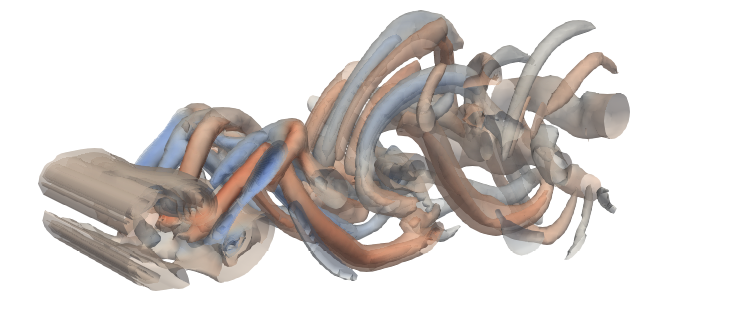}
  \end{subfigure}
  \begin{subfigure}[b]{0.24 \linewidth}
  \centering
    \includegraphics[width=0.96\linewidth, height=0.48\linewidth]{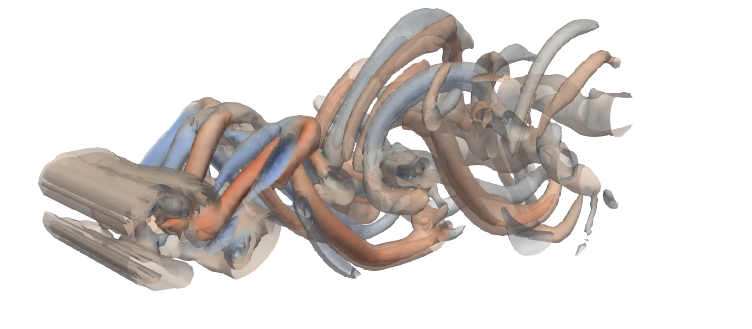}
  \end{subfigure}
 \begin{subfigure}[b]{0.24 \linewidth}
 \centering
   \includegraphics[width=0.96\linewidth, height=0.48\linewidth]{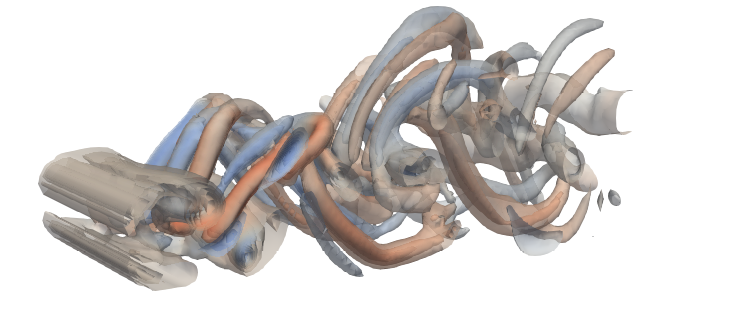}
 \end{subfigure}
 \begin{subfigure}[b]{0.24 \linewidth}
 \centering
   \includegraphics[width=0.96\linewidth, height=0.48\linewidth]{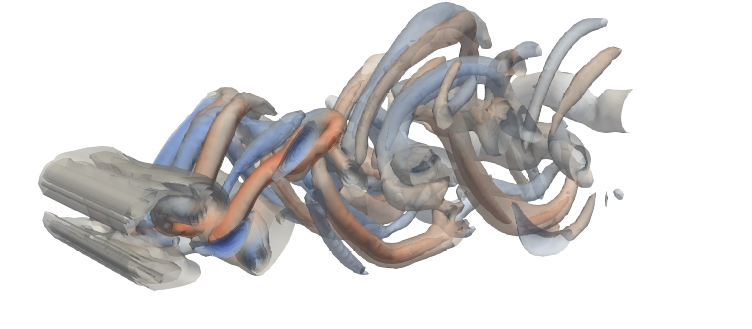}
 \end{subfigure}\vspace{1em}

  \begin{subfigure}[b]{0.24 \linewidth}
  \centering
    \includegraphics[width=0.96\linewidth, height=0.48\linewidth]{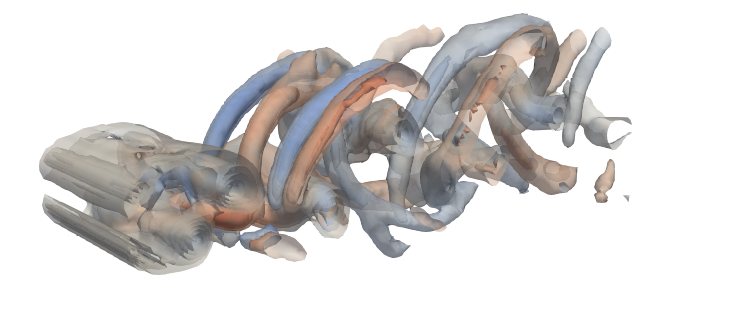}
  \end{subfigure}
 \begin{subfigure}[b]{0.24 \linewidth}
 \centering
   \includegraphics[width=0.96\linewidth, height=0.48\linewidth]{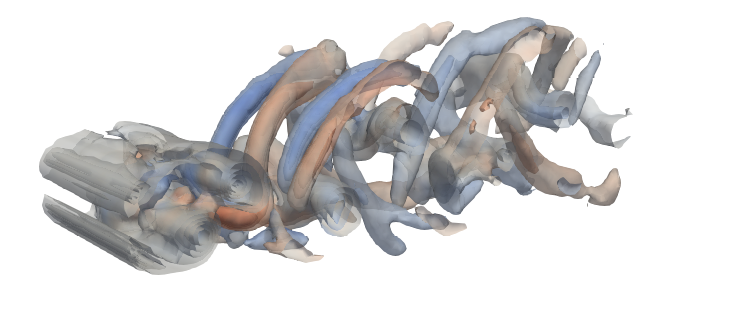}
 \end{subfigure}
 \begin{subfigure}[b]{0.24 \linewidth}
 \centering
   \includegraphics[width=0.96\linewidth, height=0.48\linewidth]{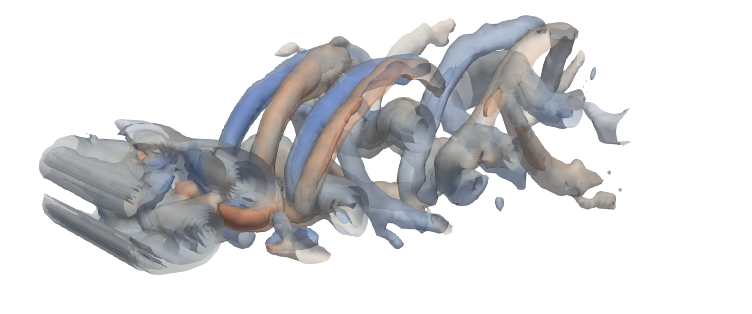}
 \end{subfigure}
  \begin{subfigure}[b]{0.24 \linewidth}
  \centering
    \includegraphics[width=0.96\linewidth, height=0.48\linewidth]{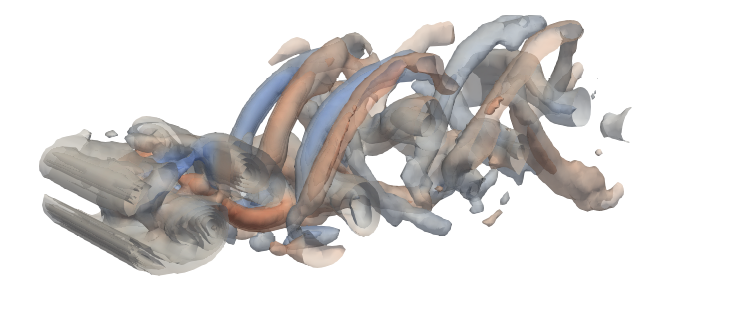}
  \end{subfigure}
  \caption{$\text{Re} = 300$}
\end{subfigure}
\caption[Visual comparison of $Q$-values over time at various flow conditions]{Comparison of iso-surfaces of $Q$-values, colored according to downstream vorticity, over time at four flow conditions. CFD solutions are shown in the top row, while generated solutions are shown in the bottom row.}
\label{fig: 3d_q}
\end{figure}

\begin{figure}
\centering
\begin{subfigure}[h]{0.9\linewidth}
    \begin{subfigure}[b]{0.24 \linewidth}
  \centering
    \includegraphics[width=0.96\linewidth, height=0.48\linewidth]{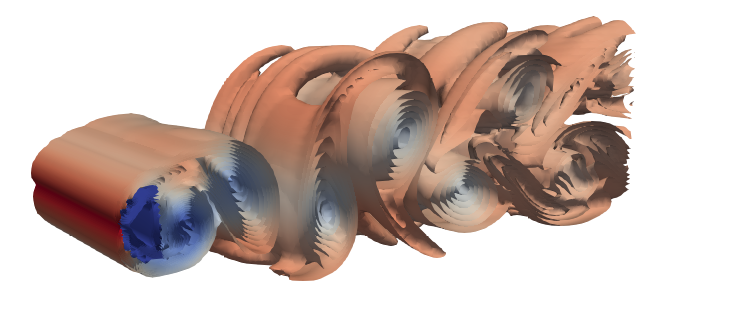}
  \end{subfigure}
  \begin{subfigure}[b]{0.24 \linewidth}
  \centering
    \includegraphics[width=0.96\linewidth, height=0.48\linewidth]{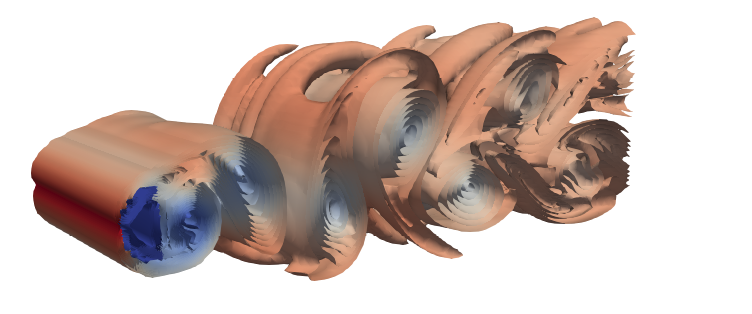}
  \end{subfigure}
 \begin{subfigure}[b]{0.24 \linewidth}
 \centering
   \includegraphics[width=0.96\linewidth, height=0.48\linewidth]{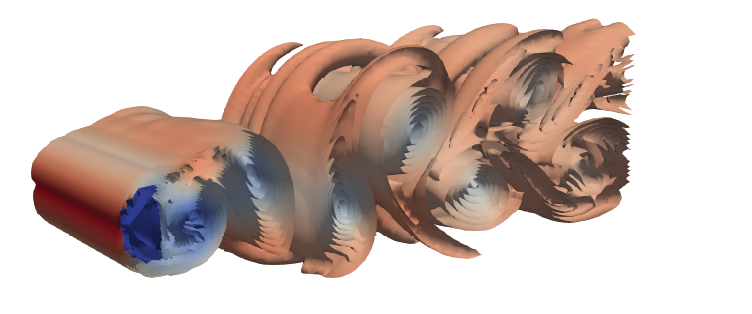}
 \end{subfigure}
 \begin{subfigure}[b]{0.24 \linewidth}
 \centering
   \includegraphics[width=0.96\linewidth, height=0.48\linewidth]{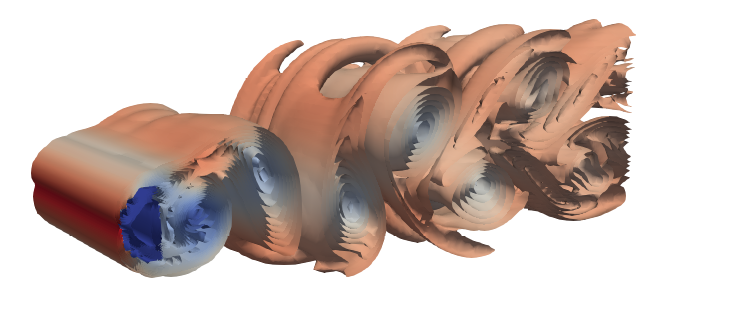}
 \end{subfigure}\vspace{1em}

  \begin{subfigure}[b]{0.24 \linewidth}
  \centering
    \includegraphics[width=0.96\linewidth, height=0.48\linewidth]{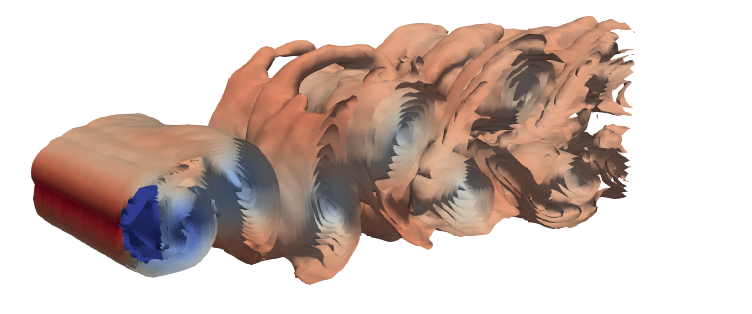}
  \end{subfigure}
 \begin{subfigure}[b]{0.24 \linewidth}
 \centering
   \includegraphics[width=0.96\linewidth, height=0.48\linewidth]{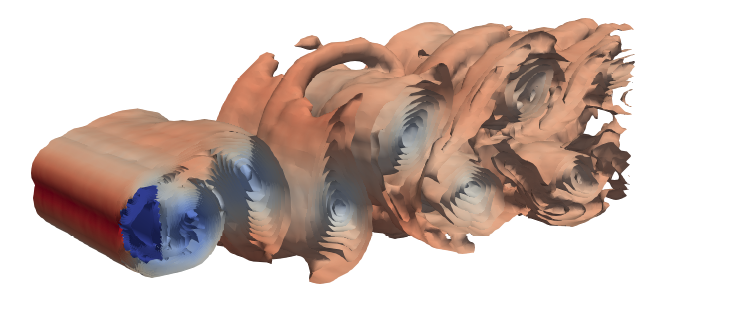}
 \end{subfigure}
 \begin{subfigure}[b]{0.24 \linewidth}
 \centering
   \includegraphics[width=0.96\linewidth, height=0.48\linewidth]{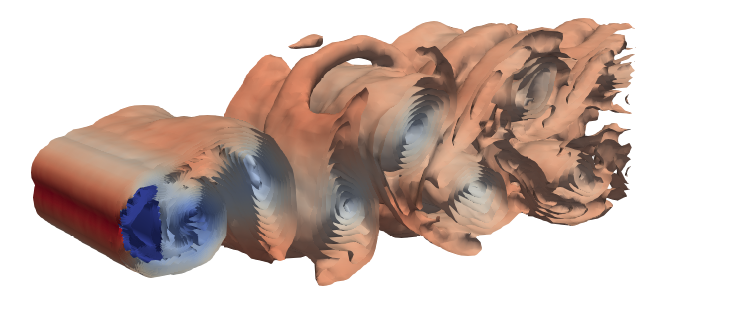}
 \end{subfigure}
  \begin{subfigure}[b]{0.24 \linewidth}
  \centering
    \includegraphics[width=0.96\linewidth, height=0.48\linewidth]{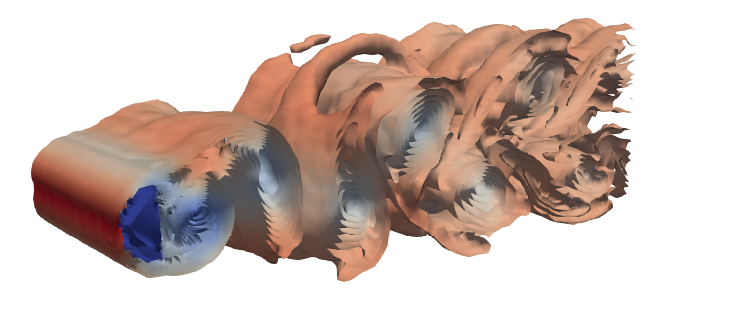}
  \end{subfigure}
  \caption{$\text{Re} = 175$}
\end{subfigure}\vspace{1em}

\begin{subfigure}[h]{0.9\linewidth}
    \begin{subfigure}[b]{0.24 \linewidth}
  \centering
    \includegraphics[width=0.96\linewidth, height=0.48\linewidth]{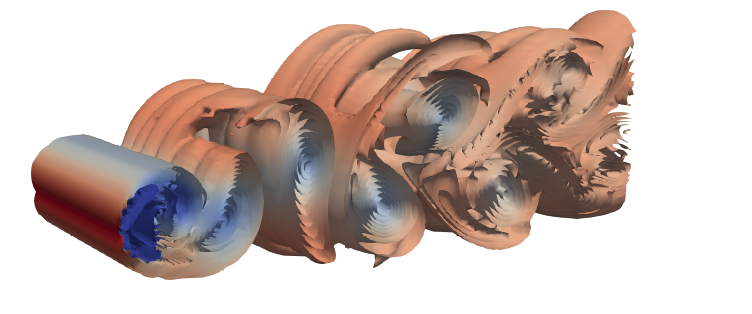}
  \end{subfigure}
  \begin{subfigure}[b]{0.24 \linewidth}
  \centering
    \includegraphics[width=0.96\linewidth, height=0.48\linewidth]{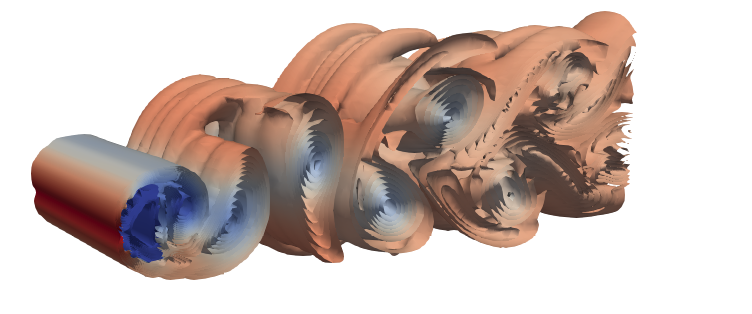}
  \end{subfigure}
 \begin{subfigure}[b]{0.24 \linewidth}
 \centering
   \includegraphics[width=0.96\linewidth, height=0.48\linewidth]{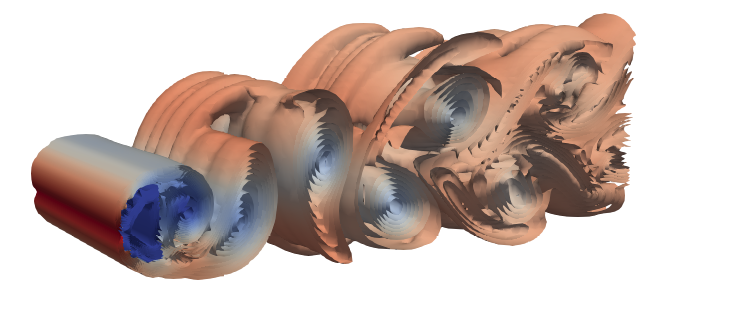}
 \end{subfigure}
 \begin{subfigure}[b]{0.24 \linewidth}
 \centering
   \includegraphics[width=0.96\linewidth, height=0.48\linewidth]{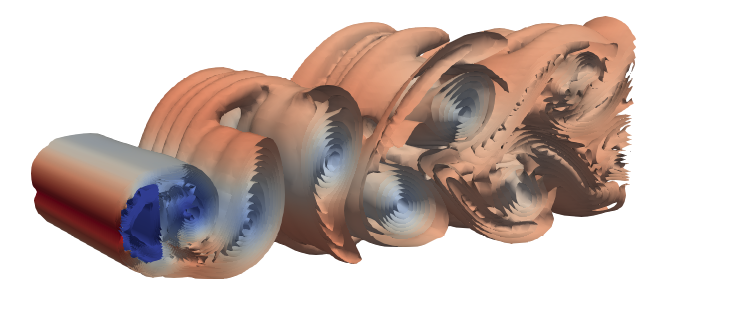}
 \end{subfigure}\vspace{1em}

  \begin{subfigure}[b]{0.24 \linewidth}
  \centering
    \includegraphics[width=0.96\linewidth, height=0.48\linewidth]{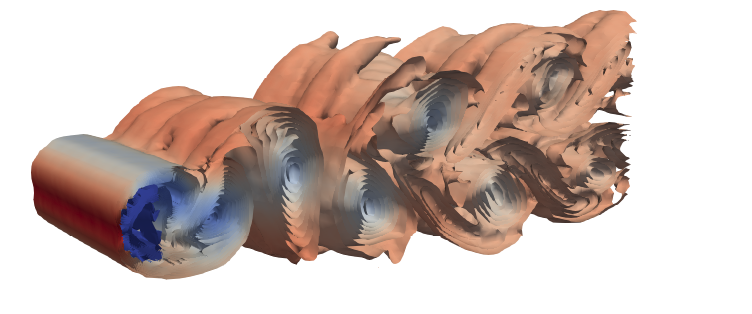}
  \end{subfigure}
 \begin{subfigure}[b]{0.24 \linewidth}
 \centering
   \includegraphics[width=0.96\linewidth, height=0.48\linewidth]{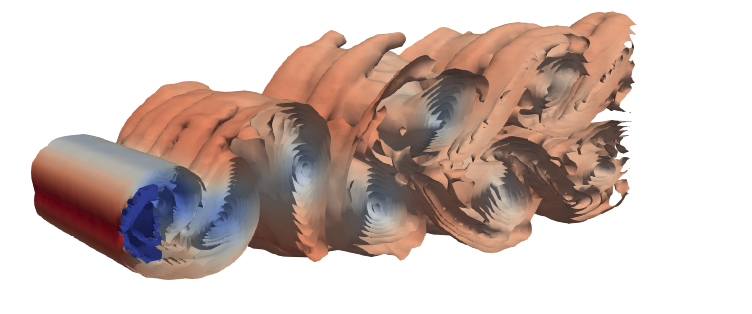}
 \end{subfigure}
 \begin{subfigure}[b]{0.24 \linewidth}
 \centering
   \includegraphics[width=0.96\linewidth, height=0.48\linewidth]{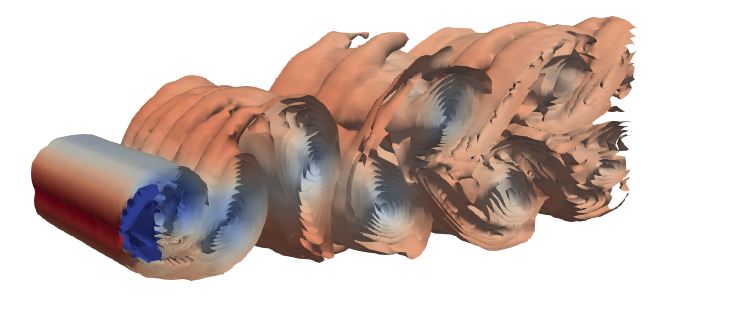}
 \end{subfigure}
  \begin{subfigure}[b]{0.24 \linewidth}
  \centering
    \includegraphics[width=0.96\linewidth, height=0.48\linewidth]{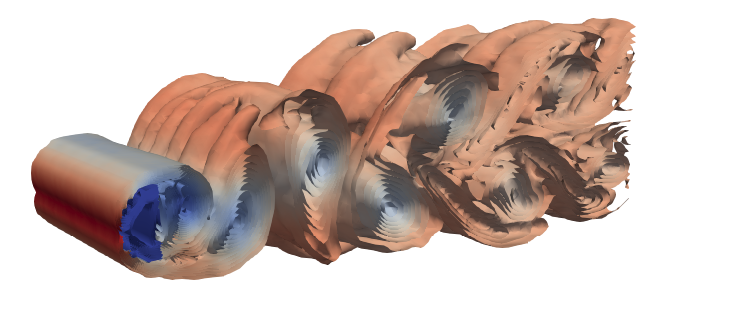}
  \end{subfigure}
  \caption{$\text{Re} = 200$}
\end{subfigure}\vspace{1em}

\begin{subfigure}[h]{0.9\linewidth}
    \begin{subfigure}[b]{0.24 \linewidth}
  \centering
    \includegraphics[width=0.96\linewidth, height=0.48\linewidth]{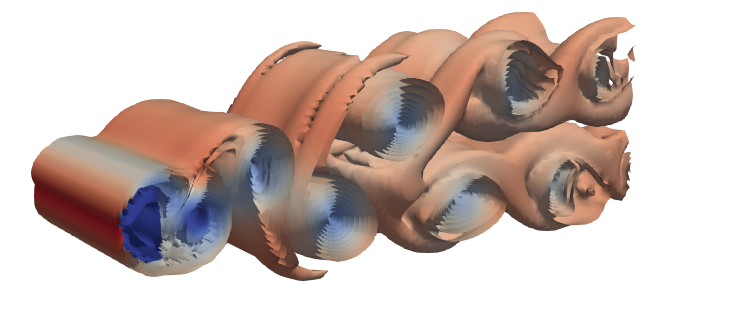}
  \end{subfigure}
  \begin{subfigure}[b]{0.24 \linewidth}
  \centering
    \includegraphics[width=0.96\linewidth, height=0.48\linewidth]{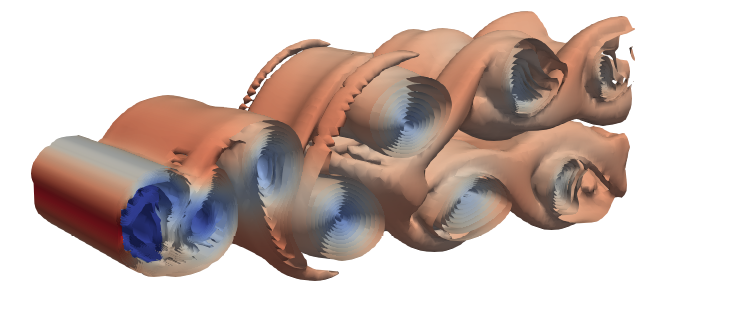}
  \end{subfigure}
 \begin{subfigure}[b]{0.24 \linewidth}
 \centering
   \includegraphics[width=0.96\linewidth, height=0.48\linewidth]{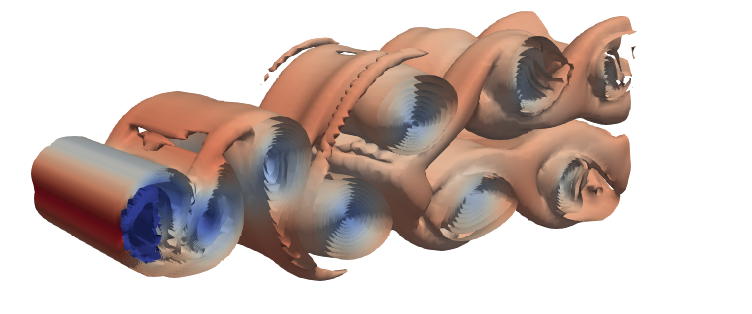}
 \end{subfigure}
 \begin{subfigure}[b]{0.24 \linewidth}
 \centering
   \includegraphics[width=0.96\linewidth, height=0.48\linewidth]{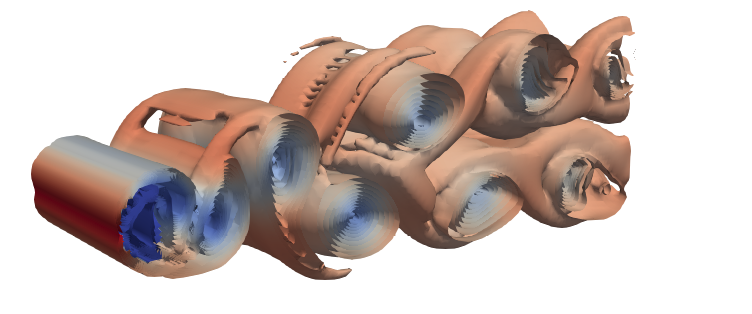}
 \end{subfigure}\vspace{1em}

  \begin{subfigure}[b]{0.24 \linewidth}
  \centering
    \includegraphics[width=0.96\linewidth, height=0.48\linewidth]{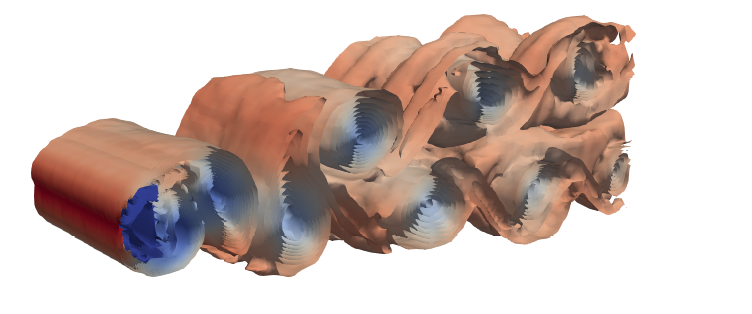}
  \end{subfigure}
 \begin{subfigure}[b]{0.24 \linewidth}
 \centering
   \includegraphics[width=0.96\linewidth, height=0.48\linewidth]{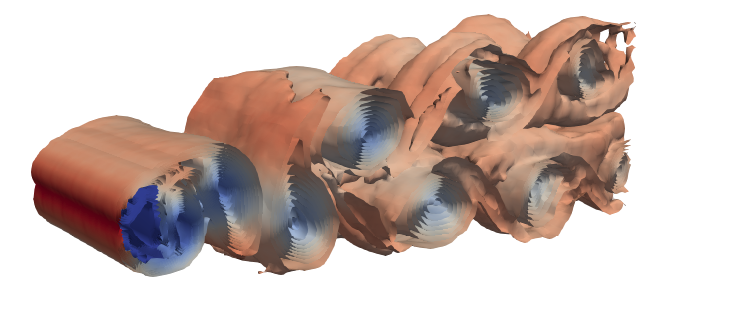}
 \end{subfigure}
 \begin{subfigure}[b]{0.24 \linewidth}
 \centering
   \includegraphics[width=0.96\linewidth, height=0.48\linewidth]{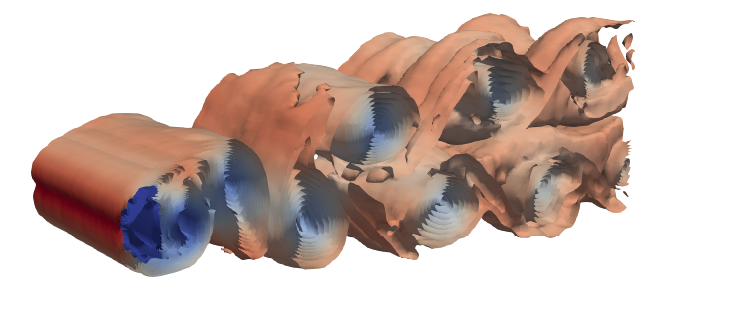}
 \end{subfigure}
  \begin{subfigure}[b]{0.24 \linewidth}
  \centering
    \includegraphics[width=0.96\linewidth, height=0.48\linewidth]{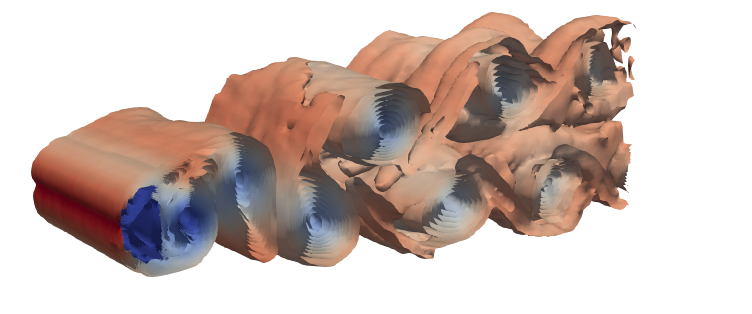}
  \end{subfigure}
  \caption{$\text{Re} = 250$}
\end{subfigure}\vspace{1em}

\begin{subfigure}[h]{0.9\linewidth}
    \begin{subfigure}[b]{0.24 \linewidth}
  \centering
    \includegraphics[width=0.96\linewidth, height=0.48\linewidth]{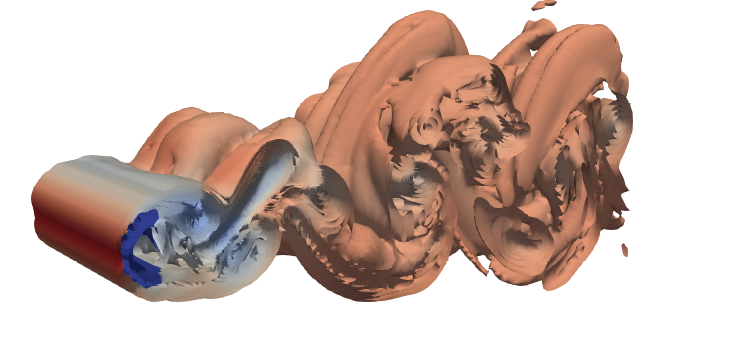}
  \end{subfigure}
  \begin{subfigure}[b]{0.24 \linewidth}
  \centering
    \includegraphics[width=0.96\linewidth, height=0.48\linewidth]{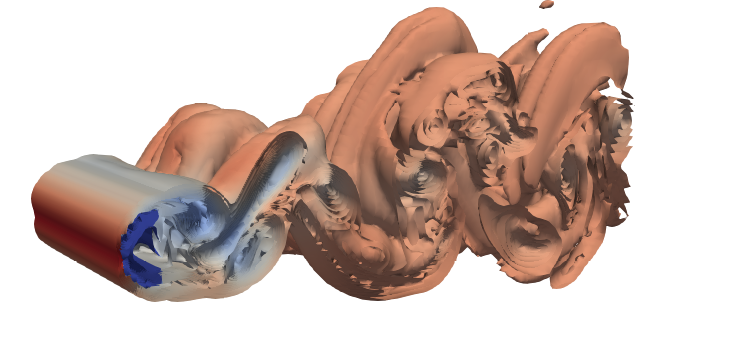}
  \end{subfigure}
 \begin{subfigure}[b]{0.24 \linewidth}
 \centering
   \includegraphics[width=0.96\linewidth, height=0.48\linewidth]{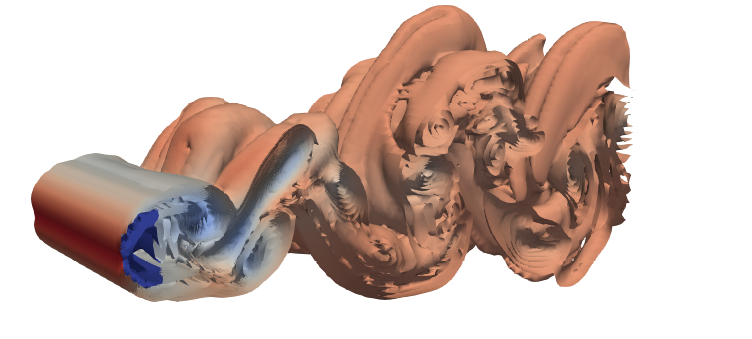}
 \end{subfigure}
 \begin{subfigure}[b]{0.24 \linewidth}
 \centering
   \includegraphics[width=0.96\linewidth, height=0.48\linewidth]{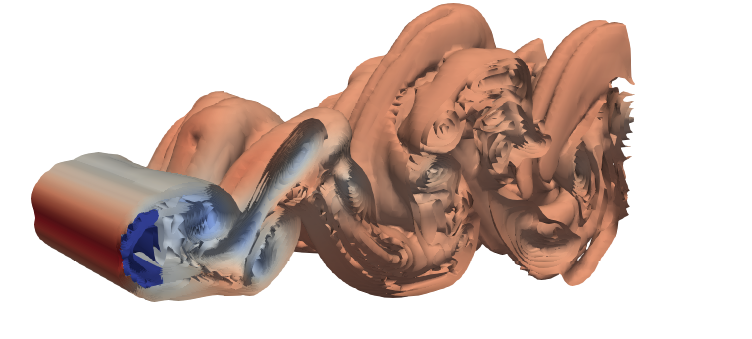}
 \end{subfigure}\vspace{1em}

  \begin{subfigure}[b]{0.24 \linewidth}
  \centering
    \includegraphics[width=0.96\linewidth, height=0.48\linewidth]{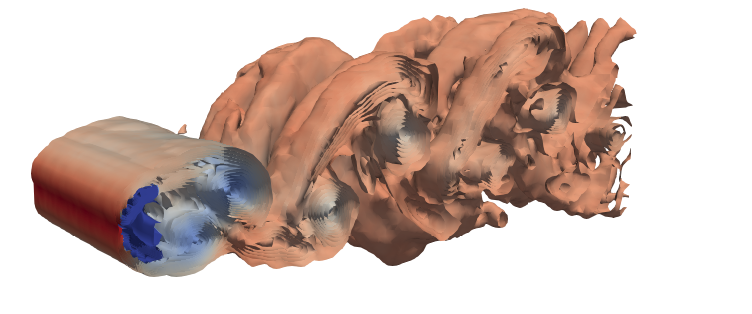}
  \end{subfigure}
 \begin{subfigure}[b]{0.24 \linewidth}
 \centering
   \includegraphics[width=0.96\linewidth, height=0.48\linewidth]{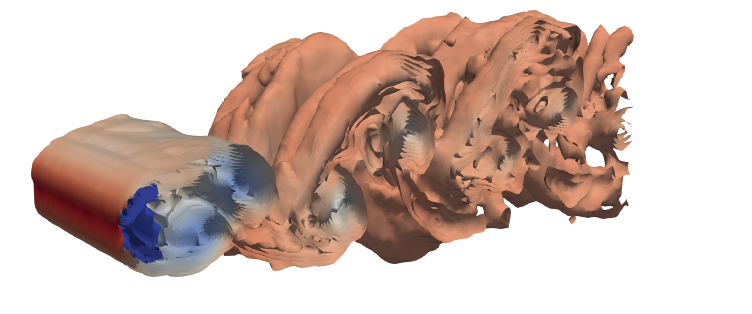}
 \end{subfigure}
 \begin{subfigure}[b]{0.24 \linewidth}
 \centering
   \includegraphics[width=0.96\linewidth, height=0.48\linewidth]{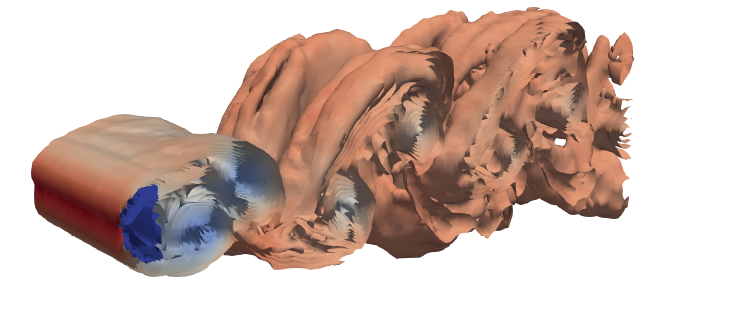}
 \end{subfigure}
  \begin{subfigure}[b]{0.24 \linewidth}
  \centering
    \includegraphics[width=0.96\linewidth, height=0.48\linewidth]{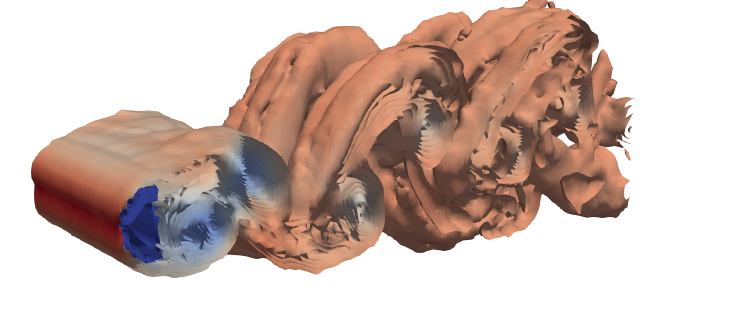}
  \end{subfigure}
  \caption{$\text{Re} = 300$}
\end{subfigure}
\caption[Visual comparison of vorticity over time at various flow conditions]{Comparison of iso-surfaces of vorticity magnitude, colored according to pressure, over time at four flow conditions. CFD solutions are shown in the top row, while generated solutions are shown in the bottom row.}
\label{fig: 3d_vort}
\end{figure}

\subsection{Three-Dimensional Flow Results}

The results presented here are solely meant to demonstrate that the generative modeling techniques can scale to larger problems; an in-depth quantitative evaluation of the generated results is reserved for future work.
Therefore, only a qualitative evaluation of the generated solutions is presented in this section.
\Cref{fig: 3d_q,fig: 3d_vort} present a visual comparison between the CFD and generated solutions at the four flow conditions contained within the training data.
For each flow condition, the top row of images contains results from the CFD simulations, while the bottom row contains results sampled from the generative model.
\Cref{fig: 3d_q} shows iso-surfaces of $Q$, the second invariant of the velocity gradient tensor, colored according to vorticity in the downstream direction.
\Cref{fig: 3d_vort} shows iso-surfaces of vorticity magnitude, colored according to pressure.

A strong visual similarity can be observed between the CFD solutions and the solutions sampled from the generative model.
In particular, the generative model seems effective at capturing the variation in the quantity and nature of vortical structures with Reynolds number.
It is worth emphasizing that the iso-surfaces in all images are functions of spatial derivatives of the flow velocities.
These spatial derivatives are estimated through finite difference, and thus the presence of the iso-surfaces is strongly reliant upon accurately modeling the manner in which the velocity field varies in the cylinder wake.
Hence, it is noteworthy that the trained model is capable of generating flow fields that not only contain these vortical structures, but also retain them across successive time steps while propagating them downstream.

\section{Conclusions}
This paper presented a method for learning parameter-conditioned sequential generative models.
The central element of the proposed method is a variational inference procedure that enables the discovery of low-dimensional latent states capable of accurately capturing the dynamics of a modeled system.
These inferred latent representations can be conditioned on parameters that are assumed to govern the modeled system's behavior, thereby enabling the generative modeling of dynamical systems, where the modeled dynamics are a function of prescribed parameters.

The proposed method was first evaluated based on its ability to simulate two-dimensional airflow over a pair of counter-rotating cylinders.
Extensive qualitative and quantitative experiments demonstrated that learned generative models were capable of effectively modeling both local and global properties of the flow field at a wide range of flow conditions.
Furthermore, it was found that, relative to a CFD solver, a speedup of approximately $120\times$ can be obtained by running simulations with a trained generative model.
A final set of experiments with data from three-dimensional, turbulent flow simulations demonstrated that the generative modeling techniques can scale to complex and high-dimensional problems.

There are some limitations to the proposed approach that should be addressed in future work.
First, in all experiments it was assumed that the flow geometries were fixed.
However, many aerospace design problems, such as airfoil or wing design, focus predominantly on how changes in geometry affect the resulting flow.
At the moment it is unclear how well generative models would perform if, for example, the separation between the cylinders were allowed to vary, thereby altering the location of the cylinders within the flow field.
Hence, more work is needed to identify and address potential shortcomings in this regard.
One further limitation is that the employed neural network architectures require the flow quantities to be sampled at uniformly spaced points.
Future work should be dedicated to overcoming this restriction, which will make the techniques better suited to modeling fluid flows from simulations with unstructured grids.

\section{Acknowledgments}
This material is based upon work supported by the Stanford Strategic Energy Alliance.
The authors would like to thank Antony Jameson for valuable guidance and feedback.


\printbibliography

\end{document}